\documentclass[twocolumn,Astrosymb]{aastex63}

\usepackage{amsmath}

\newcommand{\kms} {\rm{km~s}$^{-1}$} 
\newcommand{\ci} {\rm{cm}$^{-1}$}  
\newcommand{\csi} {\rm{cm}$^{-2}$}  
\newcommand{\si} {\rm{s}$^{-1}$}  

\newcommand{\acet} {$\mathrm{C}_{2}\mathrm{H}_{2}$}
\newcommand{\acetiso} {$^{13}\mathrm{CCH}_{2}$}
\newcommand{\amm}{$\mathrm{NH}_{3}$}
\newcommand{\meth} {$\mathrm{CH}_{4}$}
\newcommand{\sotwo}{$\mathrm{SO}_{2}$}
\newcommand{\htwo} {$\mathrm{H}_{2}$}
\newcommand{\nhtwo}{$N_{\mathrm{H}_2}$}
\newcommand{\water}{$\mathrm{H}_{2}\mathrm{O}$}
\newcommand{\hcniso}{$\mathrm{H}^{13}\mathrm{CN}$}

\newcommand{\vfwhm}{$v_{\mathrm{FWHM}}$}
\newcommand{\vlsr}{$v_{\mathrm{LSR}}$}

\shorttitle{Orion molecular inventory}
\shortauthors{Nickerson et al.}
\begin{document}


\title{The mid-infrared molecular inventory towards Orion IRc2}

\correspondingauthor{Sarah Nickerson}
\email{sarah.nickerson@nasa.gov}

\author[0000-0002-7489-3142]{Sarah Nickerson}
\affiliation{Space Science and Astrobiology Division, NASA Ames Research Center, Moffett Field, CA, 94035 USA}
\affiliation{Bay Area Environmental Research Institute, Moffett Field, CA, 94035, USA}

\author[0000-0001-9920-7391]{Naseem Rangwala}
\affiliation{Science Directorate, NASA Ames Research Center, Moffett Field, CA, 94035 USA}

\author[0000-0001-6275-7437]{Sean W. J. Colgan}
\affiliation{Space Science and Astrobiology Division, NASA Ames Research Center, Moffett Field, CA, 94035 USA}

\author[0000-0002-6528-3836]{Curtis DeWitt}
\affiliation{USRA, SOFIA, NASA Ames Research Center MS 232-11, Moffett Field, CA 94035, USA}

\author[0000-0002-9986-4604]{Jose S. Monzon}
\affiliation{Department of Astronomy,Yale University, P.O. Box 208101, New Haven, CT 06520-8101, USA}

\author[0000-0003-2458-5050]{Xinchuan Huang}
\affiliation{Space Science and Astrobiology Division, NASA Ames Research Center, Moffett Field, CA, 94035 USA}
\affiliation{SETI Institute, 339 Bernardo Ave, Suite 200, Mountain View, CA, 94043 USA}

\author[0000-0002-0603-8777]{Kinsuk Acharyya}
\affiliation{Planetary Science Division, Physical Research Laboratory, Ahmedabad, 380009, India}

\author[0000-0001-7479-4948]{Maria N. Drozdovskaya}
\affiliation{Center for Space and Habitability, University of Bern, Gesellschaftsstrasse 6, CH-3012 Bern, Switzerland}

\author[0000-0003-4716-8225]{Ryan C. Fortenberry}
\affiliation{Department of Chemistry and Biochemistry, University of Mississippi, 38677 USA}

\author[0000-0002-4649-2536]{Eric Herbst}
\affiliation{Departments of Chemistry and Astronomy, University of Virginia, McCormick Rd, Charlottesville, VA, 22904 USA}

\author[0000-0002-2598-2237]{Timothy J. Lee}
\affiliation{Space Science and Astrobiology Division, NASA Ames Research Center, Moffett Field, CA, 94035 USA}

\begin{abstract}


We present the first high spectral resolution mid-infrared survey in the Orion BN/KL region, covering 7.2 to 28.3 \micron. With SOFIA/EXES we target the enigmatic source Orion IRc2. While this is in the most prolifically studied massive star-forming region, longer wavelengths and molecular emission lines dominated previous spectral surveys.  The mid-infrared observations in this work access different components and molecular species in unprecedented detail. We unambiguously identify two new kinematic components, both chemically rich with multiple molecular absorption lines. The ``blue clump'' has \vlsr$\;=-7.1\pm0.7$ \kms\ and the ``red clump'' $1.4\pm0.5$ \kms. While the blue and red clumps have similar temperatures and line widths, molecular species in the blue clump have higher column densities. They are both likely linked to pure rotational \htwo\ emission also covered by this survey. This work provides evidence for the scenario that the blue and red clumps are distinct components unrelated to the classic components in the Orion BN/KL region. Comparison to spectroscopic surveys towards other infrared targets in the region show that the blue clump is clearly extended. We analyze, compare, and present in depth findings on the physical conditions of \acet, \acetiso, \meth, CS, \water, HCN, \hcniso, HNC, \amm, and \sotwo\ absorption lines  and an \htwo\ emission line associated with the blue and red clumps. We also provide limited analysis of \water\ and SiO molecular emission lines towards Orion IRc2 and the atomic forbidden transitions [FeII], [SI], [SIII], and [NeII]. 


\end{abstract}

\section{Introduction} \label{sec:intro}

\begin{figure}
\centering
\includegraphics[height=0.29\textheight]{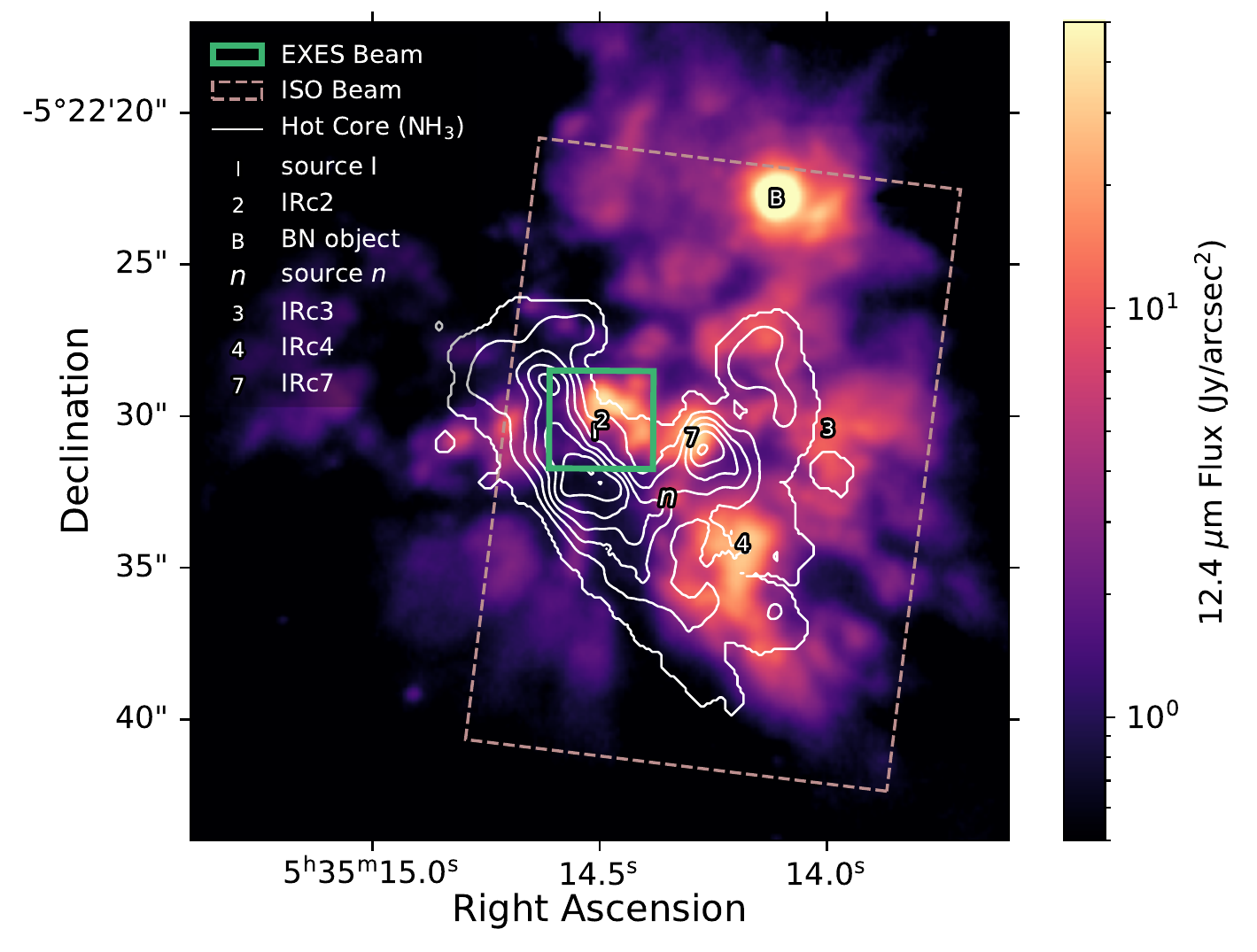}
\caption{Mid-infrared image of the BN/KL region. Colour map is the 12.4 \micron\ flux \citep[SUBARU/COMICS,][]{Okumura2011}. White contours are the hot core traced by the \amm\ inversion transition $(J,K)=(7,7)$ \citep[NRAO/EVLA,][]{Goddi2011}, and each level is 75 mJy beam$^{-1}$\kms. The thick green box is the EXES beam for the 7.6a \micron\ setting. The dashed brown box is the smallest beam used in the \textit{ISO} survey \citep{VanDishoeck1998}. Symbols refer to regional features discussed in this work. The position of ``2'' for IRc2 is at the EXES beam centre for all settings in this survey.  The location of BN is from \citet{Gomez2005}. All other symbols, source I, source \textit{n}, IRc3, IRc4, and IRc7, are placed according to \citet{Okumura2011}. Refer to Figure \ref{fig:mapzoom} for a closeup of the IRc2 region. \label{fig:bigmap}}
\end{figure}

The closest and best studied massive star-forming region is the Orion Molecular Cloud 1 \citep[OMC-1,][]{Genzel1989} with a distance of $418\pm6$ pc \citep{Kim2008}. The Becklin-Neugebauer/Kleinmann-Low region \citep[Orion BN/KL,][]{Becklin1967,Kleinmann1967} within OMC-1 has been the subject of numerous molecular emission line surveys across several decades, ranging from the far infrared (FIR) to radio spectroscopy \citep[e.g.][refer to \citealt{Gong2015} for a more complete list]{Johansson1984,Blake1987,Sutton1995,Schilke1997,Schilke2001,Comito2005,Lerate2006,Olofsson2007,Tercero2010,Crockett2014,Feng2015,Gong2015,Rizzo2017,Luo2019,Peng2019}.

The emission lines in these previous FIR to radio surveys are typically divided into four classic components that each have distinct central velocities and line widths. The extended ridge is the quiescent, ambient gas of the molecular cloud. The compact ridge is a smaller region of denser, hotter quiescent gas separate from the extended ridge. The plateau is an outflow of shocked gas that may be further subdivided into a high and a low velocity flow. Finally, the hot core is the hot, dense gas rich in molecular species  \citep{Blake1987,Genzel1989}. 

Figure \ref{fig:bigmap} provides an overview of the Orion BN/KL region, indicating objects discussed in the following paragraphs and throughout this work. The region's complex morphology may be due to a recent explosive event. About 500 years ago, a multi-body encounter between the massive protostars radio source I, BN, and source \textit{n} ejected the three objects and launched a massive outflow of gas from the Orion BN/KL region \citep{Bally2011,Bally2015,Bally2017}. Source I is a heavily embedded protostar candidate with no infrared counterpart \citep{Churchwell1987,Greenhill2004}, that drives regional outflows and masers \citep{Menten1995,Greenhill1998,Plambeck2009,Hirota2017,Wright2020,Wright2022}. 

Concurrent with sources I and \textit{n} lies the Orion hot core. Despite being the first discovered and eponymous hot core \citep{Ho1979}, the nature of the hot core is atypical. Hot cores are small regions of warm, dense, molecular-rich gas associated with massive star formation. Their molecular richness arises from the evaporation of the molecular species off the icy dust grains in the cold molecular clouds \citep{VanDishoeck1998a,Kurtz2000,VanDerTak2004,Cesaroni2005,Beltran2018}. Hot cores are most commonly internally heated by massive protostars \citep{VanDerTak2004} though externally heated hot cores also exist \citep[e.g.][]{Mookerjea2007,Qin2022}.  Evidence suggests that the Orion hot core is externally heated \citep{Blake1996,Orozco-Aguilera2017}, possibly as a preexisting dense region of gas heated by either source I or the explosive event \citep{Goddi2011,Zapata2011,Wright2017}. A few works argue that the hot core could be internally heated by an embedded protostar due to its high densities and temperatures \citep{Kaufman1998,DeVicente2002,Wilkins2022}.

Spectroscopic observations in the mid-infrared (MIR) are significantly fewer than longer wavelength surveys of Orion BN/KL. However, rovibrational transitions and molecules with no permanent dipole moment are uniquely observable in the MIR. Observations towards a MIR-bright source reveal absorption lines originating in the gas between the source and the observer. This creates a pencil-beam effect in which the gas is probed in an area equivalent to the source, which can be much smaller than the telescope's beam size. The bright MIR source Orion IRc2 probes the edge of the hot core (Figure \ref{fig:bigmap}), either illuminating it from behind or coincident with it \citep{Shuping2004}. The hot core's \amm\ column density at IRc2 is about $30\pm10$\% of the value at the hot core's peak \citep{Genzel1982,Wynn-Williams1984}.

Orion IRc2 was first identified as a compact infrared source, the second brightest in the region after BN in the MIR \citep{Rieke1973}. Its nature remains unclear. Orion BN/KL may be a hollow nebula in which IRc2 is a cavity \citep{Wynn-Williams1984}. High spatial resolution imaging at 12.5 \micron\ shows that IRc2 is U-shaped and breaks down into four point-like sources \citep{Shuping2004}. The temperature distribution reveals a gradient that peaks close to source I, suggesting that IRc2 may be externally heated by it \citep{Okumura2011}. Polarimetry, however, suggests that source \textit{n} may be illuminating IRc2 instead \citep{Simpson2006}.

Ground-based, high resolution MIR spectroscopy towards Orion IRc2 contributed to the discovery of interstellar \acet\ and \meth\ \citep{Lacy1989,Lacy1991} with IRSHELL at the NASA Infrared Telescope Facility (IRTF) \citep[resolution $R\sim\ $10,000,][]{Lacy1989}. Further IRTF/IRSHELL  observations towards Orion IRc2 detected \acet, \acetiso, HCN, OCS, CO, and \amm\ \citep{Evans1991,Carr1995}. The Short-Wavelength Spectrometer (SWS) aboard space-based \textit{ISO} \citep[$R\sim\ $1,500,][]{DeGraauw1996} covered the entire MIR spectrum towards IRc2, from 2.4 to 45.2 \micron. That work detected numerous features, including in absorption HCN and \acet, as well as in emission \htwo, \meth, and \sotwo. \water\ is observed in both emission and absorption \citep{VanDishoeck1998,Wright2000,Boonman2003}. These space-based spectral observations, while broad in coverage, were low resolution and most absorption lines were blended into larger features. Inversely, the ground-based IRSHELL observations resolved individual transitions and had a better spatial resolution, but were limited by small coverage and interference from the Earth's atmosphere.  These earlier studies had insufficient data to pinpoint the origins of the molecular MIR absorption lines towards IRc2.

In this work, we present the first high resolution MIR line survey in the Orion BN/KL region towards IRc2, with nearly continuous coverage from from 7.2 to 8 \micron\ and 12.8 to 28.3  \micron\  taken with the EXES instrument \citep[$R\sim\ $60,000,][]{Richter2018} onboard the SOFIA observatory \citep{Young2012}. We supplement this SOFIA/EXES survey with a small amount of IRTF/TEXES \citep[$R\sim\ $100,000,][]{Lacy2002} data from 11.7 to 11.9 \micron. 

\begin{figure}
\centering
\fig{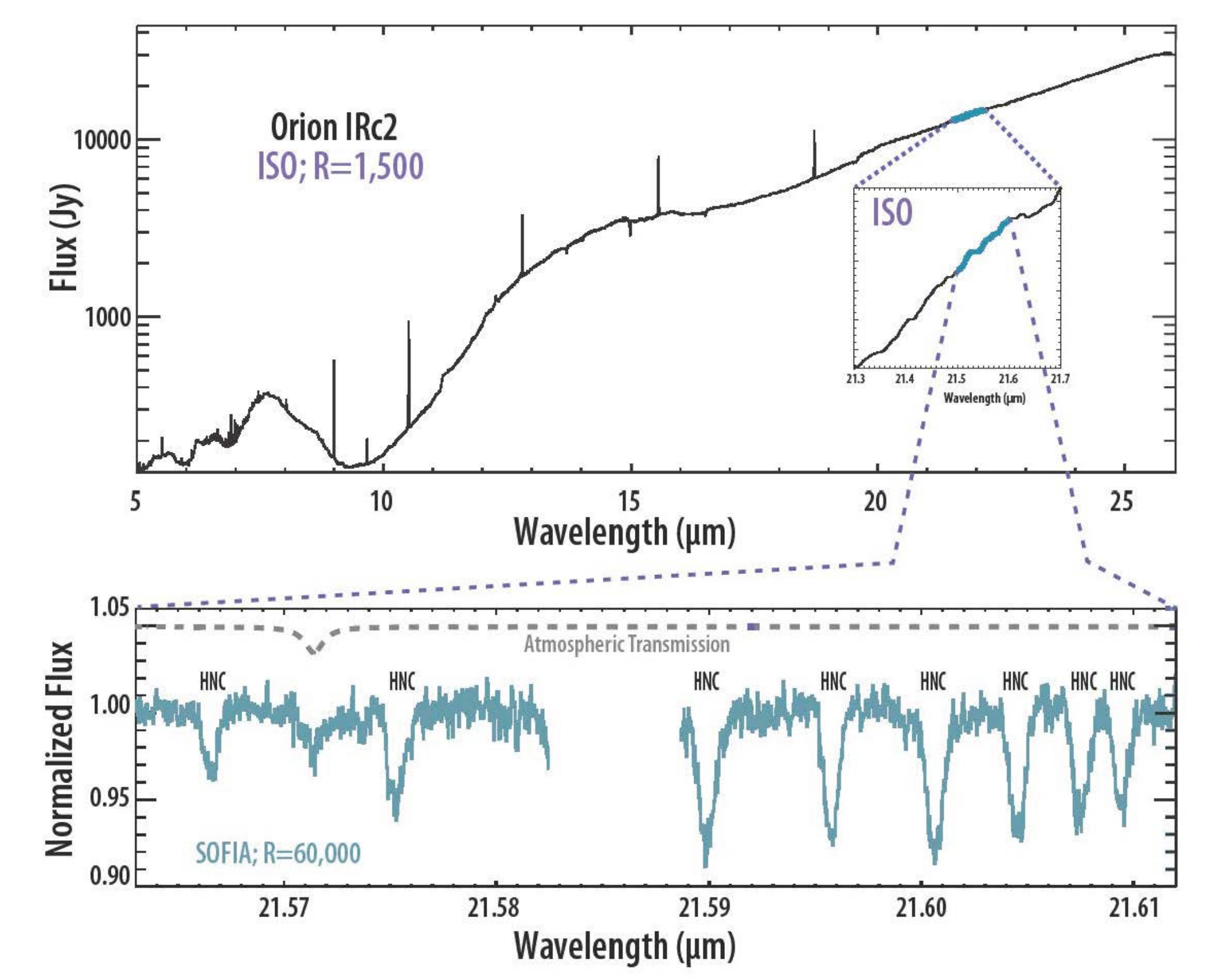}{0.49\textwidth}{}
\caption{Comparison of resolution between MIR surveys towards Orion IRc2. Top: \textit{ISO}/SWS, resolution $\sim$ 1,500 \citep{VanDishoeck1998}, which will be similar to \textit{JWST}/MIRI at this wavelength. Bottom: a segment of spectra with HNC absorption lines taken with SOFIA/EXES as part of this survey, resolution $\sim$ 60,000. With \textit{JWST}/MIRI, these lines would be indiscernible from the continuum.  \label{fig:isovssofia} Credit: NASA/SOFIA/M. Rose/N. Rangwala.}
\end{figure}

Figure \ref{fig:bigmap} shows EXES's much smaller beam size compared to that of \textit{ISO}/SWS \citep{VanDishoeck1998}. While EXES is able to isolate Orion IRc2, several MIR-bright objects fall within SWS's beam. Figure \ref{fig:isovssofia} compares resolution between EXES and SWS, illustrating that EXES has about 30 times higher spectral resolution than SWS and \textit{JWST}/MIRI.

Our results build on previous 12.96 to 13.33 \micron\ SOFIA/EXES observations towards IRc2 \citep{Rangwala2018} and a segment of this survey was previously published covering HCN and the first MIR detections of HNC and \hcniso\ in the interstellar medium \citep{Nickerson2021}. This work complements EXES and TEXES observations towards the conventional protostar-harbouring hot cores AFGL 2591, AFGL 2136 \citep[4 to 13 \micron,][]{Indriolo2015,Barr2020,Indriolo2020,Barr2022}, NGC 7538 IRS 1 \citep[7.6 to 13.7 \micron,][]{Knez2009}, and Mon R2 \citep[7.23 to 7.38 \micron,][]{Dungee2018}. Combined, these works and this present work provide a unique window into the chemistry and physical conditions in the massive star-forming regions that may shed light on our own Solar System's origins \citep{Adams2010}.

In \S \ref{sec:obs} we describe this survey's observations and \S \ref{sec:res} we detail analysis and provide results. In \S \ref{sec:dicu} we discuss the this survey's findings. \S \ref{sec:con} concludes this work.

\section{Observations} \label{sec:obs}
We observed Orion IRc2 with the EXES instrument aboard the SOFIA observatory between 2017 March 17 and 22, 2018 October 26 and November 11, and 2020 February 6 and 7, at altitudes normally above 39,000 ft. SOFIA flies above 99\% of the Earth's atmospheric water vapour, covering wavelengths that are inaccessible from the ground. 

All settings were taken in High-low mode, with the exception of the 7.3 \micron\ setting in High-medium mode. Spectra were acquired in the cross-dispersed high-resolution mode with a slit width of $3\farcs2$, giving a resolving power of about 60,000 ($\sim$ 5 \kms). We used the cross-disperser grating in 1st order to obtain the broadest simultaneous wavelength coverage per spectral setting. The slit length varied between $2\farcs2$ and $12\farcs5$, depending on the spectral setting. Table \ref{tab:obs} gives the details for these settings, each of which is divided into several orders. For all observations, we nodded the telescope to an off-source position relatively free of emission 15\arcsec\ East  and 25\farcs9 North of IRc2, at 1 minute intervals, in order to remove sky emission and thermal background from the telescope system. Two settings appear twice: 7.6a/b \micron\ and 17.7a/b \micron. The 7.6a \micron\ setting was  observed in 2018 October and we noticed asymmetries in the atmospheric \meth\ lines. We revisited this setting in 2020 February (7.6b \micron) with a Doppler shift favourable to capturing the \meth\ lines. Accordingly, we use only 7.6b for \meth\ analysis. The 17.7 \micron\ setting was observed on two separate nights, but because no molecular lines were observed we did not go further with analysis to combining the two nights. 

\begin{deluxetable*}{cccccccc}
\tablecaption{Specifications for each setting\label{tab:obs}}
\tablehead{\colhead{Setting} & \colhead{Species} & \colhead{Min $\lambda$} & \colhead{Max $\lambda$} &
\colhead{Date} & \colhead{Configuration} & \colhead{Slit Length} & \colhead{Integration Time} \\
\colhead{(\micron)} & \colhead{} & \colhead{(\micron)} & \colhead{(\micron)} &
\colhead{(yyyy-mm-dd)} & \colhead{} & \colhead{(\arcsec)} & \colhead{(s)}}
\startdata
\multicolumn{8}{c}{SOFIA/EXES (slit width: 3.2\arcsec)}\\
7.3 & \acet, \water $^*$, HCN, \sotwo & 7.2 & 7.3 & 2020-02-06 & High-med & $8.3$ &  5888 \\
7.6a & \acet, \water $^*$& 7.5 & 7.7 & 2018-10-26 & High-low & $3.4$ & 8196 \\
7.6b & \acet, \meth, \water $^*$& 7.5 & 7.7 & 2020-02-07 & High-low & $3.4$ & 7296 \\
7.8 & CS, \water $^*$ & 7.7 & 7.9 & 2018-10-31 & High-low & $3.6$ &  8064\\
7.9 & CS, \water $^*$, SiO$^*$ & 7.8 & 8.0 & 2018-11-01 & High-low & $3.8$ & 4608  \\
13.2 & \acet, \acetiso, HCN, [NeII]$^{*\dagger}$ & 12.8 & 13.6 & 2018-10-31 & High-low & $2.2$ & 2560 \\
13.9 & \acet, \acetiso, HCN, \hcniso & 13.5 & 14.3 & 2018-10-30 & High-low & $2.4$ & 2880 \\
16.3 & --- & 15.9 & 16.7 & 2018-10-30 & High-low & $3.3$ &  2816 \\
17.0 & \htwo $^*$ & 16.6 & 17.4 & 2018-10-30 & High-low & $3.7$ & 1024 \\
17.7a & --- & 17.2 & 18.0 & 2018-10-27 & High-low & $4.3$ & 512 \\
17.7b & --- & 17.2 & 18.0 & 2018-10-30 & High-low & $4.5$ &  768 \\
18.4 & \sotwo & 17.9 & 18.7 & 2018-10-27 & High-low & $4.7$ &  768 \\
19.1 & \sotwo, [SIII]$^{*\dagger}$ & 18.7 & 19.4 & 2018-10-27 & High-low & $5.3$ & 1024 \\
19.8 & \sotwo & 19.4 & 20.1 & 2018-10-27 & High-low & $5.4$ & 704 \\
20.5 & HNC, \sotwo & 20.1 & 20.8 & 2018-10-27 & High-low & $6.0$ & 576 \\
21.2 & HNC & 20.8 & 21.5 & 2018-10-27 & High-low & $6.4$ &  512 \\
21.9 & HNC & 21.5 & 22.2 & 2018-10-27 & High-low & $7.0$ & 512 \\
22.6 & HNC & 22.2 & 22.9 & 2018-10-27 & High-low & $7.7$ & 512 \\
23.3 & --- & 22.9 & 23.6 & 2018-10-27 & High-low & $8.3$ & 384 \\
23.9 & --- & 23.5 & 24.2 & 2018-10-26 & High-low & $8.7$ & 352 \\
24.7 & OH$^?$ & 24.3 & 24.9 & 2018-10-26 & High-low & $9.4$ & 448 \\
25.3 & [SI]$^{*\dagger}$ & 24.9 & 25.7 & 2017-03-22 & High-low & $9.6$ & 1920 \\
26.0 & \water, [FeII]$^{*\ddagger}$ & 25.6 & 26.3 & 2017-03-22 & High-low & $10.2$ & 1856 \\
26.7 & --- & 26.2 & 26.9 & 2017-03-17 & High-low & $10.4$ & 1280 \\
27.4 & --- & 26.9 & 27.6 & 2017-03-17 & High-low & $11.5$ & 1280 \\
28.1 & --- & 27.6 & 28.3 & 2017-03-17 & High-low & $12.5$ & 1728 \\
\hline
\multicolumn{8}{c}{IRTF/TEXES (slit width: 1.4 \arcsec)}\\
11.76 & \amm\ & 11.71 & 11.81 &2018-02-08& High-med&$8$& 2331\\
11.83 & \amm\ & 11.78 & 11.89 &2018-02-11& High-med&$8$& 2331\\
\enddata
\tablecomments{$^*$ denotes emission lines. All other lines are in absorption. $^?$ denotes a tentative detection. $^{\dagger}$ denotes lines observed in both the on- and off-source positions; $^{\ddagger}$ denotes a line observed only in the off-source position. All other lines are observed only in the on-source position towards IRc2. --- denotes no detected lines in that setting.}
\end{deluxetable*}

The EXES data were reduced by the SOFIA Redux pipeline \citep{Clarke2015}. Wavelength scales were calibrated using sky emission line spectra produced for each setting by omitting the nod subtraction step and then adjusting the scale to match observed sky emission line wavelengths to their values in the HITRAN database \citep{Gordon2017}. The wavelength uncertainty is 0.3 \kms, estimated by comparing atmospheric emission lines to their wavelengths in the HITRAN database. Figures \ref{fig:flux1} to \ref{fig:flux3} in Appendix \ref{ap:galflux} show examples of normalized flux with molecular lines for each species with atmospheric models.   

The flux peak of IRc2 is known to shift in position at different MIR wavelengths \citep[e.g.][]{Gezari1992,Greenhill2004,Okumura2011}. To ensure the survey had a consistent centre, all EXES observations began by acquiring flux peak as seen at $~7.8$ \micron\ and maintaining that pointing during the observation legs. Figure \ref{fig:beams} in Appendix \ref{ap:beams} gives the EXES beam location, size, and orientation for each setting in this work. The edge of the EXES beam fell over nearby IR source, IRc7, for settings 20.5 \micron\ through to 24.7 \micron\ and 7.3 \micron. After examining spectra split along the slit width, we confirm that the spectral lines are strong over the slit centre at IRc2, and not the edge over IRc7. Furthermore, with ground-based MIR spectroscopy \citet{Evans1991} found that \acet\ and HCN have 2 and 3 times higher column densities, respectively, towards IRc2 than IRc7. Therefore, these observations are centred over IRc2 and it is highly unlikely that any lines come from IRc7.

The SOFIA point spread function size is about 3--3.5\arcsec\ and rises to 4\arcsec\ at the long wavelength limit for EXES. Thus for settings (Table \ref{tab:obs}) with slit sizes near this size or smaller, it is sensible to extract the spectrum over the entire available slit length, because any spatial information gets averaged together by the observatory seeing. This alone does not ensure that the exact same material is probed because the brightest part of the continuum of IRc2 moves depending on wavelength as explained above, especially for observations with longer slits. However, we inspected the spectra as extracted from three sectors along the slit and found little impact on the absorption lines, even for the longer slit lengths (the one exception is the sole \water\ absorption line, which will be discussed in \S \ref{sec:h2o}). Because of this we choose to sum over the entire slit for each setting in order to improve signal-to-noise for a better parameter estimation. Furthermore, as discussed in \S \ref{sec:kin}, the kinematic components probed by our observations may be spatially extended beyond IRc2 and in this case, the slightly differing location of our pencil beam is still probing the same kinematic component.

We supplement this EXES survey towards Orion IRc2 with two settings at 11.76 and 11.83 \micron, taken with the TEXES \citep{Lacy2002} instrument on the nights of 2018 February 8 and 11 at the NASA Infrared Telescope Facility (IRTF) on Mauna Kea. This small amount of data fell within an atmospheric window in which SOFIA is comparable to ground-based observations and contains \amm. We used high-medium mode with a resolution of about 100,000 ($\sim$ 3 \kms) and a beam sized $8\arcsec \times1\farcs4$ oriented north-south. The beam was centred on the continuum peak of IRc2 for the wavelength of each setting. The telescope nodded 3\arcsec\ along the slit to enable background subtraction. The flux was extracted by weighting the signal by the continuum flux distribution. 

\section{Analysis and Results} \label{sec:res}

\subsection{Flux Preparation} \label{ssec:ana}

We normalize the EXES data following the procedure detailed in \citet{Nickerson2021}. Here we summarize it briefly. In the raw EXES flux and an unsmoothed ATRAN atmospheric transmission model \citep{Lord1992}\footnote{\url{https://atran.arc.nasa.gov/cgi-bin/atran/atran.cgi}}, we identify baselines, noise, atmospheric lines, and molecular lines towards IRc2 both with algorithms and by hand where algorithms are insufficient.  We fit the baseline and atmospheric line segments in the raw flux and the atmospheric model to find the constant by which to normalize the flux in each order and the sigma in each setting by which to Gaussian smooth the atmospheric model's lines. This normalized flux will match the smoothed atmospheric model as closely as possible, aside from noise in the flux and the atmospheric model's limitations. Some molecular lines fall near to atmospheric lines and we divide the flux by the atmospheric model to recover molecular lines. The exceptions are the HNC and \sotwo\ lines, which do not require the division. 

Additionally, a number of orders in settings 13.2, 20.5, 21.9, and 22.6 \micron\ exhibit uneven baselines from standing waves and require division by a polynomial in order to flatten the baseline. The upper right panel in Figure \ref{fig:flux1} is an example of an uncorrected standing wave in the baseline, while the bottom left panel in Figure \ref{fig:flux1} is an example of a baseline post polynomial correction. 

The small quantity of TEXES data included in this paper has already been normalized by the instrument team from division by an atmospheric model. We do not need to correct it further.

\subsection{Single Lines}
\label{sec:singlelines}

For all species, except \sotwo\ (\S \ref{sec:crowd}), we fit individual lines to Gaussians to extract their column densities for rotation diagram analysis. 

The majority of this survey's species are absorption lines, analyzed following the procedure detailed in \citet{Nickerson2021}. Briefly summarized here, we fit with a Gaussian profile following \citet{Indriolo2015}:
\begin{equation}
    I = I_0 e^{-\tau_0 G},
\label{eqn:gausabs}
\end{equation}
where,
\begin{equation}
    G = \mathrm{exp}\Big[-\frac{(v-v_{\mathrm{LSR}})^2}{2\sigma^2_v}\Big],
\label{eqn:g}
\end{equation}
$I_0$ is the normalized continuum level (close to unity), $\tau_0$ is the line centre optical depth, $v$ is the velocity in the local standard of rest (LSR) frame, \vlsr\ is the LSR velocity at the line's centre, and $\sigma_v$ is the velocity dispersion (full-width half maximum \vfwhm$\;=2\sqrt{2\ln2}\sigma_v$). 

One Gaussian is the optimal fit for CS, \water, HNC, \hcniso, and \amm, resulting in only one apparent velocity component. The species \acet, \acetiso, \meth, and HCN, have two apparent velocity components and were best fit by double Gaussians: 
\begin{equation}
    I = I_0 e^{-(\tau_{01} G_1 + \tau_{02} G_2)}. 
\label{eqn:gauss2}
\end{equation}
It is not possible, however, to fit every single line from these species to double Gaussians. Some have only single Gaussians, particularly if the line is weak, suffers from atmospheric interference, or blends with another molecular line. 

\begin{figure*}
\centering
\gridline{\fig{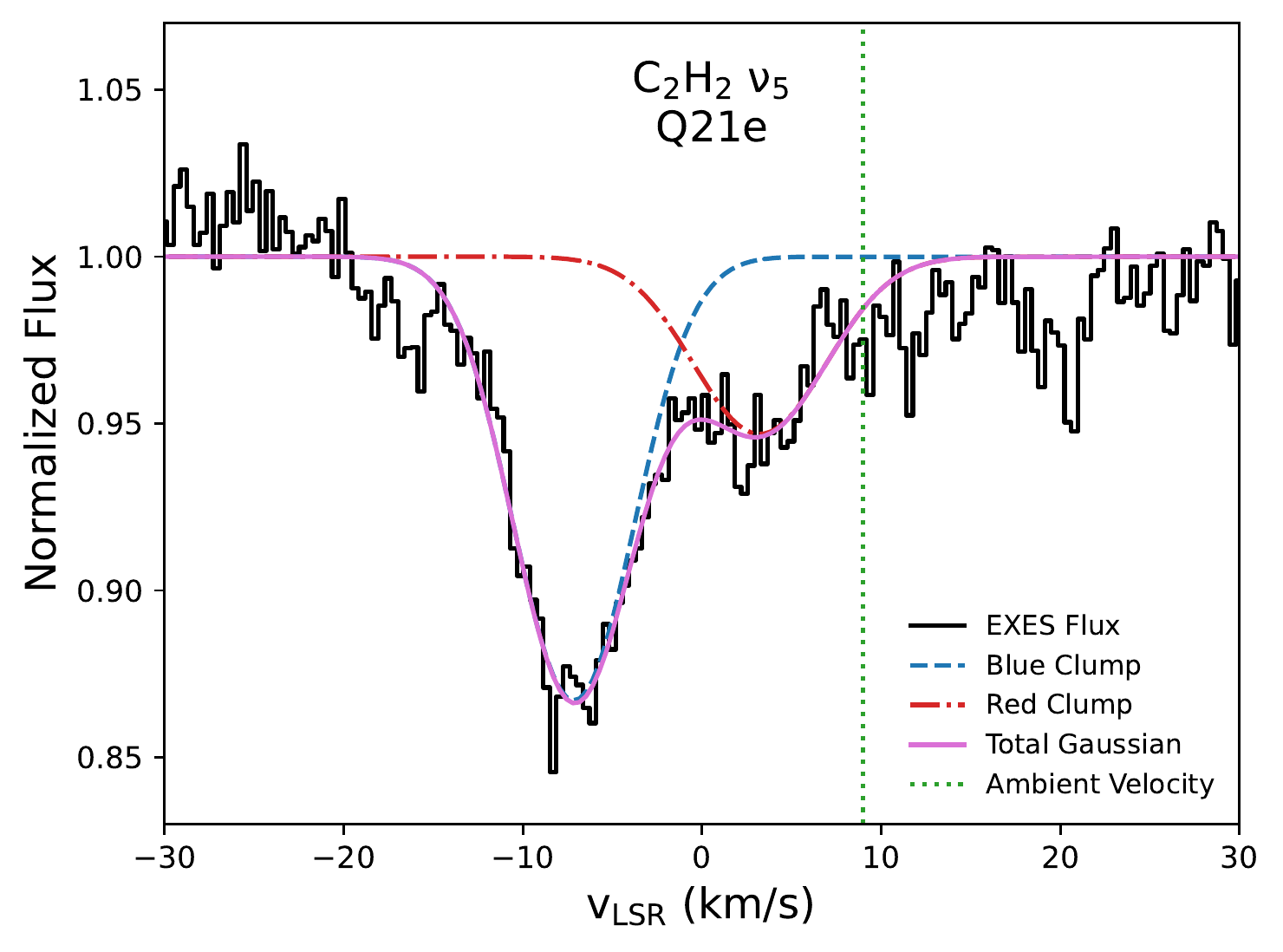}{0.32\textwidth}{}
          \fig{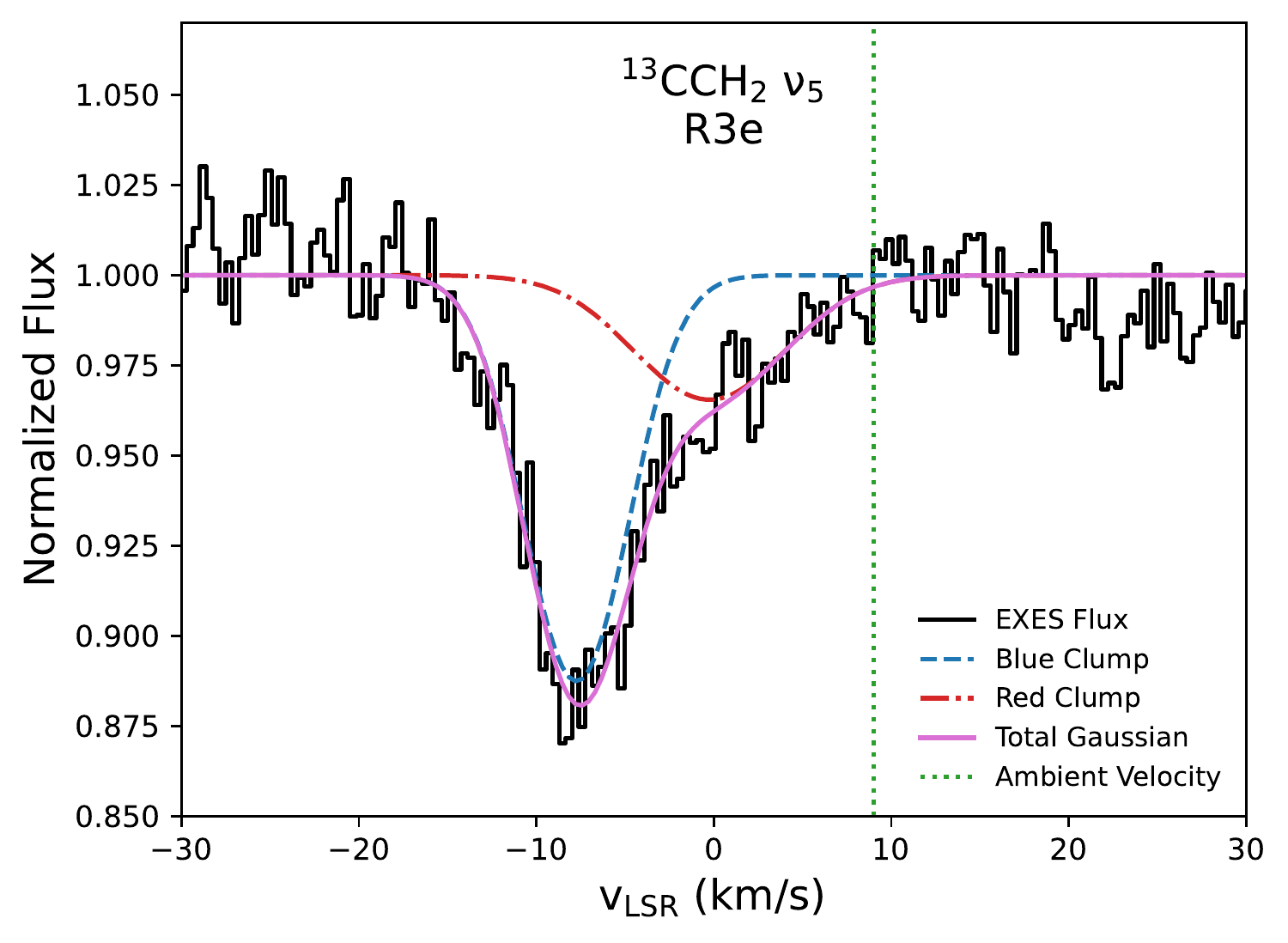}{0.32\textwidth}{}
          \fig{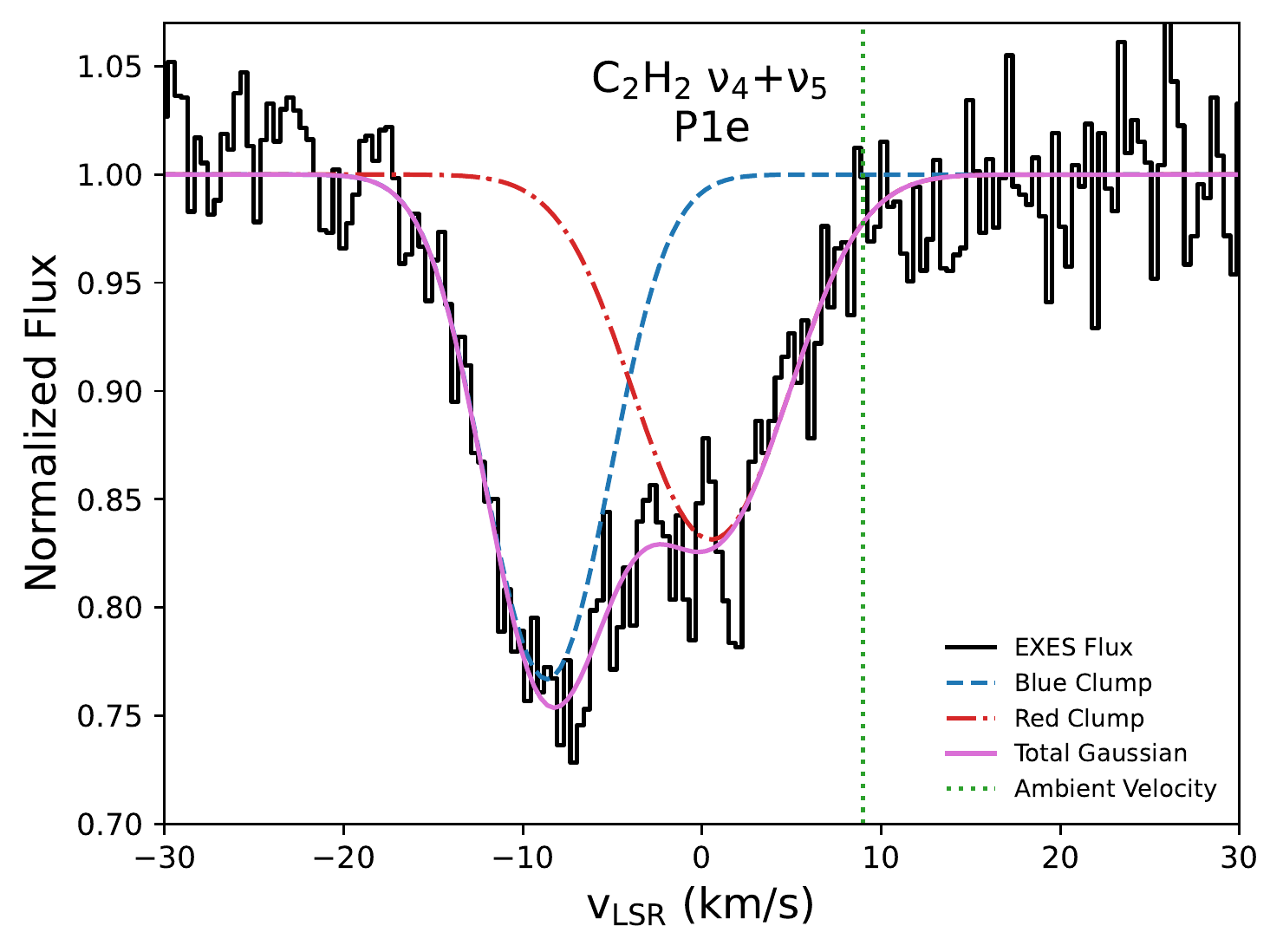}{0.32\textwidth}{}}
\vspace{-10mm}
\gridline{\fig{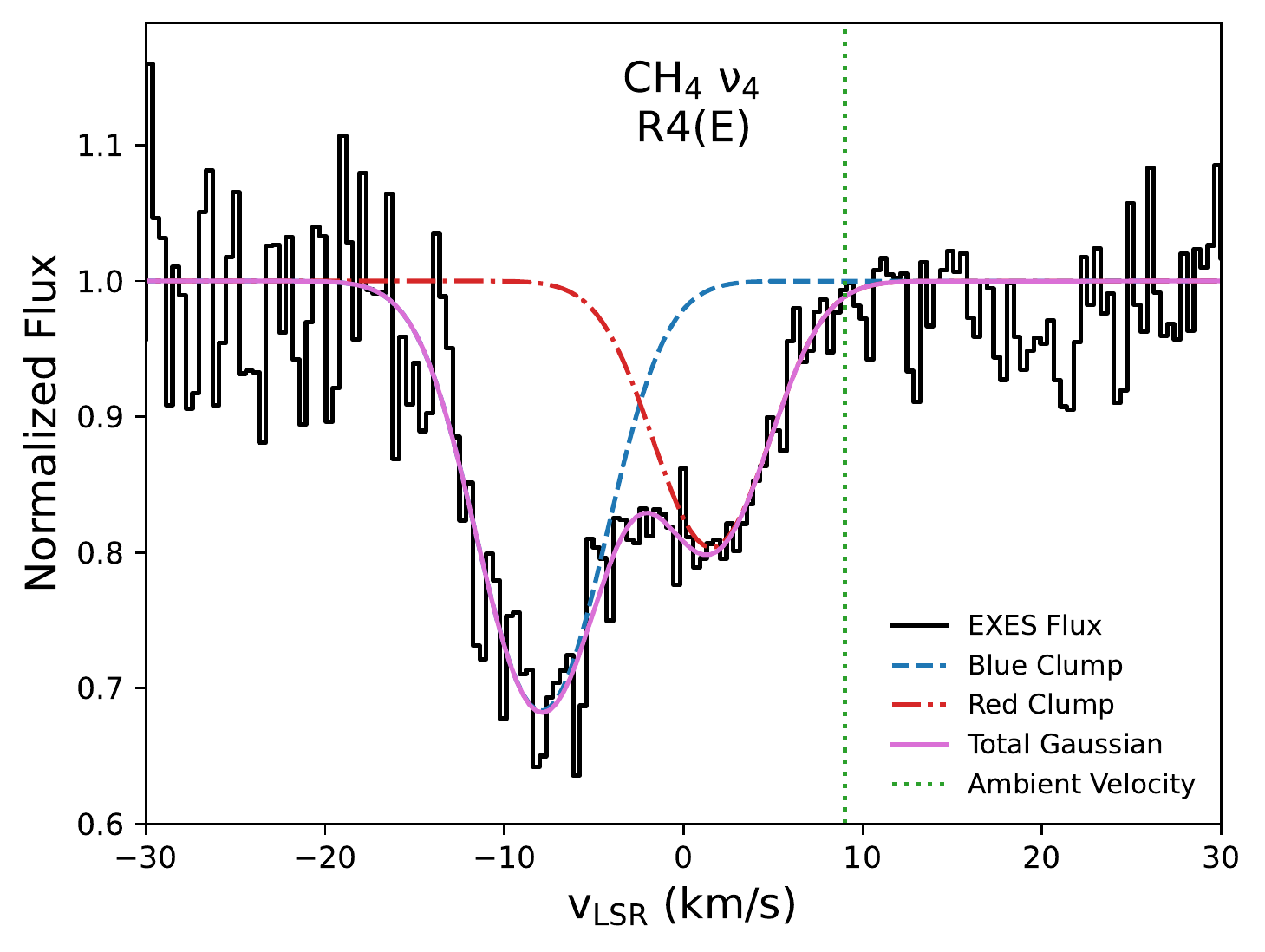}{0.32\textwidth}{}
\fig{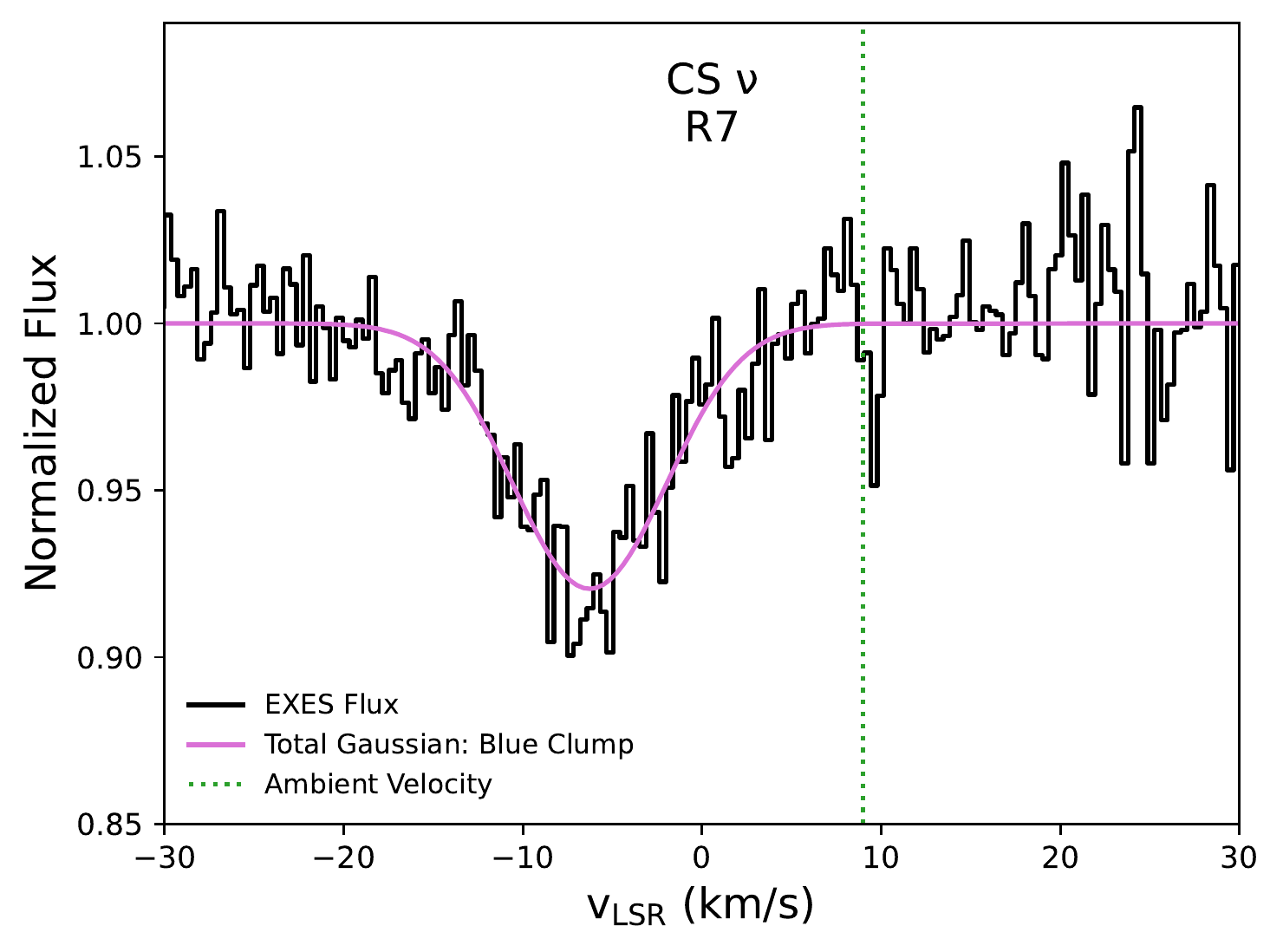}{0.32\textwidth}{}
\fig{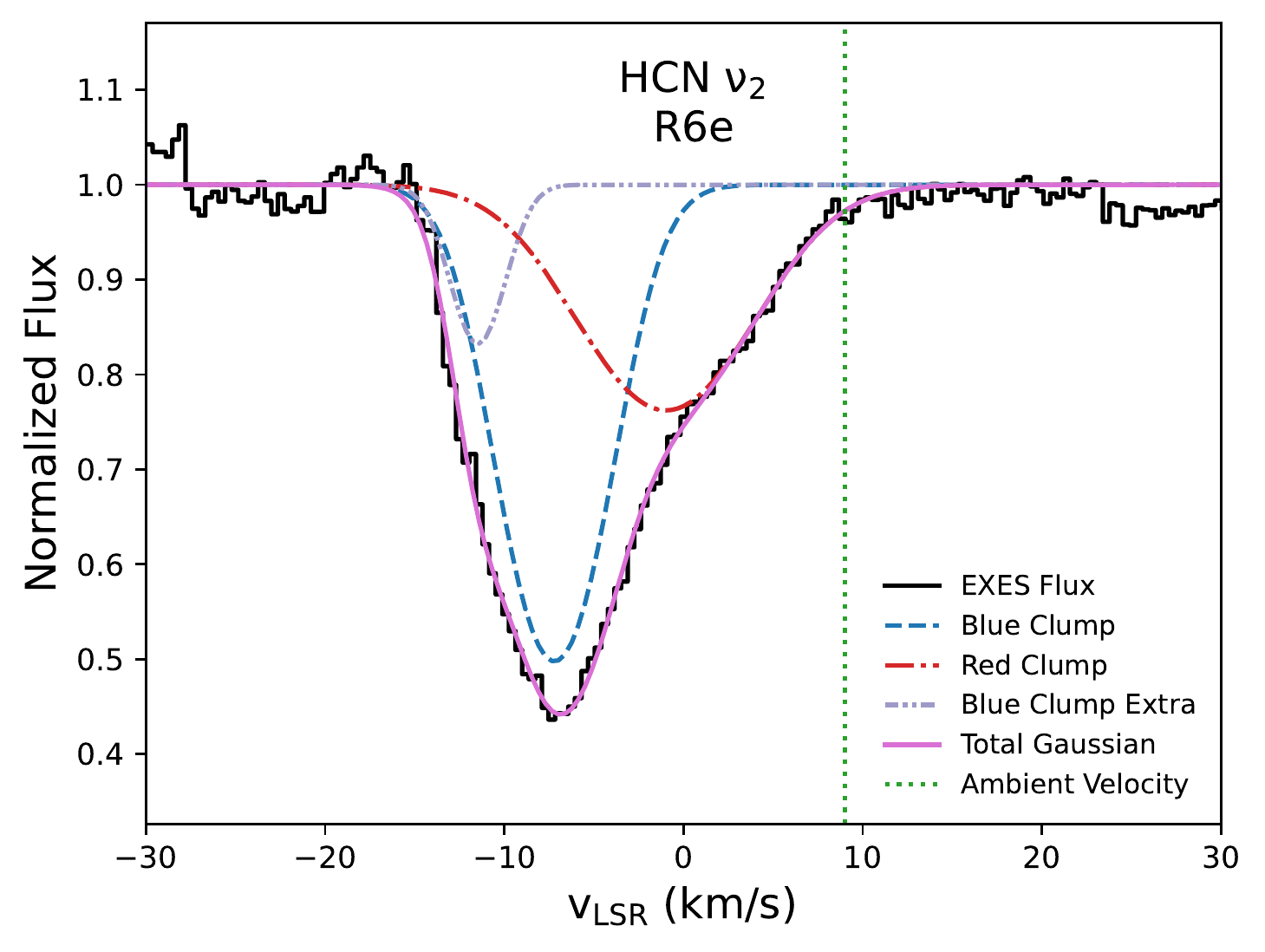}{0.32\textwidth}{}}
\vspace{-10mm}
\gridline{\fig{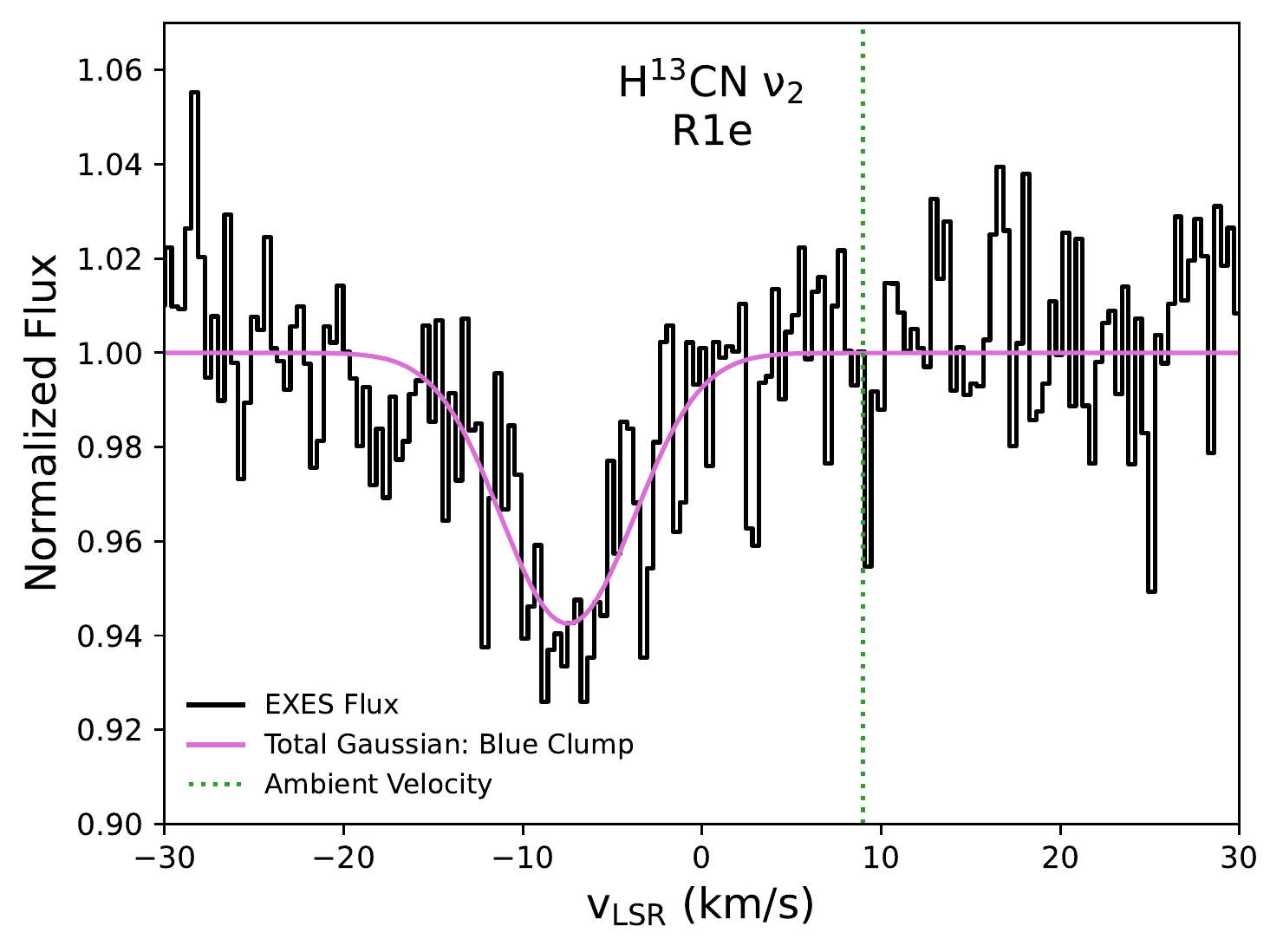}{0.32\textwidth}{}
\fig{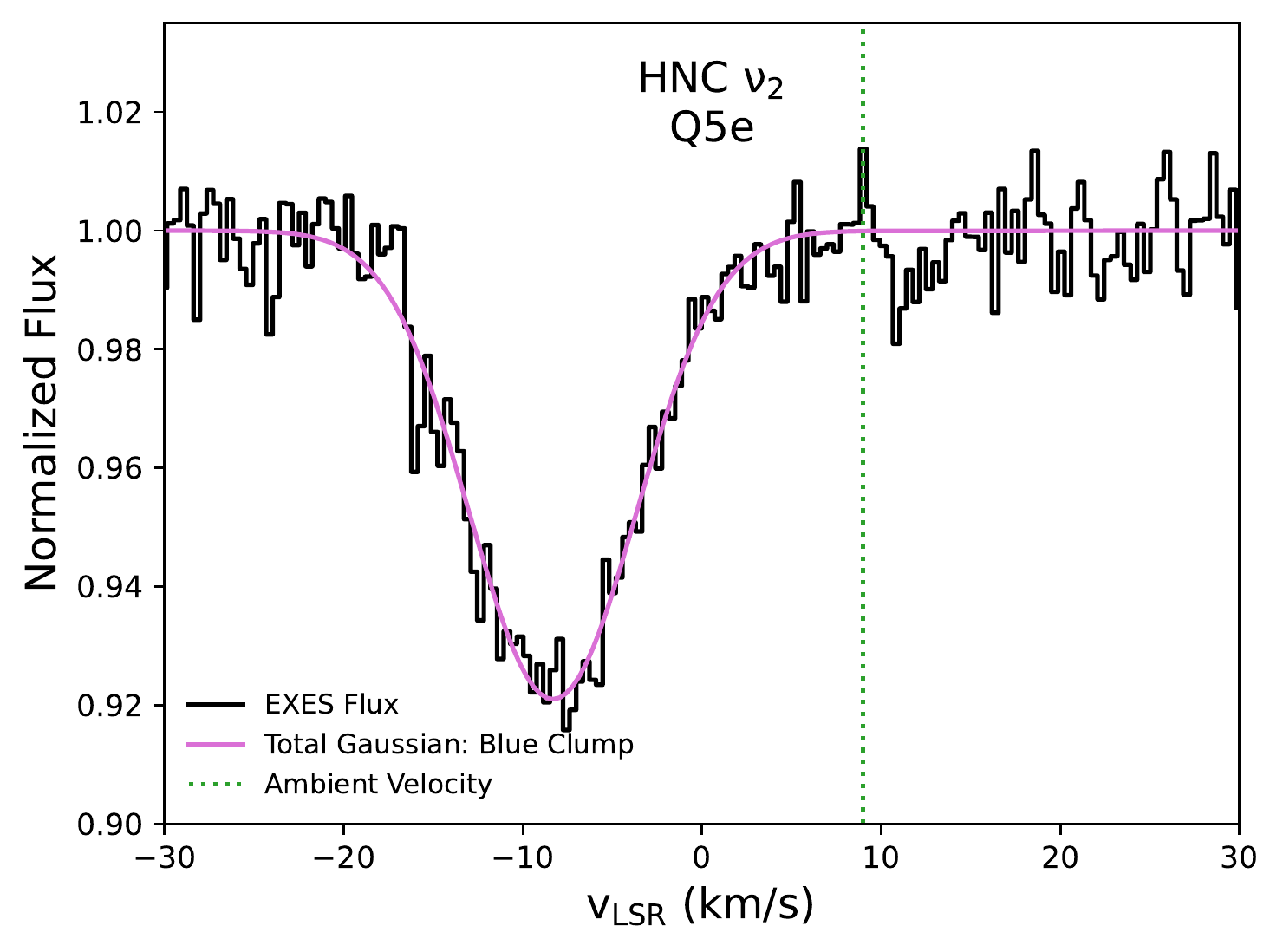}{0.32\textwidth}{}
\fig{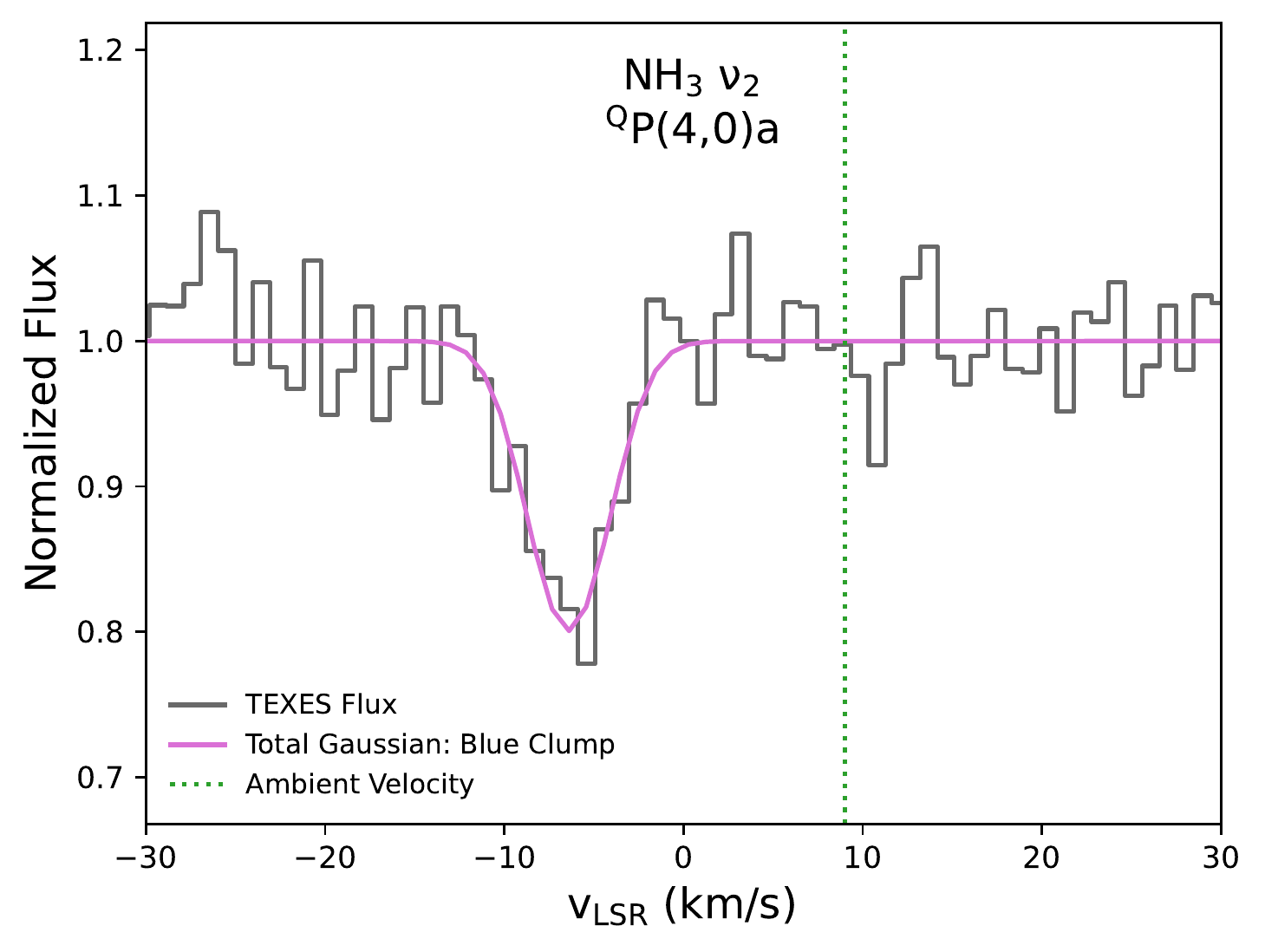}{0.32\textwidth}{}}
\vspace{-10mm}
\gridline{\fig{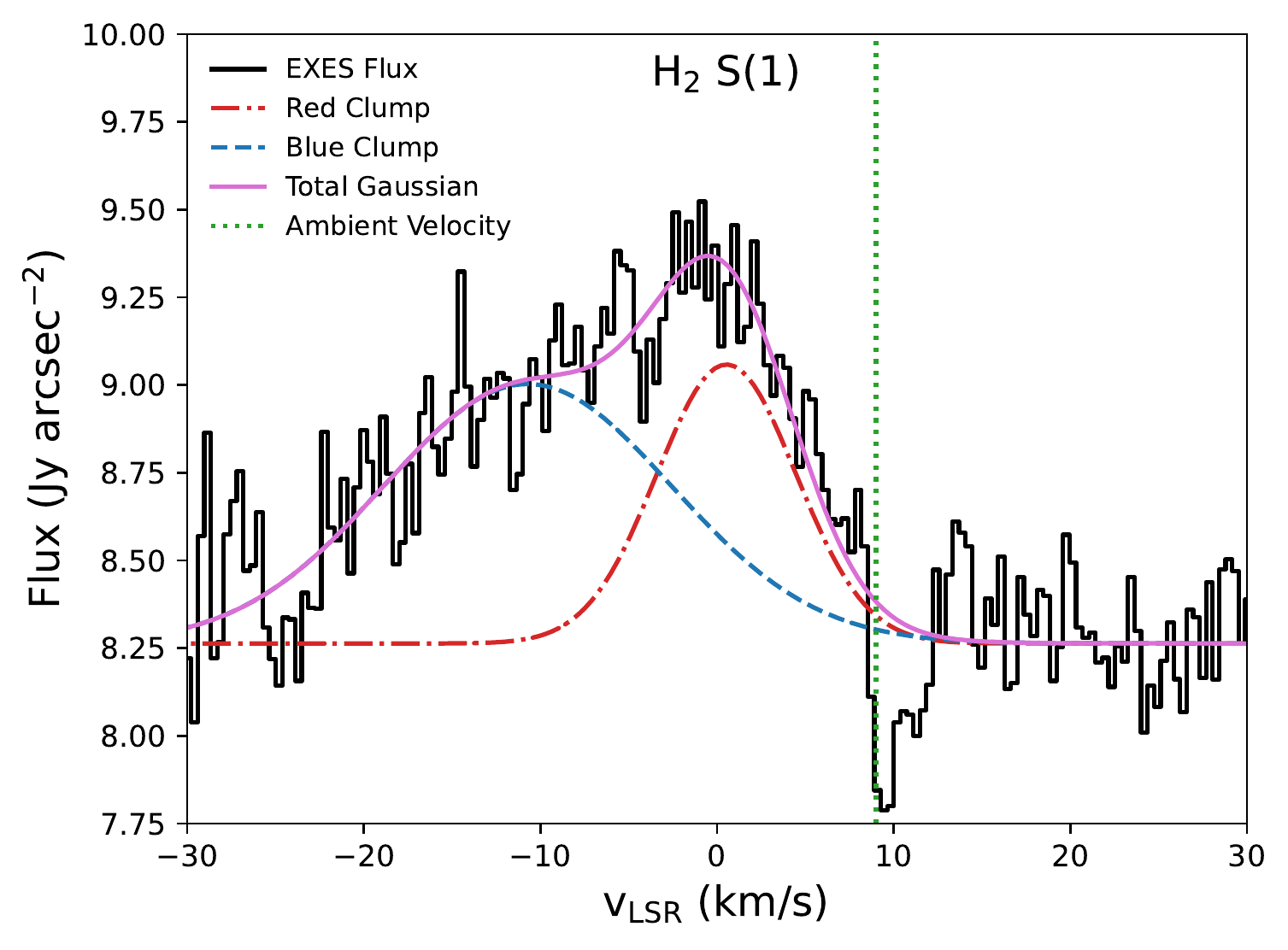}{0.32\textwidth}{}
\fig{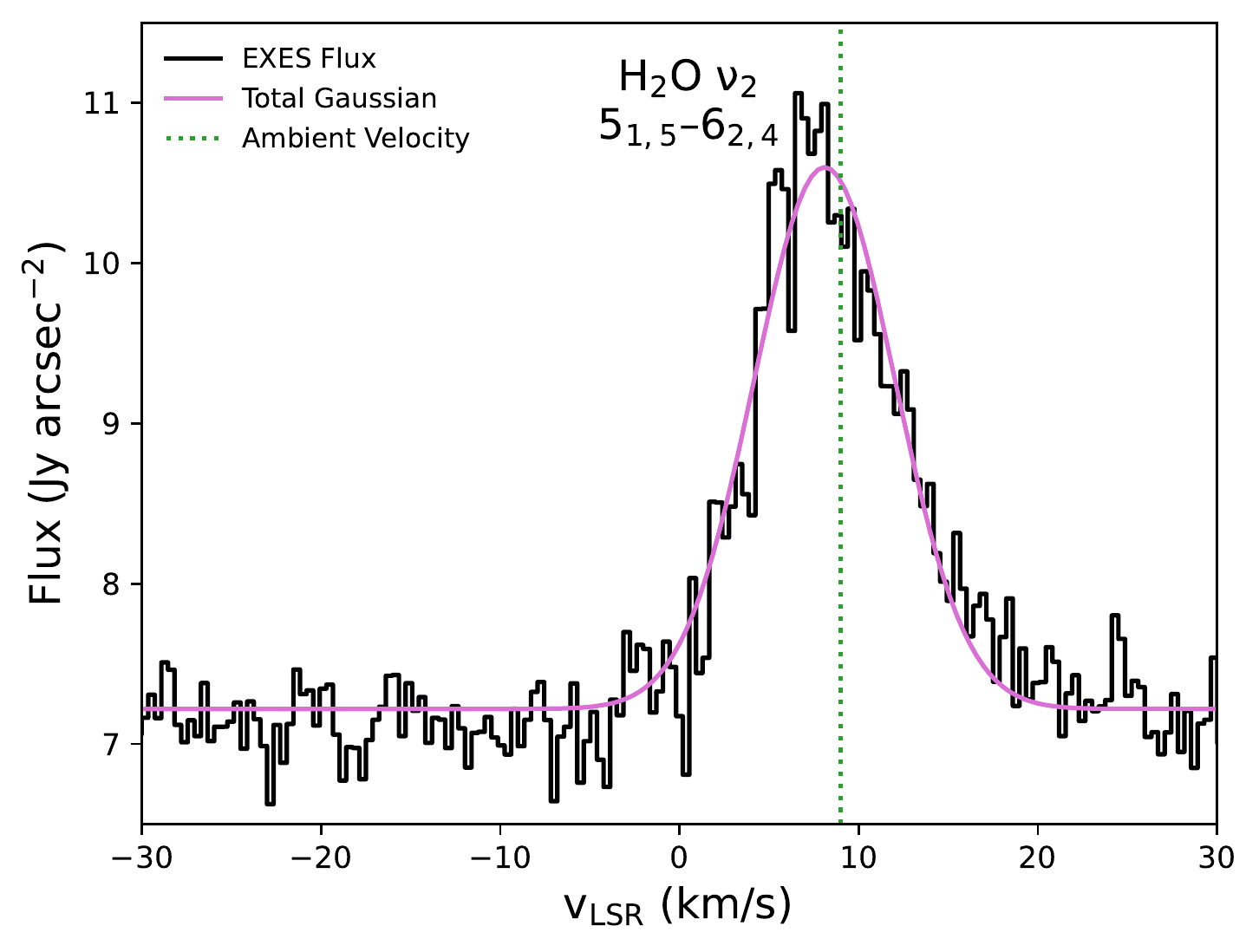}{0.32\textwidth}{}
\fig{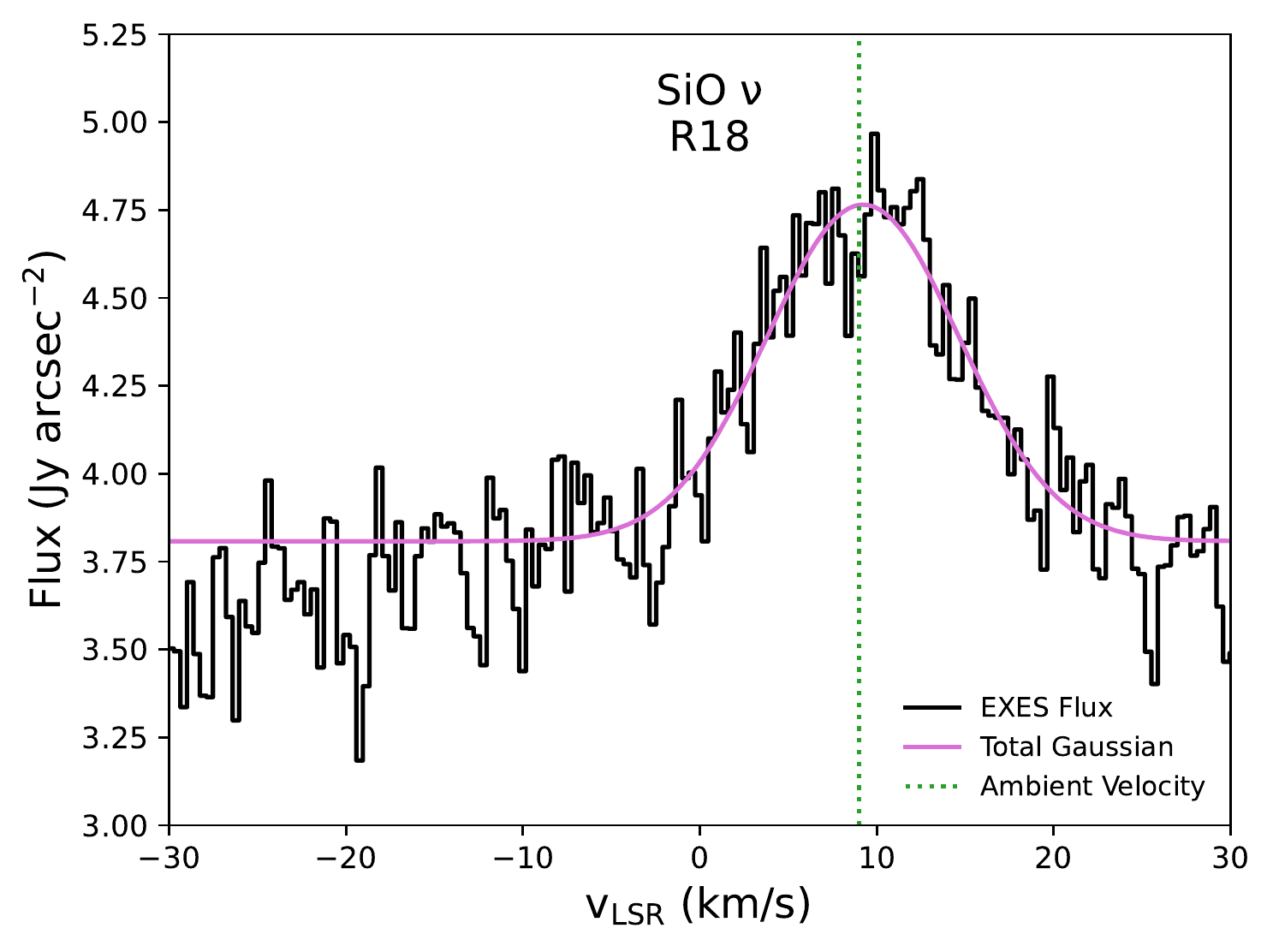}{0.32\textwidth}{}}
\vspace{-10mm}
\caption{Sample Gaussian fits, left to right and top to bottom, for the $\nu_5$ band of \acet, \acetiso, the $\nu_5+\nu_4$ band of \acet, \meth, CS, HCN, \hcniso, HNC, \amm, \htwo, \water, and SiO. Black is EXES flux and grey is TEXES flux. Fluxes are normalized for absorption lines and in Jy arcsec$^{-2}$ for  emission lines. Magenta is the total Gaussian fit, while for double and triple Gaussians the blue, red, and lavender dash-dot lines belong respectively to the blue clump, the red clump, and in this example extra for the blue clump. Single component fits belong to the blue clump for the absorption lines. The vertical, green dotted line indicates the systematic, ambient cloud velocity 9 \kms\ \citep{Zapata2012}. \label{fig:gauss}}
\end{figure*}

We give the examples of these fits for each species in Figure \ref{fig:gauss}. The absorption species fit two distinct kinematic components, with average \vlsr$\;=-7.1 \pm 0.7$ \kms\ and $1.4\pm0.5$ \kms, which we refer to for convenience as the ``blue clump'' and the ``red clump''  due to their blue- and red-shifted velocities relative to the LSR. Both are blue shifted with respect to the ambient cloud velocity, 9 \kms\ \citep{Zapata2012}. The single Gaussian species (CS, \water, HNC, \hcniso, and \amm) all belong to the blue clump, while the species with transitions best fit to double Gaussians (\acet, \acetiso, \meth, and HCN) have a deeper Gaussian that belongs to the blue clump and a shallower Gaussian that belongs to the red clump. This is the first work that unambiguously identifies the blue and red clumps in detail. We discuss these two components further in \S \ref{sec:kin} and compare their properties in Table \ref{tab:comp}.

Some lines for HCN, \acet, and  \acetiso\ that coincide with the telluric features are distorted even after the atmospheric division. Many of these require fits to a triple Gaussian, which follows similarly from Equation \ref{eqn:gauss2}. The third Gaussian components are either discarded or absorbed into the red or blue clump component, depending on the line shape. For these lines we take the line with the deepest $\tau_0$ to provide the LSR velocity for the combined line. We also do not include these lines for calculating the mean \vfwhm\ for the species in Table \ref{tab:rot}, because the Gaussians overlap. See the HCN  Gaussian fit in Figure \ref{fig:gauss} for an example of this situation.

\begin{deluxetable*}{lllrrrrr}
\tablecaption{Overview of Species Properties\label{tab:rot}}
\tablehead{
\colhead{Species} & \colhead{Band} & \colhead{Component} & \colhead{\#}&\colhead{\vlsr} & \colhead{\vfwhm} &  \colhead{$T$} &
\colhead{$N$} \\
\colhead{} & \colhead{(\micron)} & \colhead{}& \colhead{} & \colhead{(\kms)} & \colhead{(\kms)} & \colhead{(K)} &\colhead{\csi}}
\startdata
\hline
\acet\ Ortho &$\nu_5\ 13.5$ &Blue Clump&24&$-7.3\,\pm\,$0.1&8.5$\,\pm\,$0.1&175$\,\pm\,$12&(1.50$\,\pm\,$0.15)$\times 10^{16}$\\
&&Red Clump&17&1.4$\,\pm\,$0.2&7.5$\,\pm\,$0.2&229$\,\pm\,$27&(3.58$\,\pm\,$0.71)$\times 10^{15}$\\
\phantom{\acet\ }Para &$\nu_5\ 13.5$ &Blue Clump&20&$-7.5\,\pm\,$0.0&8.1$\,\pm\,$0.1&145$\,\pm\,$9&(1.23$\,\pm\,$0.15)$\times 10^{16}$\\
&&Red Clump&12&1.8$\,\pm\,$0.2&7.0$\,\pm\,$0.3&158$\,\pm\,$16&(3.09$\,\pm\,$0.57)$\times 10^{15}$\\
\hline
\acetiso\ & $\nu_5$ 13.5 &Blue Clump&10&$-7.4\,\pm\,$0.1&7.2$\,\pm\,$0.4&91$\,\pm\,$9&(2.56$\,\pm\,$0.18)$\times 10^{15}$\\
&&Red Clump&4&2.1$\,\pm\,$1.0&8.2$\,\pm\,$1.7&64$\,\pm\,$6&(6.74$\,\pm\,$0.64)$\times 10^{14}$\\
\hline
\acet\ Ortho& $\nu_4+\nu_5$ 7.6&Blue Clump&8&$-7.5\,\pm\,$0.1&8.2$\,\pm\,$0.2&124$\,\pm\,$13&(8.39$\,\pm\,$1.44)$\times 10^{16}$\\
&&Red Clump&8&1.3$\,\pm\,$0.2&7.5$\,\pm\,$0.3&111$\,\pm\,$14&(4.73$\,\pm\,$1.06)$\times 10^{16}$\\
\phantom{\acet\ }Para& $\nu_4+\nu_5$ 7.6&Blue Clump&5&$-7.9\,\pm\,$0.2&8.2$\,\pm\,$0.4&73$\,\pm\,$14&(3.42$\,\pm\,$0.73)$\times 10^{16}$\\
&&Red Clump&5&1.3$\,\pm\,$0.3&7.3$\,\pm\,$0.6&140$\,\pm\,$18&(2.50$\,\pm\,$0.21)$\times 10^{16}$\\
\hline
\meth\ & $\nu_4$ 7.6&Blue Clump&6&$-8.0\,\pm\,$0.1&7.6$\,\pm\,$0.3&193$\,\pm\,$42&(1.99$\,\pm\,$0.28)$\times 10^{17}$\\
&&Red Clump&6&0.6$\,\pm\,$0.2&7.9$\,\pm\,$0.4&141$\,\pm\,$33&(8.80$\,\pm\,$1.78)$\times 10^{16}$\\
\hline
CS& $\nu$ 7.8&Blue Clump&9&$-6.4\,\pm\,$0.1&8.6$\,\pm\,$0.5&175$\,\pm\,$34&(6.97$\,\pm\,$0.58)$\times 10^{15}$\\
\hline
\htwo $^*$&17&Blue Clump&1&$-10.7\,\pm$2.6&19.2$\,\pm$5.1&---&---\\
&&Red Clump&1&$0.5\,\pm$0.5&9.3$\,\pm$1.8&---&---\\
\hline
\water & 26 & Blue Clump&1&$-8.0\,\pm0.4$&17.0$\,\pm\,$1.0&---&---\\
\water $^*$\tablenotemark{a} & $\nu_2$ 7.6 &---&13&$8.8\,\pm\,$0.1&$10.6\,\pm\,$0.2&---&---\\
\hline
HCN &$\nu_2$ 13.5&Blue Clump&22&$-7.3\,\pm\,$0.0&8.7$\,\pm\,$0.1&135$\,\pm\,$9&(5.44$\,\pm\,$0.43)$\times 10^{16}$\\
&&Red Clump&15&1.0$\,\pm\,$0.2&8.3$\,\pm\,$0.4&182$\,\pm\,$34&(1.87$\,\pm\,$0.39)$\times 10^{16}$\\
 &$2\nu_2$ 7.0&Blue Clump&2&$-5.8\,\pm\,$0.6&8.7$\,\pm\,$1.0&---&---\\
&&Red Clump&2&1.8$\,\pm\,$0.9&6.4$\,\pm\,$1.4&---&---\\
\hline
\hcniso & $\nu_2$ 13.5&Blue Clump&3&$-6.6\,\pm\,$0.2&7.9$\,\pm\,$0.5&99$\,\pm\,$16&(4.36$\,\pm\,$0.65)$\times 10^{15}$\\
\hline
HNC &$\nu_2$ 22&Blue Clump\tablenotemark{b}&24& $-7.7\,\pm\,$0.1&11.3$\,\pm\,$0.2&97$\,\pm\,$8&(7.41$\,\pm\,$0.62)$\times 10^{14}$\\
\hline
\amm & $\nu_2$ 11 &Blue Clump&5&$-6.7\,\pm\,$0.4&6.7$\,\pm\,$1.3&230$\,\pm\,$86&(1.58$\,\pm\,$0.77)$\times 10^{16}$\\
\hline
SiO$^*$\tablenotemark{a}& $\nu$ 7.9 &---&12&$9.8\,\pm\,$0.1&13.6$\,\pm\,$0.4&---&---\\
\hline
\sotwo & $\nu_2$ 19 &Blue Clump\tablenotemark{b}&---&$-6.1_{-0.5}^{+0.5}$&$12.9_{0.9}^{0.9}$&$94_{-6}^{+7}$&$(6.17_{-0.42}^{+0.44})\times10^{16}$\\
 & $\nu_3$ 7.2 &Blue Clump\tablenotemark{b}&---&$-6.0_{-0.3}^{+0.3}$&$11.2_{-0.5}^{+0.5}$&$128_{-5}^{+5}$&$(1.10_{-0.03}^{+0.03})\times10^{17}$\\
\hline
\enddata
\tablecomments{For each species, band, and velocity component: \# is the number of lines, \vlsr\ is the average central local standard rest of velocity, \vfwhm\ is the average full-width half-maximum, $T$ is the temperature, and $N$ is the total column density. Note that the \water\ and \htwo\ transitions with no band are pure rotational. $^*$ denotes emission line. \tablenotemark{a} The temperatures and column densities of the \water\ and SiO emission will appear in Monzon et al., in preparation.\tablenotemark{b} HNC and \sotwo\ have wide enough line widths that there may be an unresolved red clump component in the data, as discussed in \S \ref{sec:kin}.}
\end{deluxetable*}

For absorption lines, we calculate the column density, $N_l$, in the lower state of an observed transition as follows for the optically thin limit:
\begin{equation}
    N_l = \sqrt{2\pi} \frac{g_l}{g_u}\frac{8\pi}{A\lambda^3} \tau_0 \sigma_v,
\label{eqn:nl}
\end{equation}
where $g_l$ and $g_u$ are the lower and upper statistical weights respectively, $A$ is the Einstein coefficient for spontaneous emission, and $\lambda$ is the rest wavelength of the transition. 

Emission lines are analyzed following the procedure detailed and derived in Monzon et al., in preparation. There we will also present the full analysis of the \water\ and SiO emission lines. In this work, we present the results for the two-component \htwo\ line. We obtained the \htwo\ spectrum in units of ergs/(s~cm$^{2}$~cm$^{-1}$~sr) by averaging along the row in the co-added image, and multiply by a conversion factor, $S_{\mathrm{Ja}}$, to obtain the spectrum in units of Jy arcsec$^{-2}$. We fit the \htwo\ line to the following double Gaussian:
\begin{align}
    S_{\nu}(v) = B_{\nu} + S_{\nu 01} \ \mathrm{exp}(\frac{-(v - v_{\mathrm{LSR1}})^2}{2\sigma_{v1}^{2}}) \\ 
    + S_{\nu 02} \ \mathrm{exp}(\frac{-(v - v_{\mathrm{LSR2}})^2}{2\sigma_{v2}^{2}}),
\label{eqn:gaussemi}
\end{align}
where $B_{\nu}$ is the continuum level, and corresponding to Gaussians 1 and 2: $S_{\nu 01}$ and $S_{\nu 02}$ are the amplitudes, $v_{\mathrm{LSR1}}$ and $v_{\mathrm{LSR2}}$ are the LSR velocities of the line centres, and $\sigma_{v1}$ and $\sigma_{v2}$ are the velocity dispersions. We calculate $N_u$, the upper state column density, for each Gaussian, as follows: 
\begin{equation}
    N_u = \frac{4\pi \sqrt{2\pi} S_{\mathrm{Ja}} S_{\nu 0} \sigma_v}{hc A}.
\label{eqn:nu}
\end{equation}

The observed transitions and inferred parameters for all transition lines are given in Appendix \ref{ap:line}:  molecular absorption lines in Table \ref{tab:abslines} and molecular emission lines in Table \ref{tab:emilines}. Errors are generated by the fitting routine. We note dividing out the atmosphere from some lines does introduce uncertainly as well, but because we cannot quantify the accuracy of the ATRAN model we also cannot quantify the error introduced by this process.

The level populations of molecules in local thermal equilibrium (LTE) follow the Boltzman distribution \citep{Goldsmith1999}:
\begin{equation}
    \ln \frac{N_{l}}{g_{l}}=\ln \frac{N}{Q_R(T)}-\frac{E_{l}}{k_B T}
\label{eqn:rot}
\end{equation}
where $N_{l}$ is the column density of the lower state, $g_{l}$ is the statistical weight of the lower state, $N$ is the total column density, $Q_R$ is the rotational partition function, $E_{l}$ is the energy level of the lower state, $T$ is the temperature, and $k_B$ is the Boltzmann constant. Because we assume LTE, the temperature is equivalent to the kinetic temperature of the gas. The above applies to absorption lines where we use the lower state values, and for the \htwo\ emission line we use the upper state values instead, denoted by a subscript ``$u$''. $\lambda$, $A$, $g_{l}$, $g_{u}$, $Q_R$, $E_{l}$, and $E_{u}$ for each molecular transition come from three databases. We use GEISA \citep{Jacquinet-Husson2016} and our own calculations for $Q_R$ for HNC from levels published in ExoMol \citep{Harris2006,Barber2013}, ExoMol \citep{Tennyson2012} for SiO, and HITRAN \citep{Gordon2017} for all other species.

By linearly fitting to Equation \ref{eqn:rot}, we obtain the values of $T$ and $N$ for each species and velocity component, summarized in Table \ref{tab:rot}. Full rotation diagram analysis with the column densities and temperatures for the \water\ and SiO emission lines will appear in Monzon et al., in preparation. 

To calculate abundances (Table \ref{tab:abundance}) we use the column density range estimated by \citet{Evans1991} along the line of sight towards Orion IRc2, $N_{\mathrm{H}_2}=(1.9\pm1.1)\times10^{23}$ \csi. This upper limit was based on \amm\ mapping of the region, in which IRc2 was calculated to probe about 30\% of the material in the densest part of the hot core \citep{Wynn-Williams1984}, while the lower limit was found by calculating the depth of the silicate feature. Because this column density measures the total \htwo\ towards IRc2 and not individual components, we sum the column density of both the blue and red clumps when calculating abundances. The \htwo\ emission lines may also trace the same gas as the red and blue clumps, having similar velocities. However, we do not use them to calculate abundance ratios due to the uncertainty associated with a single transition. We discuss \htwo\ further in \S \ref{sec:h2}.

\begin{figure*}
\centering
\gridline{\fig{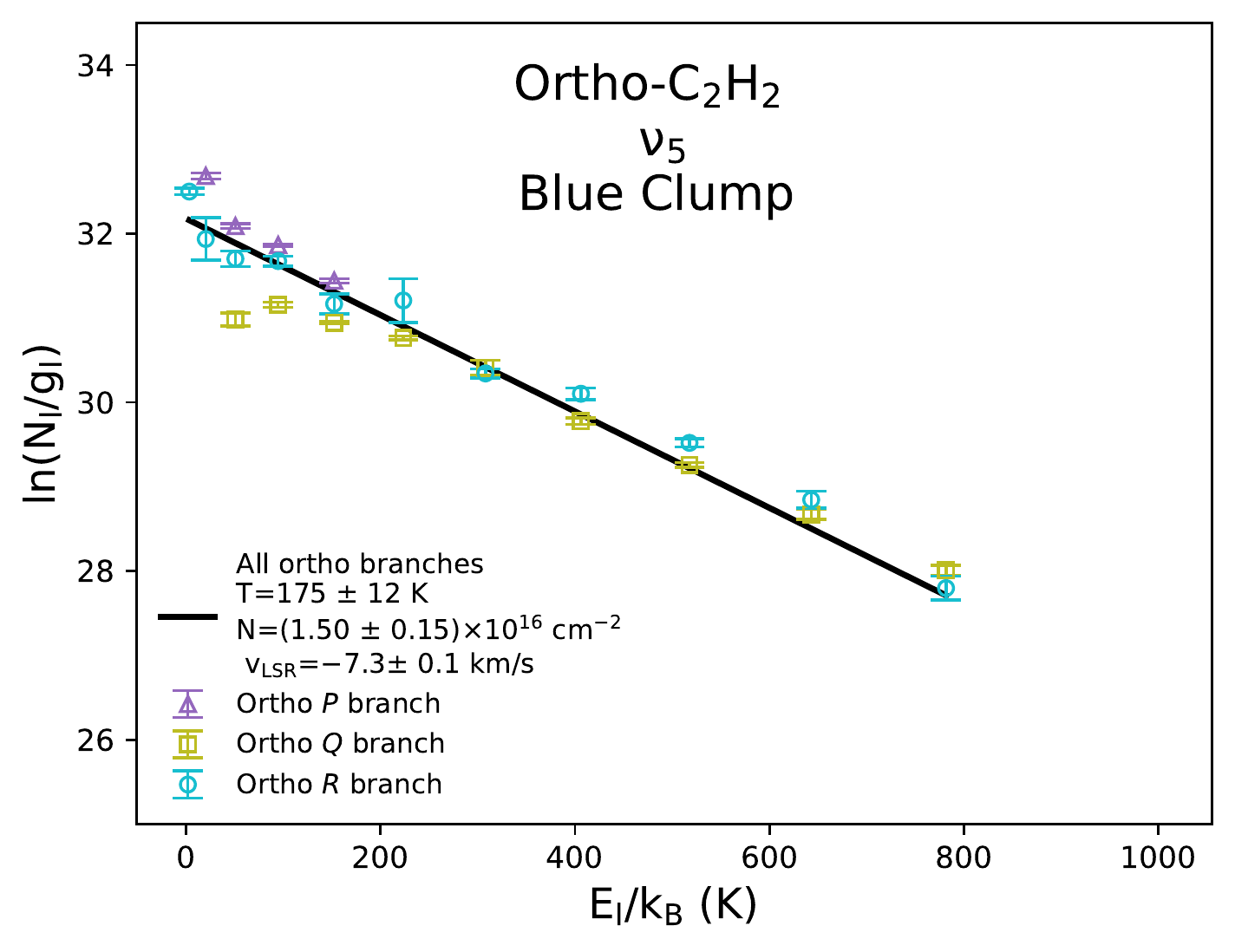}{0.49\textwidth}{}\hspace{-5mm}
          \fig{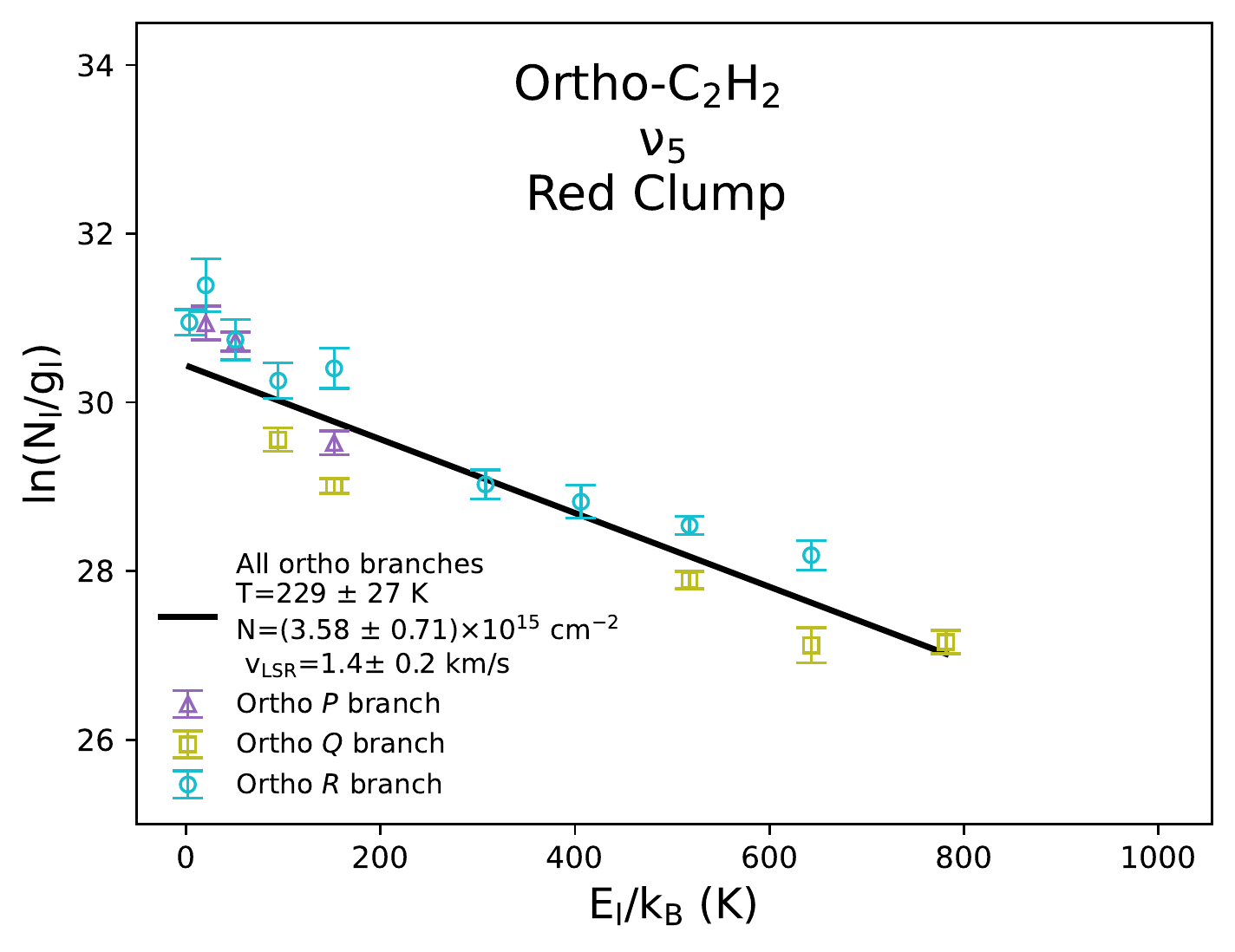}{0.49\textwidth}{}}
\vspace{-10mm}
\gridline{\fig{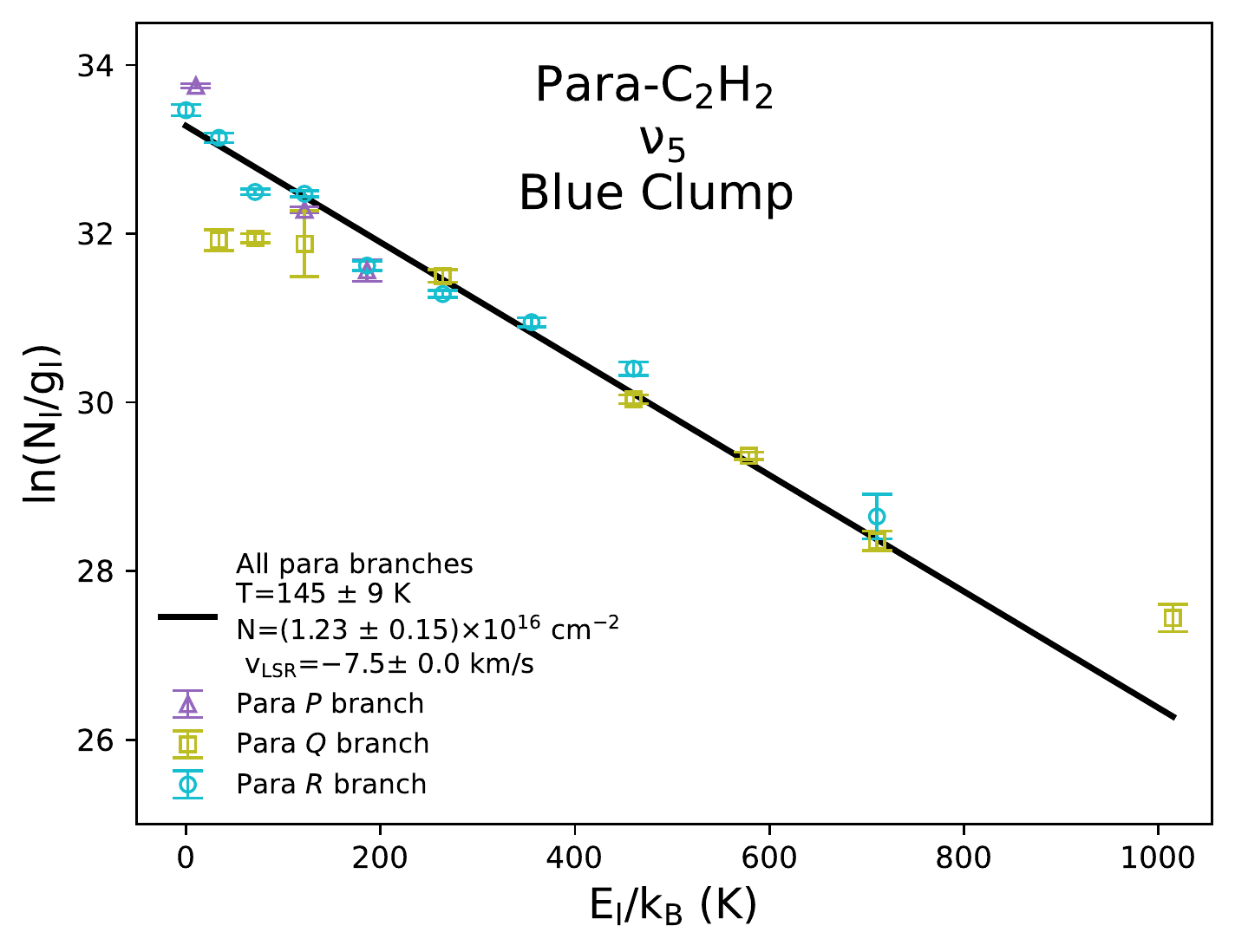}{0.49\textwidth}{}\hspace{-5mm}
          \fig{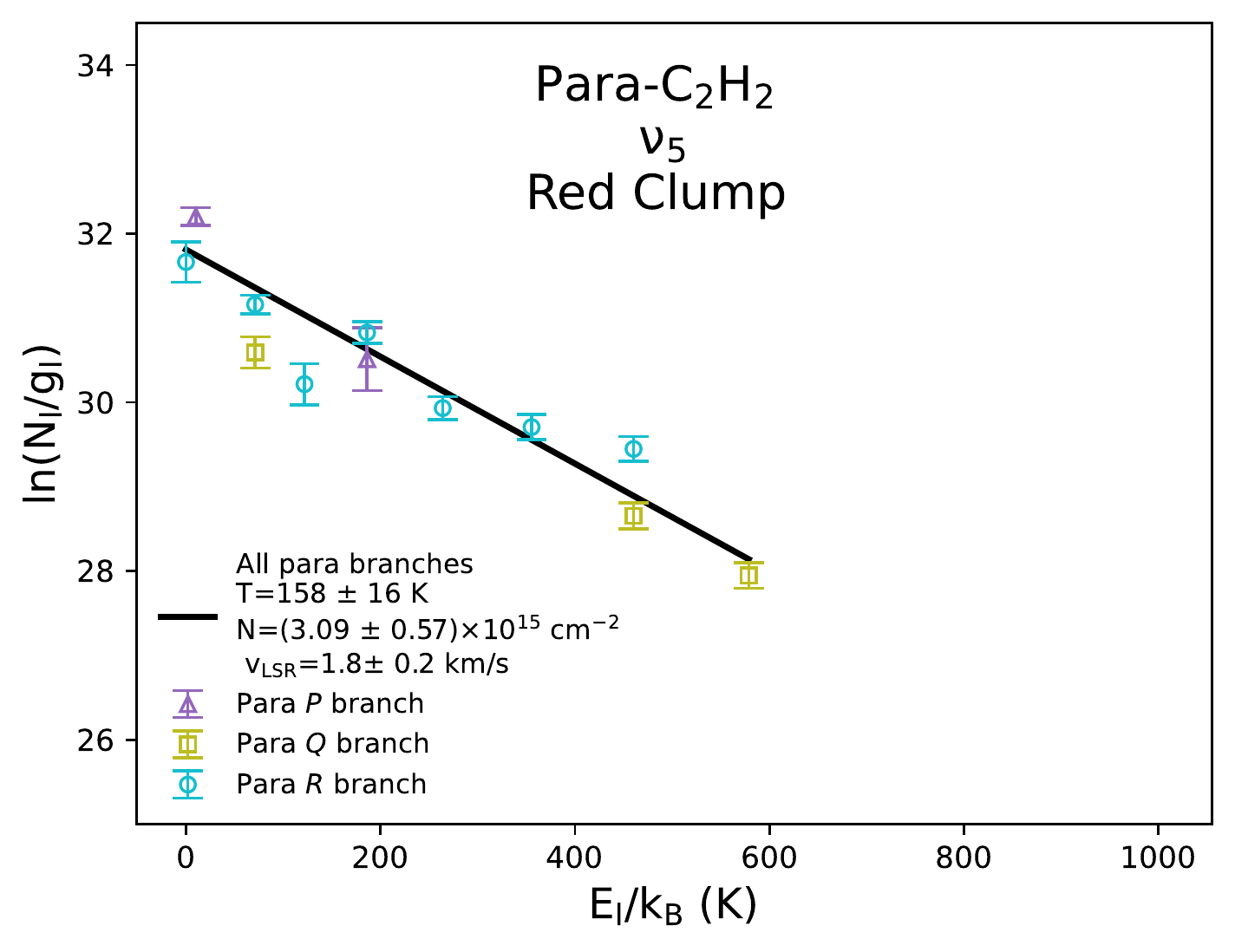}{0.49\textwidth}{}}
\vspace{-10mm}
\gridline{\fig{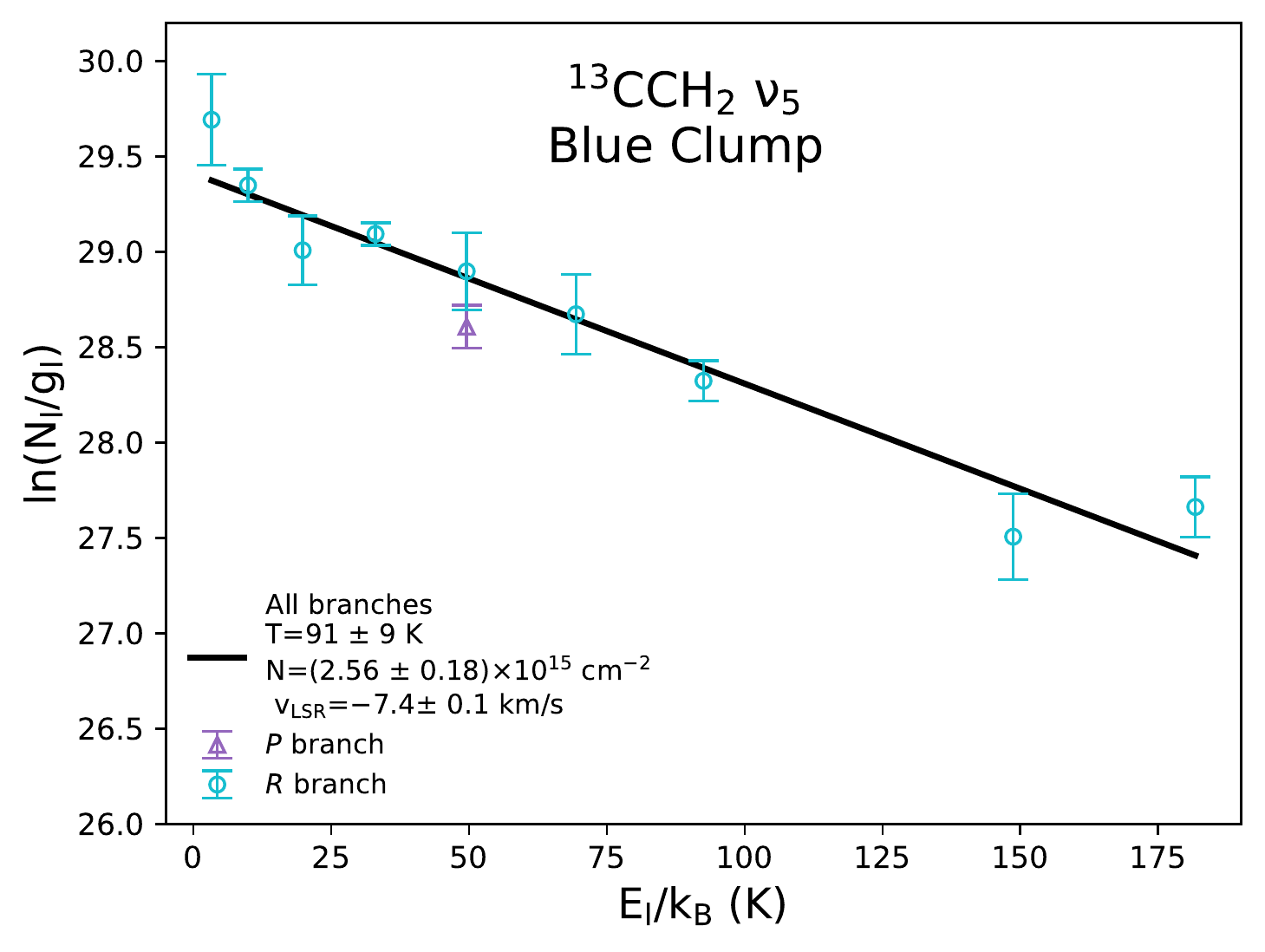}{0.49\textwidth}{}\hspace{-5mm}
          \fig{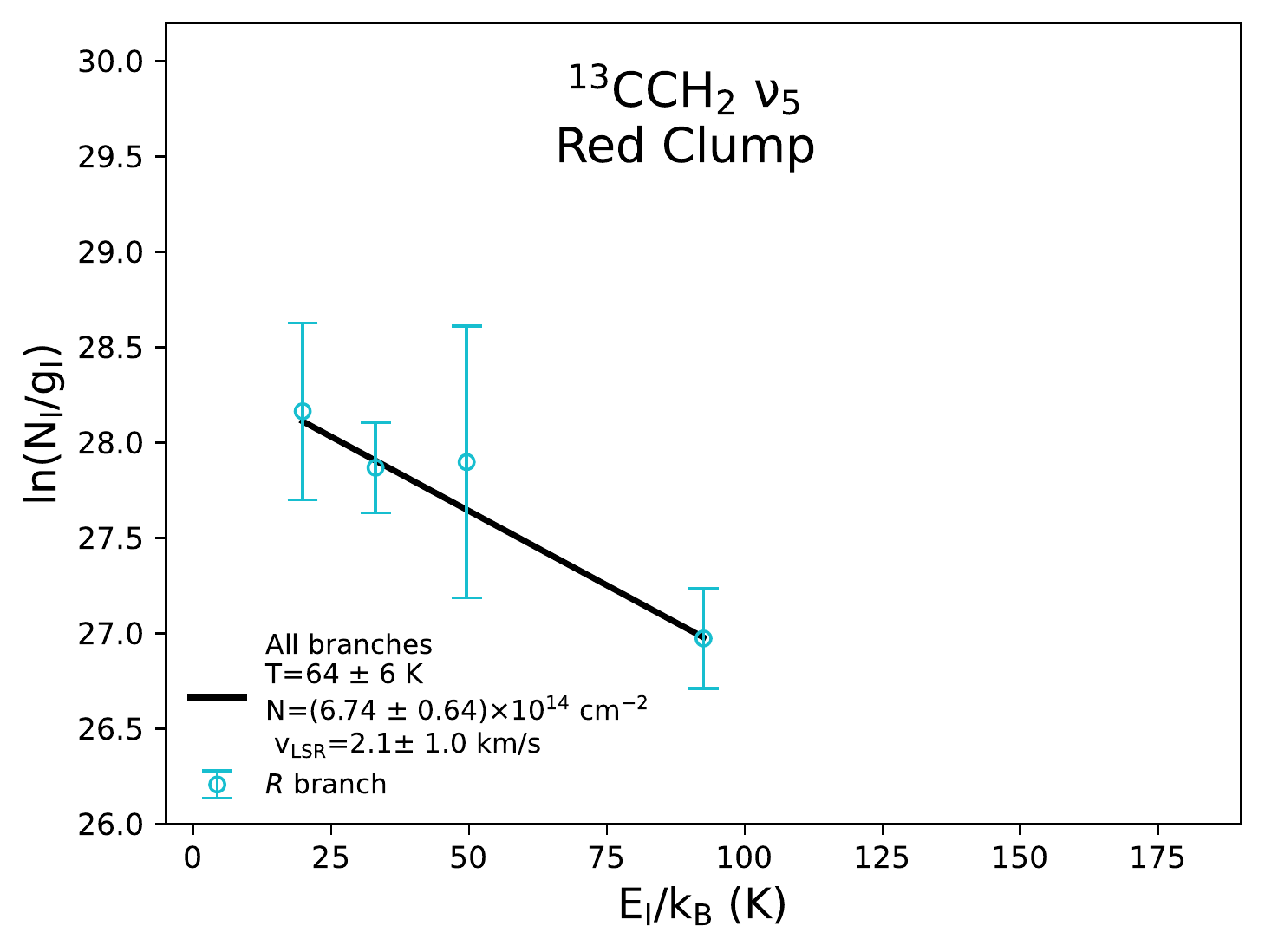}{0.49\textwidth}{}}
\caption{Rotation diagrams for the blue clump (left column) and red clump (right column): the $\nu_5$ band of ortho-\acet\ (top row), the $\nu_5$ band of para-\acet\ (middle row), and the $\nu_5$ band of \acetiso\ (bottow row). \label{fig:rot1}}
\end{figure*}

\begin{figure*}
\centering
\gridline{\fig{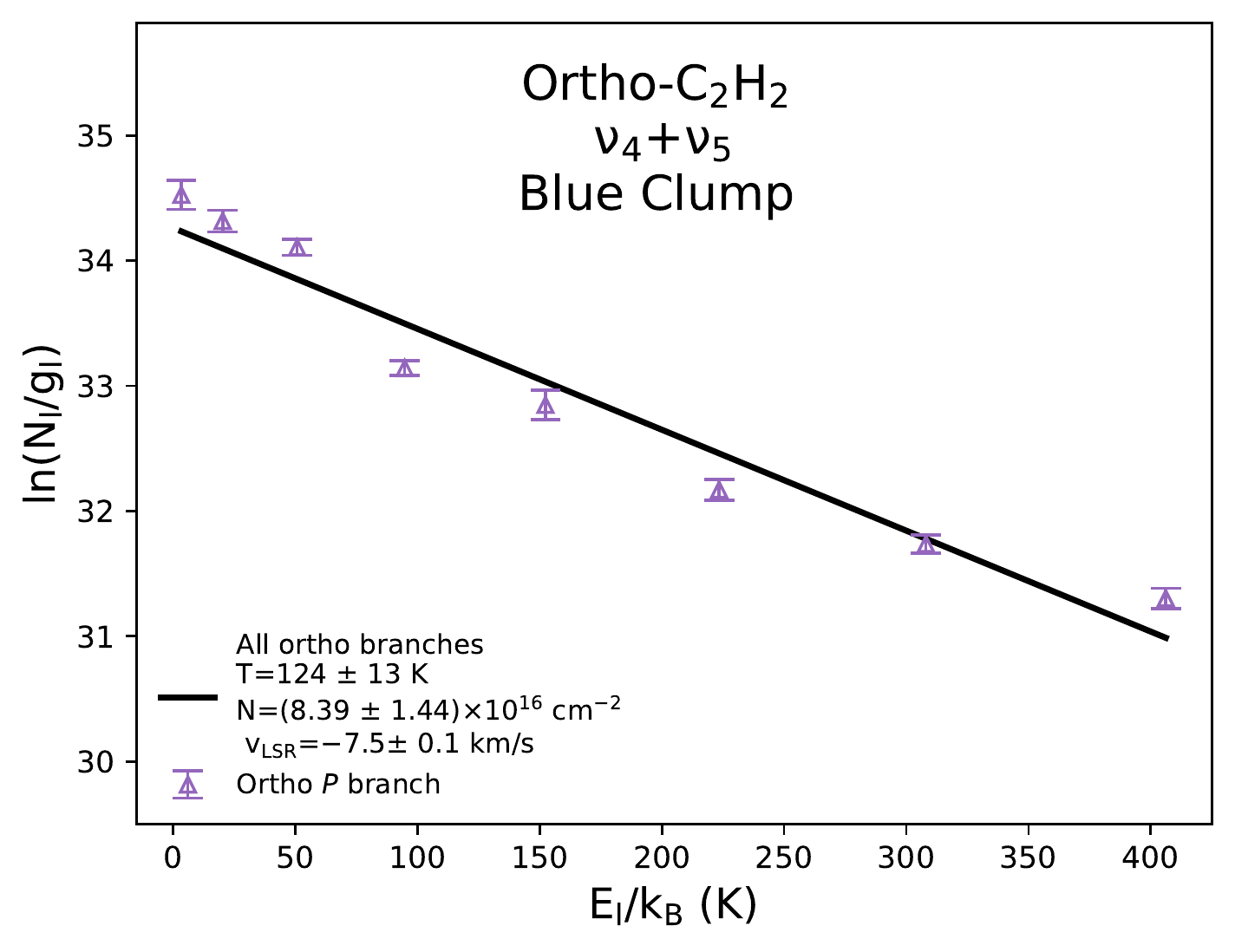}{0.49\textwidth}{}\hspace{-5mm}
          \fig{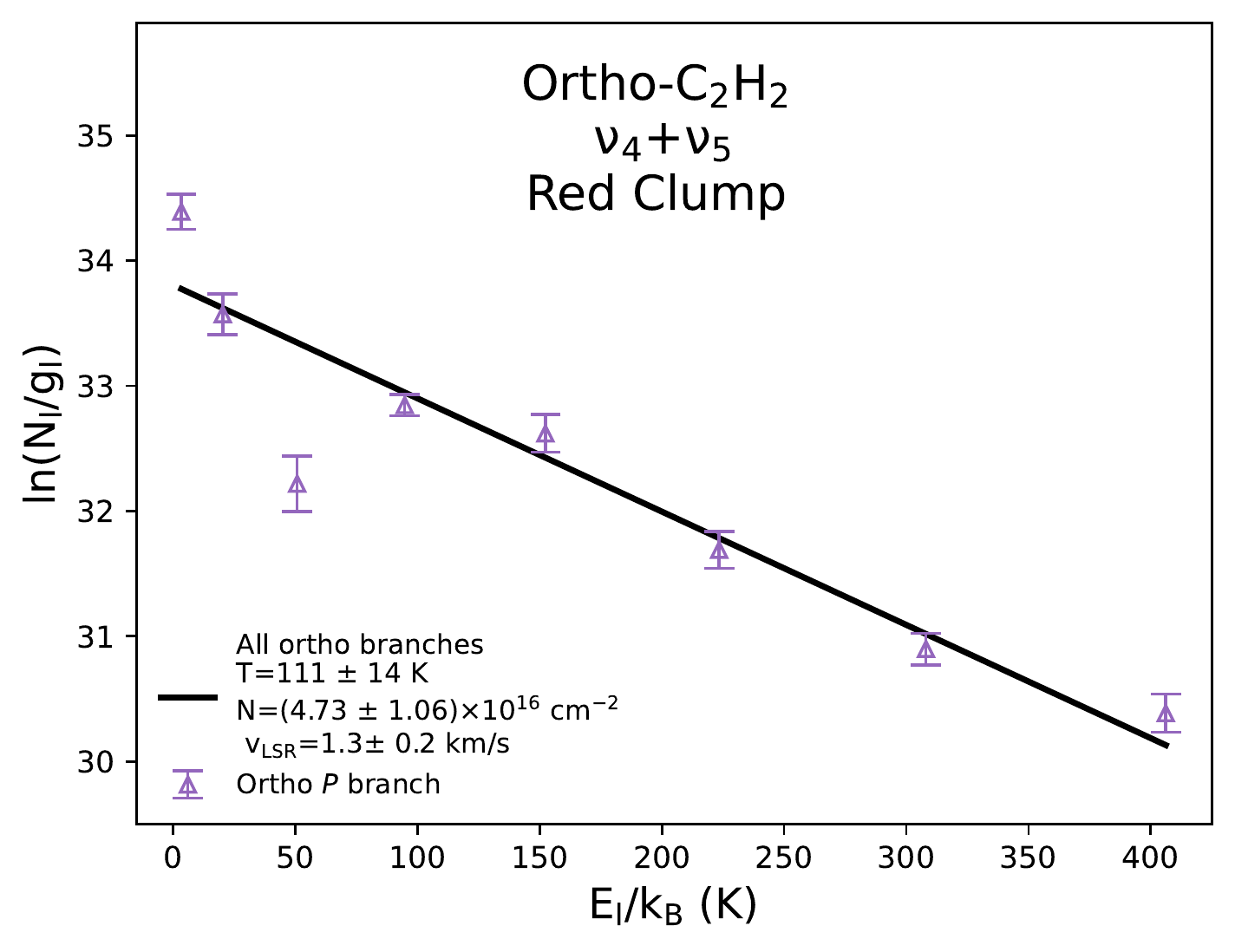}{0.49\textwidth}{}}
          \vspace{-10mm}
\gridline{\fig{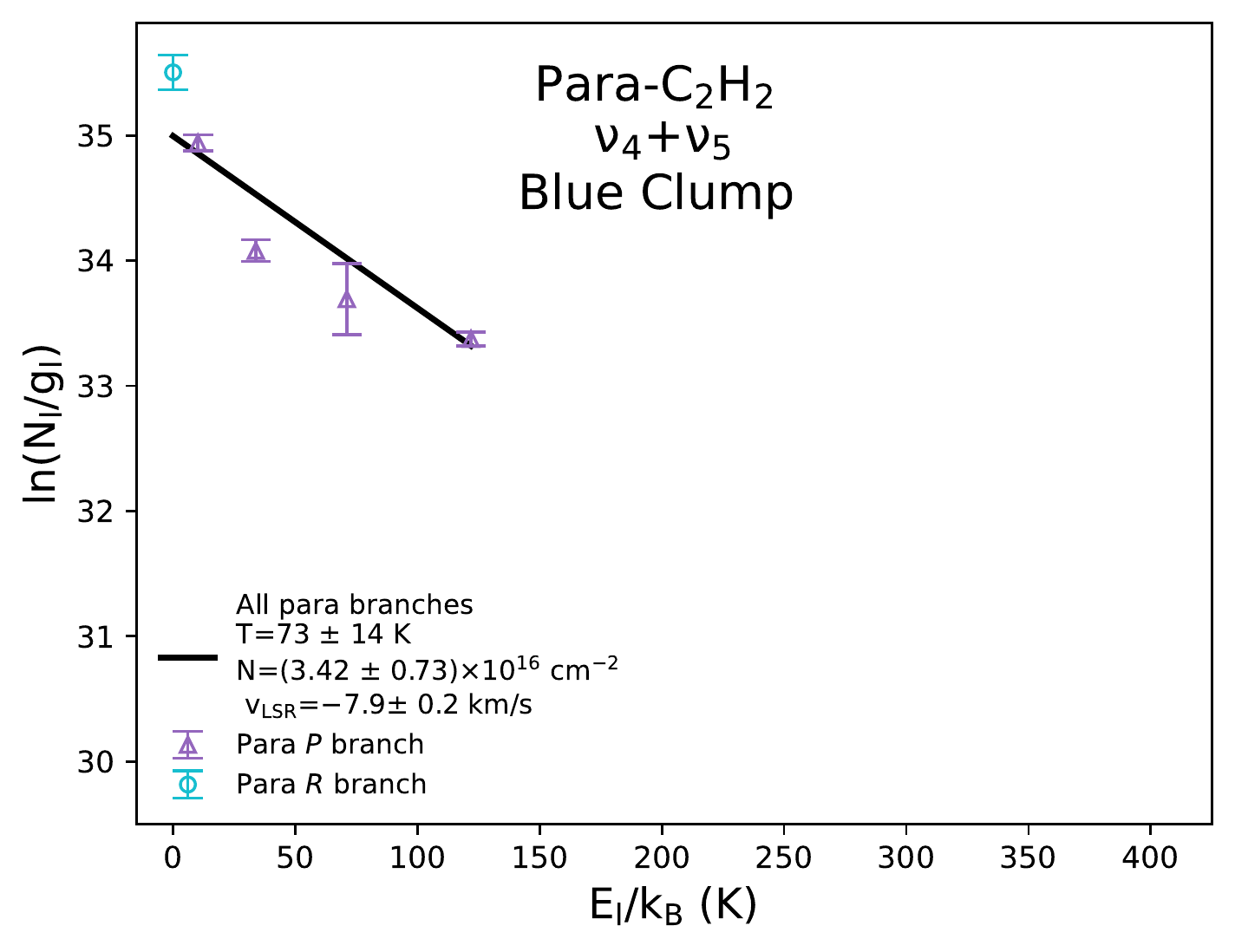}{0.49\textwidth}{}\hspace{-5mm}
          \fig{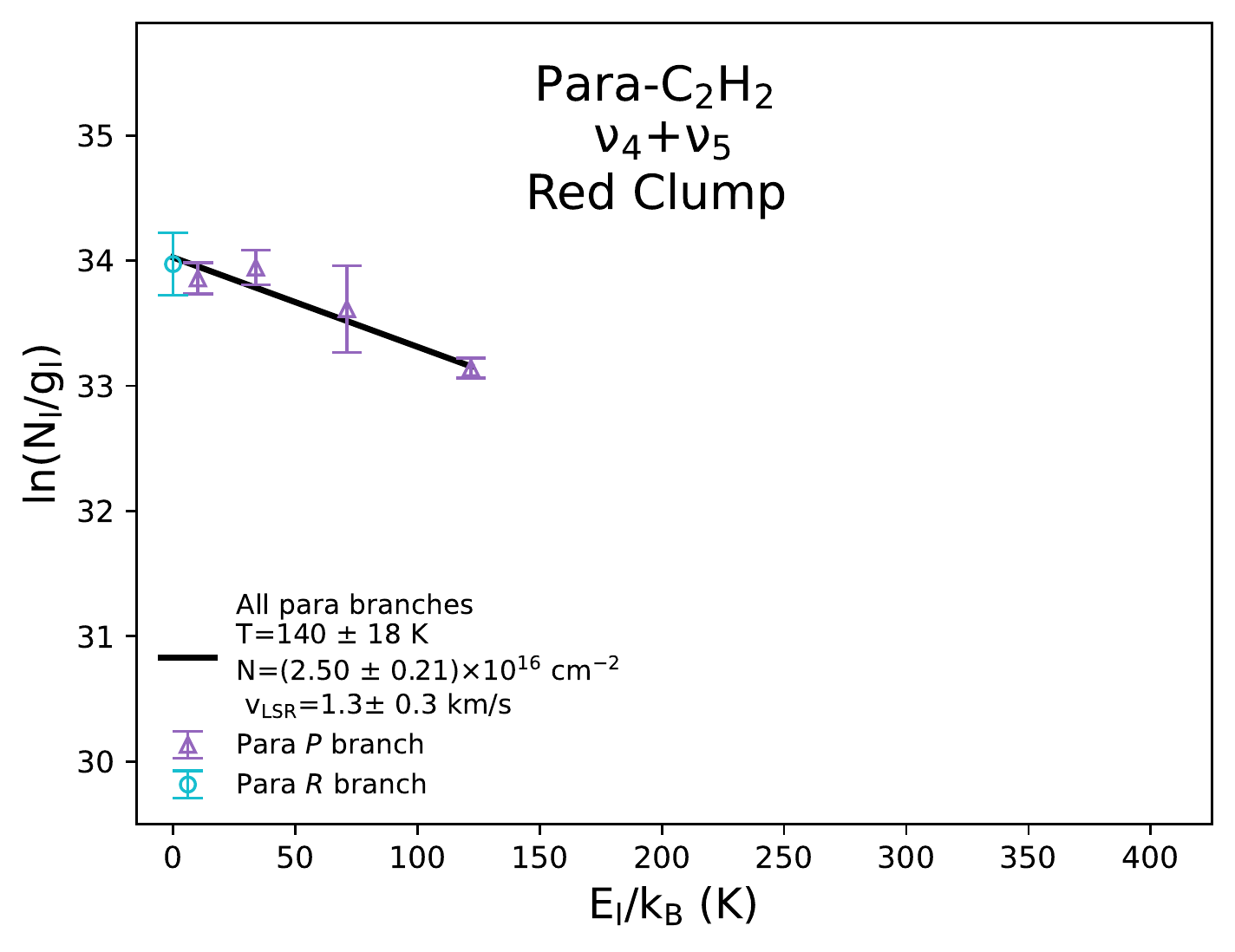}{0.49\textwidth}{}}
        \vspace{-10mm}
\gridline{\fig{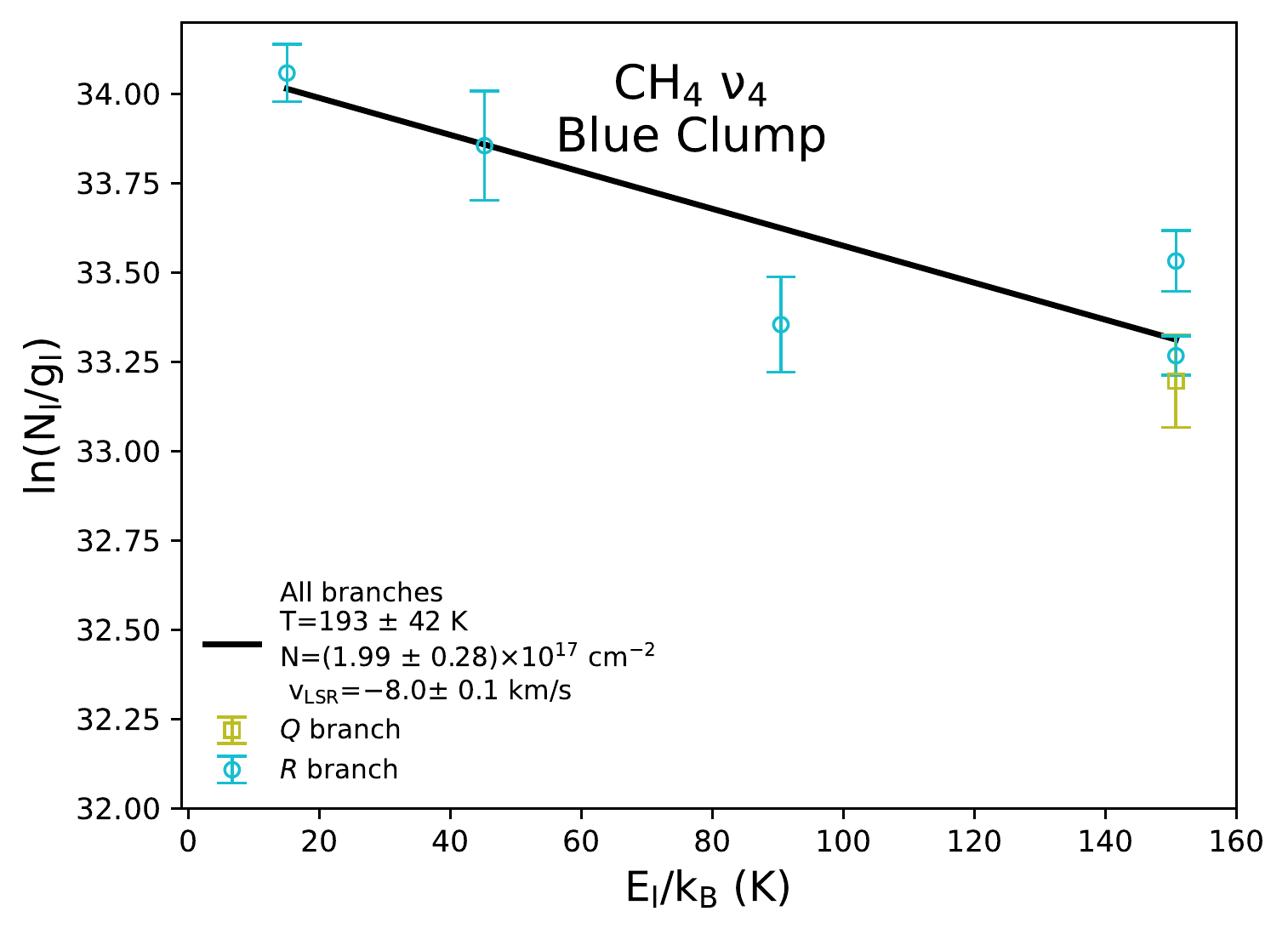}{0.49\textwidth}{}\hspace{-5mm}
          \fig{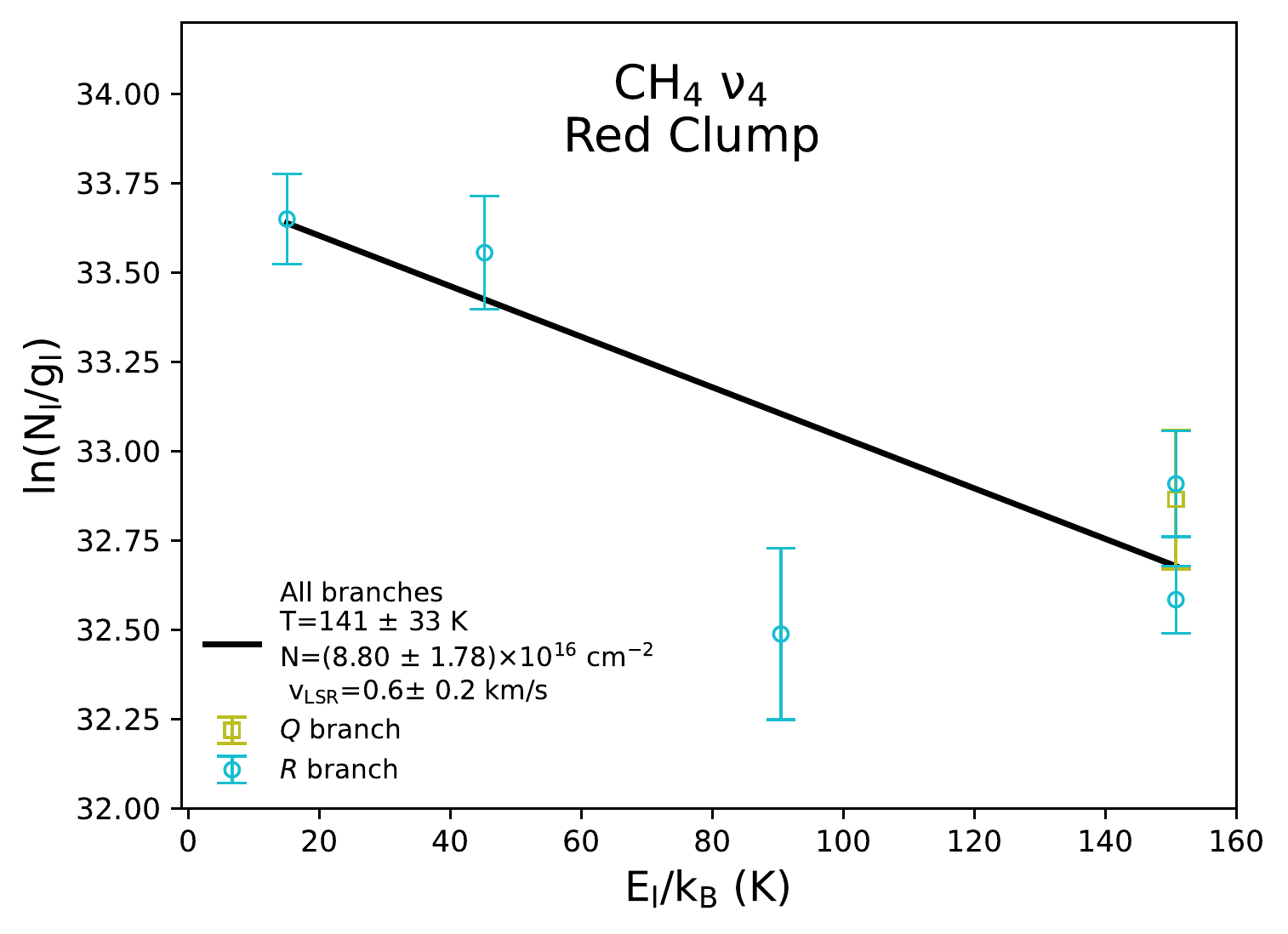}{0.49\textwidth}{}}
\caption{Rotation diagrams for the blue clump (left column) and red clump (right column): the $\nu_4+\nu_5$ band of ortho-\acet\ (top row), the $\nu_4+\nu_5$ band of para-\acet\ (middle row), and \meth\ (bottom row). \label{fig:rot2}}
\end{figure*}

\begin{figure*}
\centering
\gridline{\fig{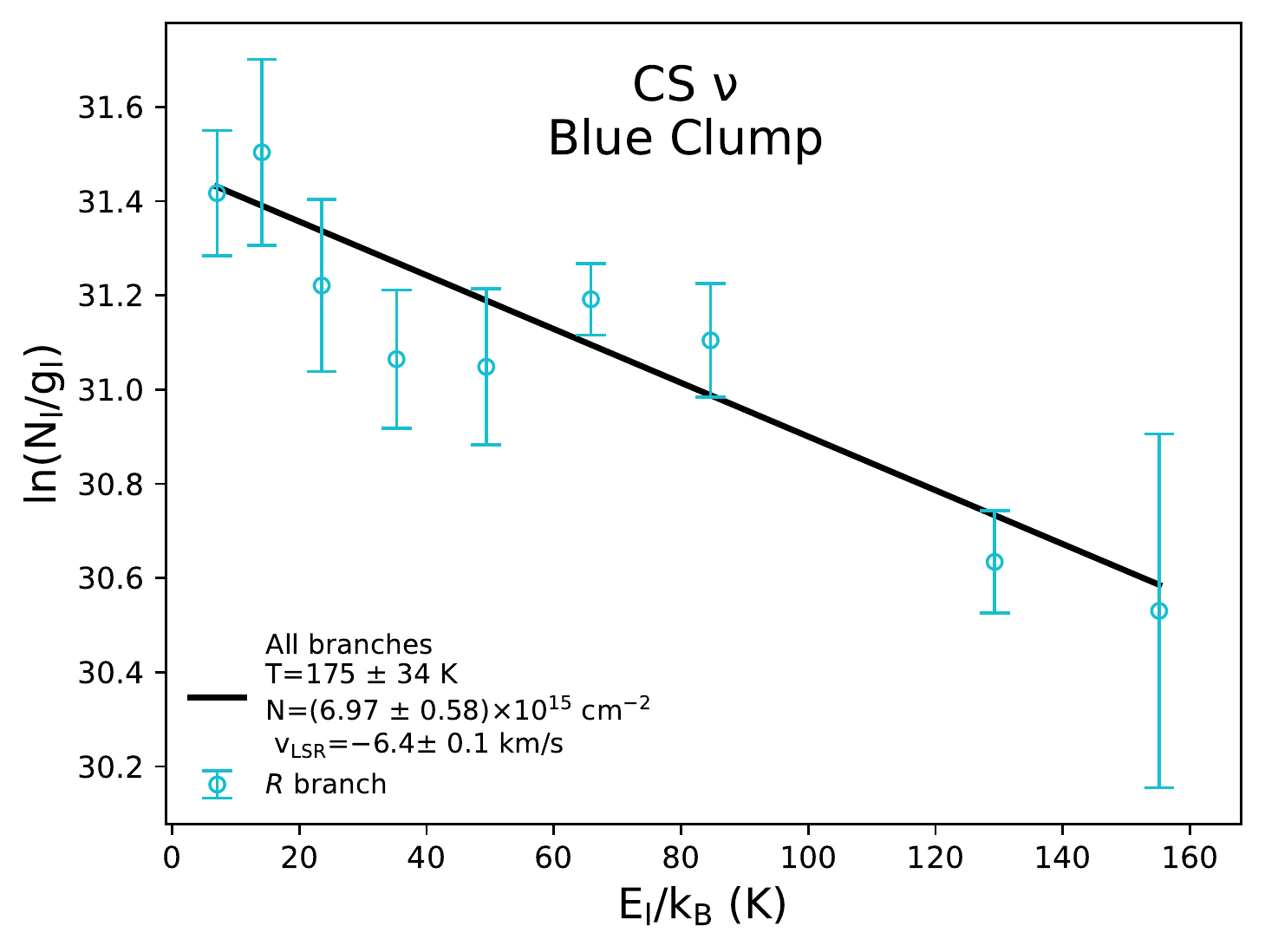}{0.49\textwidth}{}\hspace{-5mm}
          \fig{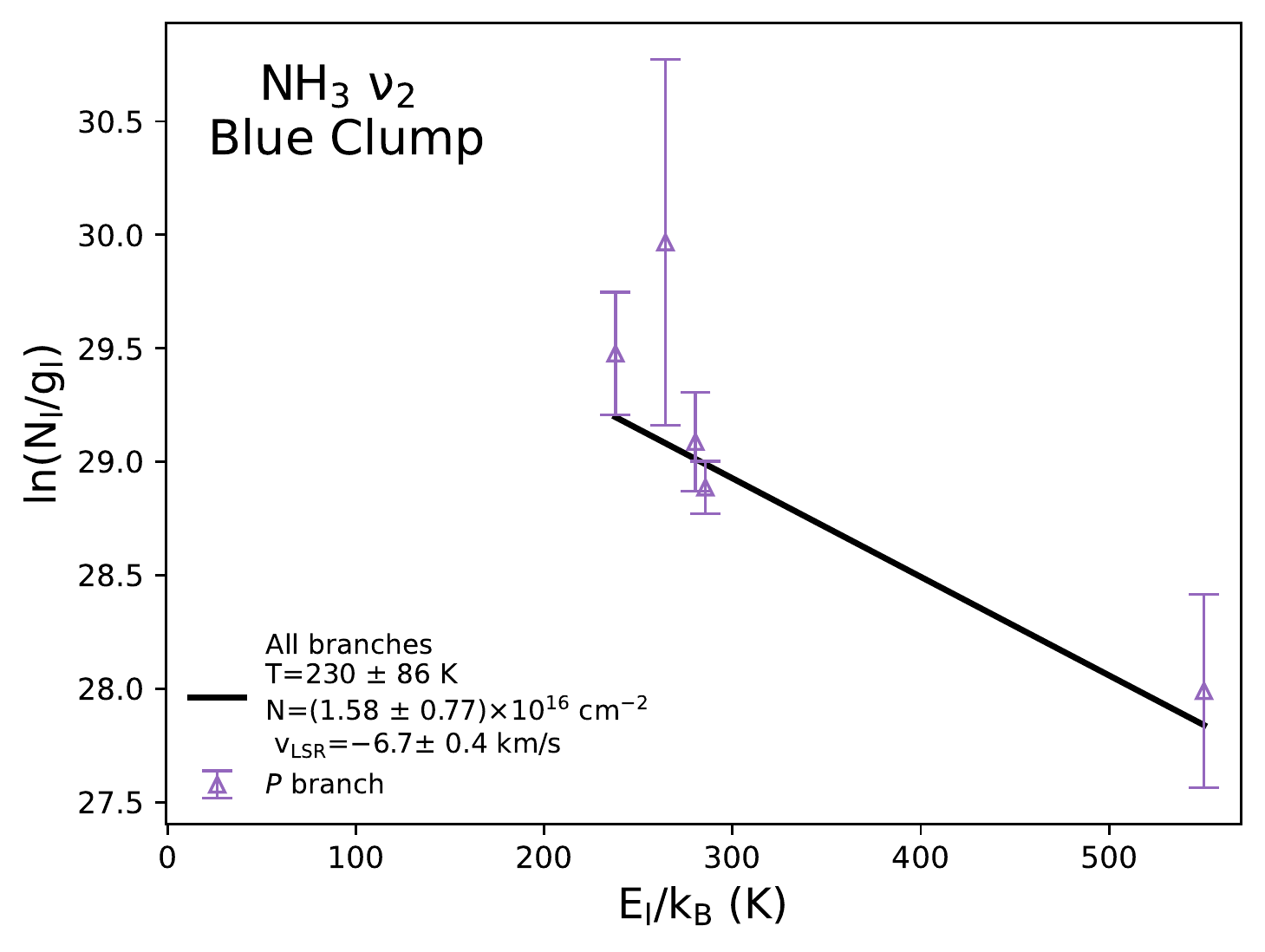}{0.49\textwidth}{}}
          \vspace{-10mm}
\gridline{\fig{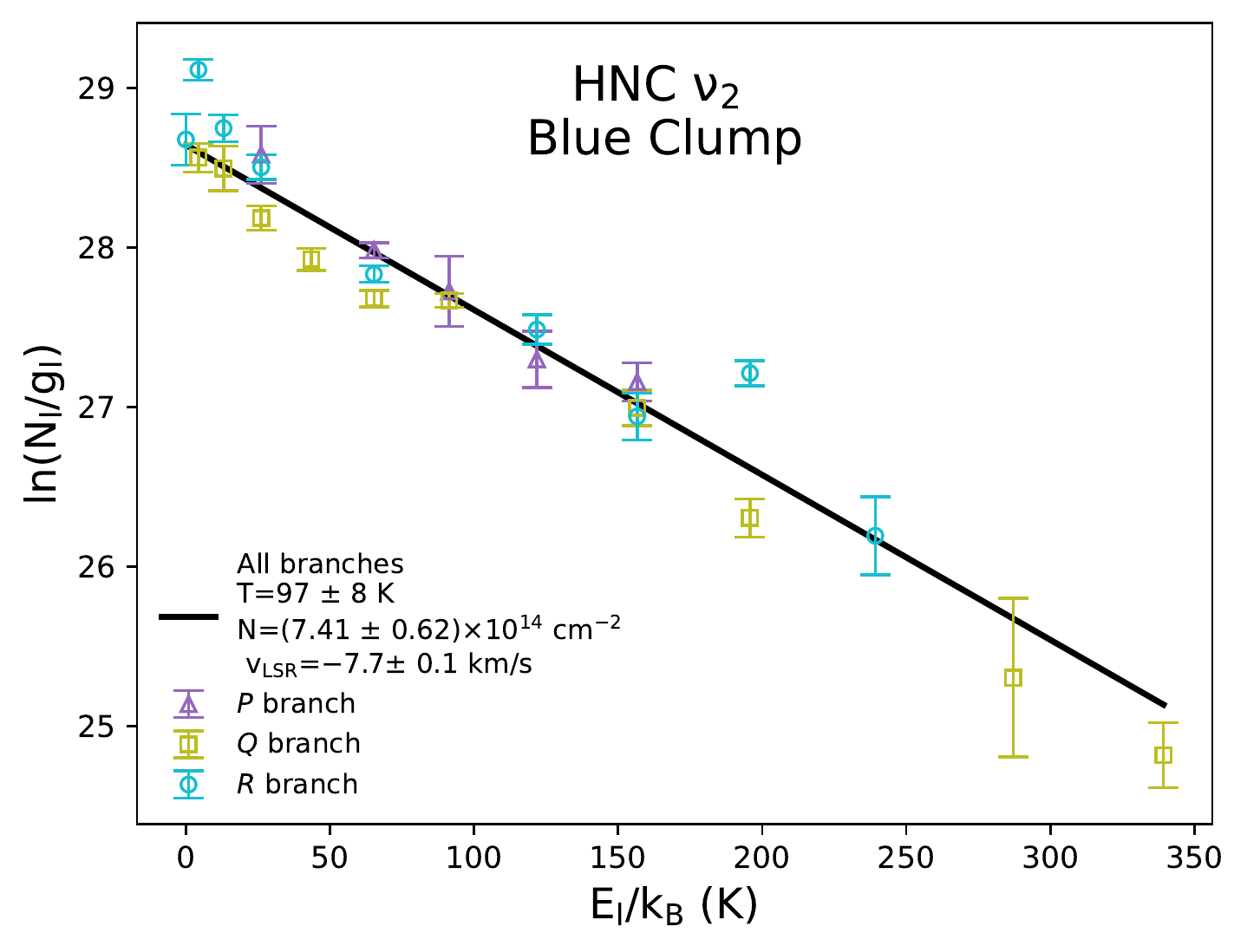}{0.49\textwidth}{}\hspace{-5mm}
          \fig{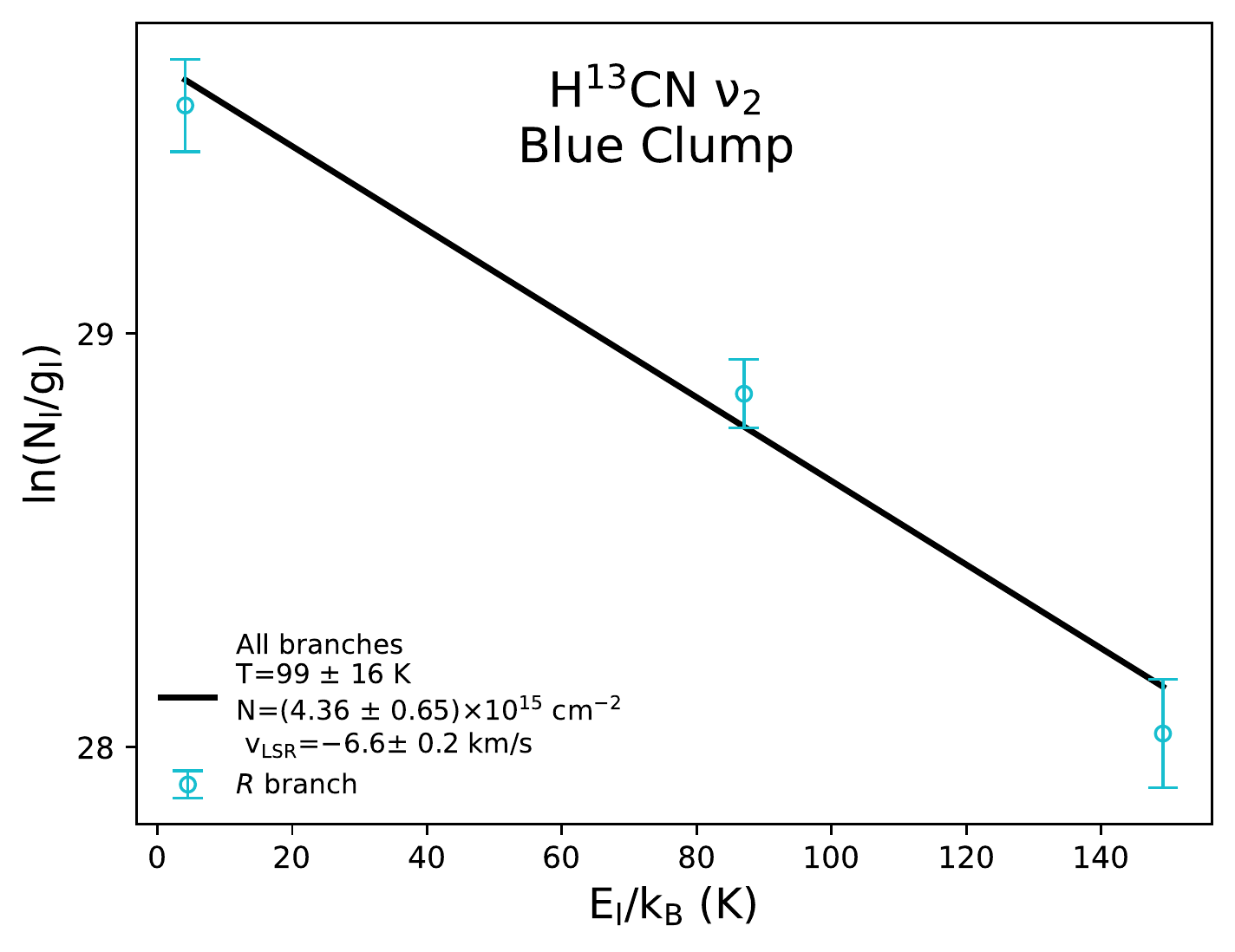}{0.49\textwidth}{}}
\vspace{-10mm}
\gridline{\fig{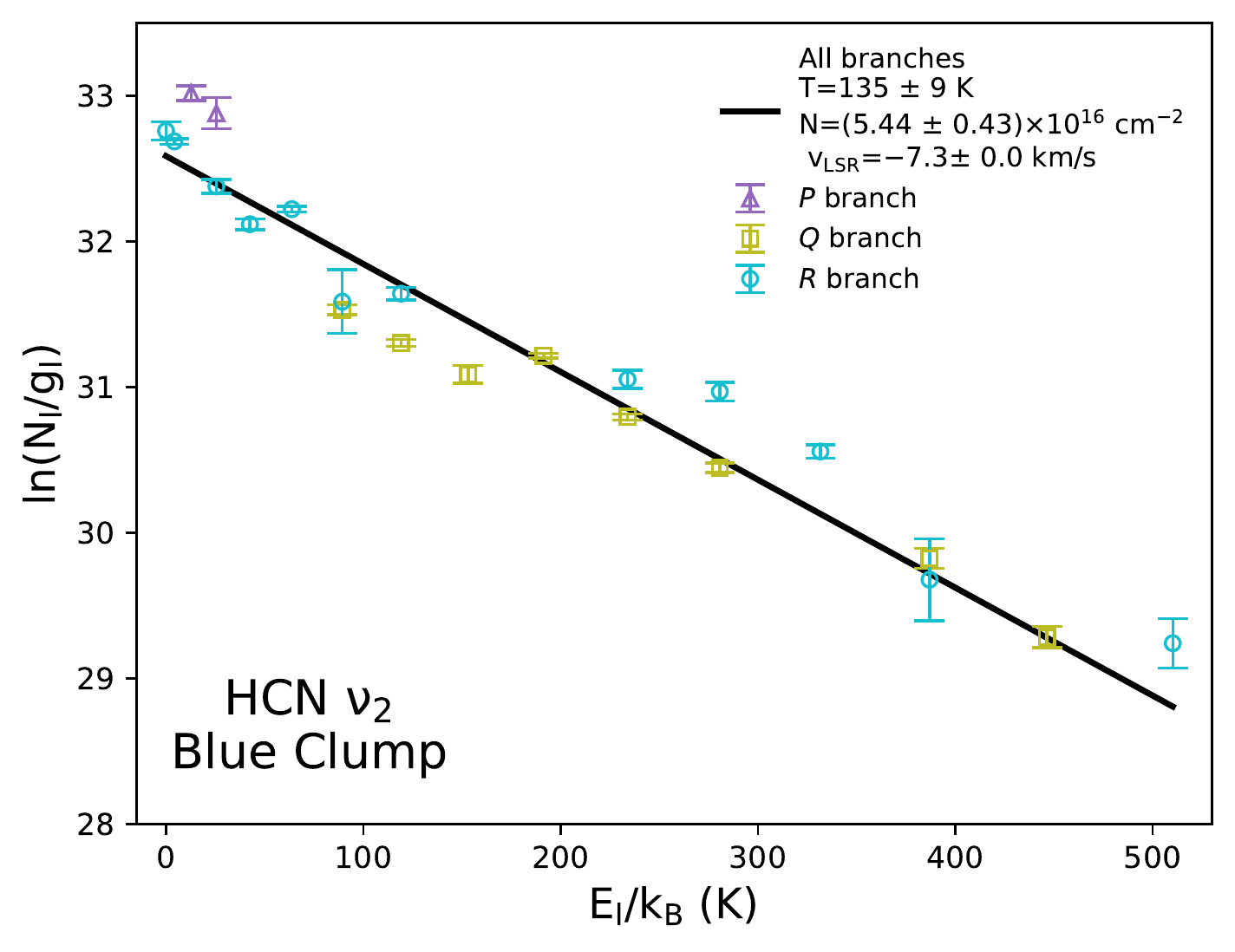}{0.49\textwidth}{}\hspace{-5mm}
          \fig{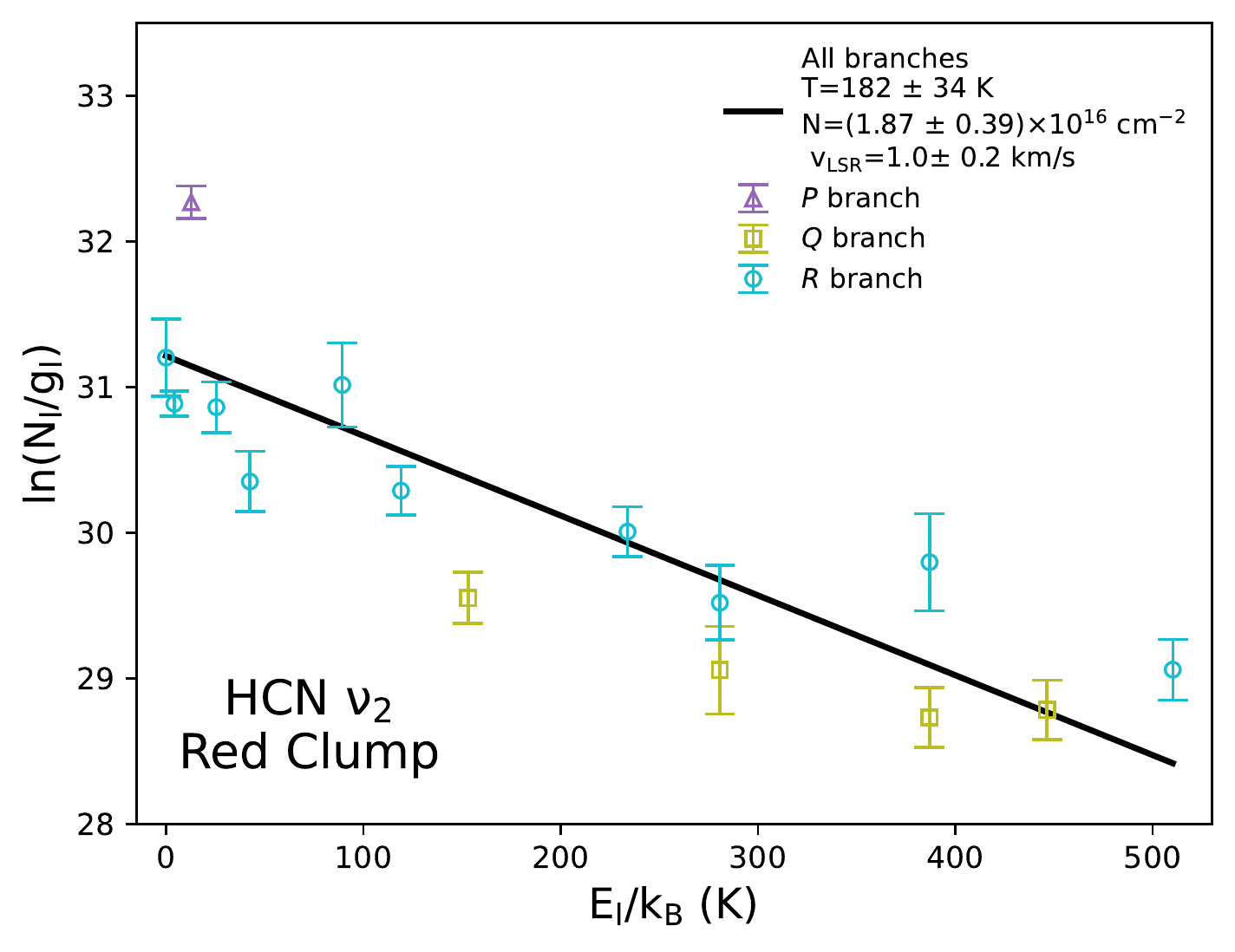}{0.49\textwidth}{}}
\caption{Rotation diagrams for: CS (top left), \amm\ (top right), HNC (middle left), \hcniso\ (middle right), blue clump HCN (bottom left), and red clump HCN (bottom right). \label{fig:rot3}}
\end{figure*}

Figures \ref{fig:rot1} to \ref{fig:rot3} give the rotation diagrams for absorption species. We fit all three branches, \textit{P}, \textit{Q}, and \textit{R}, to a single line. For \acet, we treat the ortho and para ladders as separate species fit by separate lines as in \citet{Rangwala2018}. The species generally show a linear relationship between $\ln(N_l/g_l)$ and $E_l/k_B$, implying that the LTE approximation holds. 

The one exception is that the \textit{Q} branch of the \acet\ $\nu_5$ band in the blue clump flattens towards lower transition energies. A number of these lines are compromised by proximity to atmospheric lines (see Table \ref{tab:abslines}), but another explanation might be that the \textit{Q} branch of \acet\ exhibits non-LTE behavior. As explained in \citet{Rangwala2018} we rule out the need for a covering factor, and optical depth effects because our lines do not have flat bottoms and the optical depth varies without a cap. In high resolution MIR spectra of \water\ towards the hot core AFGL 2136, \citet{Indriolo2020} noted that deeper transitions tend to be ``underpopulated'', possibly because the absorbing gas is mixed with emitting dust. We tested for this effect in our data, but found no correlation between optical depth and underpopulation.

Some rotation diagrams show much higher scatter than others (such as HCN being more scattered than HNC). This may be caused by imperfect atmospheric removal or other incompletely corrected artifacts in the data. Division with the ATRAN model cannot fully quantify how much the model deviates from observations. We do note however, that our fits to individual lines have evenly distributed residuals regardless of how impacted they were by the atmosphere. This means that the atmospheric correction had a systematic effect, either raising or lowering the flux of each affected line, thereby causing scatter in the rotation diagrams.

For HCN, HNC, and the $\nu_5$ band of \acet, the Q branch transitions appear to be slightly weaker than the R branch transitions. This may be due to an effect demonstrated by radiative transfer modelling in \citet{Lacy2013}: Q branch absorption lines are weaker compared to R branch by a factor of about 2/3. There are not enough measured P branch transitions to comment on those in our data, but \citet{Lacy2013} expects their strength to be comparable to R branch transitions.

\begin{deluxetable}{lll}
\tablecaption{Abundance Ratios \label{tab:abundance}}
\tablehead{\colhead{Species} & \colhead{Band} & \colhead{$N/N_{\mathrm{H}_2}$}}
\startdata
\acet\ Ortho&$\nu_5$&$(9.78\,\pm\,5.73)\times 10^{-8}$\\
\acet\ Para&$\nu_5$&$(8.10\,\pm\,4.76)\times 10^{-8}$\\
\acetiso&$\nu_5$&$(1.70\,\pm\,0.99)\times 10^{-8}$\\
\acet\ Ortho&$\nu_4+\nu_5$&$(6.91\,\pm\,4.11)\times 10^{-7}$\\
\acet\ Para&$\nu_4+\nu_5$&$(3.12\,\pm\,1.85)\times 10^{-7}$\\
\meth&$\nu_4$&$(1.51\,\pm\,0.89)\times 10^{-6}$\\
CS&$\nu$&$(3.67\,\pm\,2.15)\times 10^{-8}$\\
HCN&$\nu_2$&$(3.85\,\pm\,2.25)\times 10^{-7}$\\
\hcniso&$\nu_2$&$(2.29\,\pm\,1.37)\times 10^{-8}$\\
HNC&$\nu_2$&$(3.90\,\pm\,2.28)\times 10^{-9}$\\
\amm&$\nu_2$&$(8.32\,\pm\,6.29)\times 10^{-8}$\\
\sotwo&$\nu_2$&$(3.25\,\pm\,1.89)\times 10^{-7}$\\
\sotwo&$\nu_3$&$(5.79\,\pm\,3.36)\times 10^{-7}$\\
\enddata
\tablecomments{The column density, $N$, for each species is the sum of both the blue and red clumps, and $N_{\mathrm{H}_2}$ is the column density of \htwo\ along the line of sight towards IRc2, $1.9\pm1.1\times10^{23}$ \csi\ \citep{Evans1991}.}
\end{deluxetable}

\subsection{Crowded Lines}
\label{sec:crowd}
\sotwo\ transitions are characteristically a dense forest that is too crowded to allow for rotation diagram analysis via fitting individual lines to Gaussians (Figure \ref{fig:flux3}, bottom two panels). Instead, we fit simulated spectra to the normalized EXES flux using a Bayesian approach with the Markov chain Monte Carlo (MCMC) ensemble sampler \texttt{emcee} \citep{Foreman-Mackey2013}\footnote{\url{https://github.com/dfm/emcee}}.

We create the simulated spectra by working backwards from Equations \ref{eqn:rot}, \ref{eqn:nl}, and \ref{eqn:gausabs}. Treating the total column density $N$, temperature $T$, line centre LSR velocity $v_{\mathrm{LSR}}$, and line width $\sigma_{\nu}$ as input variables along with the properties of each transition from HITRAN \citep{Gordon2017}, we generate simulated spectra that are a superposition of the Gaussian profiles of all transitions in the range of interest.

To prepare the normalized flux for fitting, we exclude from the flux regions of noise, atmospheric lines, or lines of other molecules. What remains is a ``scrubbed'' flux consisting of only the baseline and \sotwo\ transitions. We minimize the following likelihood function,
\begin{equation}
    \sum_{i}^{\mathrm{pixels}} \Big\Vert F_{\mathrm{sim},i}(\log(N),T,\sigma_{\nu},v_{\mathrm{LSR}})-F_{\mathrm{scrub},i} \Big\Vert,
    \label{eqn:like}
\end{equation}
where $F_{\mathrm{sim},i}$ is the simulated flux at pixel $i$, and $F_{\mathrm{scrub},i}$ is the scrubbed flux at pixel $i$. For minimization we use the \texttt{scipy} routine \texttt{optimize.minimize} \citep{Virtanen2020} to find the values for $N$, $T$, $v_{\mathrm{LSR}}$, and $\sigma_{\nu}$ that generate a simulated spectra that most closely matches the scrubbed flux. These results of the minimization provide sensible initial values for the MCMC sampler.

To set up the MCMC sampler, we adopt priors uniform over their limits for $N$, $T$, and $v_{\mathrm{LSR}}$, and a normal distribution for $\sigma_{\nu}$ centred on the initial value, summarized in Table \ref{tab:prior}. Because $\tau_0 \propto N\sigma_{\nu}$, $N$ and $\sigma_{\nu}$ are not linearly independent they will not work as separate variables with the MCMC sampler unless we assume a non-uniform distribution for $\sigma_{\nu}$. We use Equation \ref{eqn:like} as the likelihood function and the initial values are randomized in a normal distribution close to the results of the minimization.

\begin{deluxetable}{cccc}
\tablecaption{Prior Distribution Functions for the MCMC Sampler\label{tab:prior}}
\tablehead{\colhead{Variable} & \colhead{Distribution} & \colhead{Min Value} & \colhead{Max Value}}
\startdata
log($N$)&Uniform& 1 \csi & 50 \csi\\
$T$&Uniform&1 K & $10^3$ K\\
$v_{\mathrm{LSR}}$&Uniform&$-500$ \kms& $500$ \kms\\
$\sigma_{\nu}$&$(\sigma_{\nu}-\sigma_{\nu0})^2$&0.001 \kms&50\kms\\
\enddata
\tablecomments{$\sigma_{\nu0}$ is the initial value for $\sigma_{\nu}$ as determined by minimization.}
\end{deluxetable}

The \sotwo\ transitions in the data occur at two distinct wavelengths regimes, settings 20.5 \micron\ to 18.4 \micron\ dominated by the $\nu_2$ band, and setting 7.3 \micron\ dominated by the $\nu_3$ band. We run the MCMC sampler on these bands separately, with 64 walkers and chains of 3000 stpdf each. Both bands converged after about 1000 stpdf. We excluded the 18.4 \micron\ setting from the scrubbed flux for the $\nu_2$ band due to it being nosier compared to the other settings. We nonetheless visually inspected the final result against this setting and found that the simulated spectra fit it well.  

\begin{figure*}
\centering
\fig{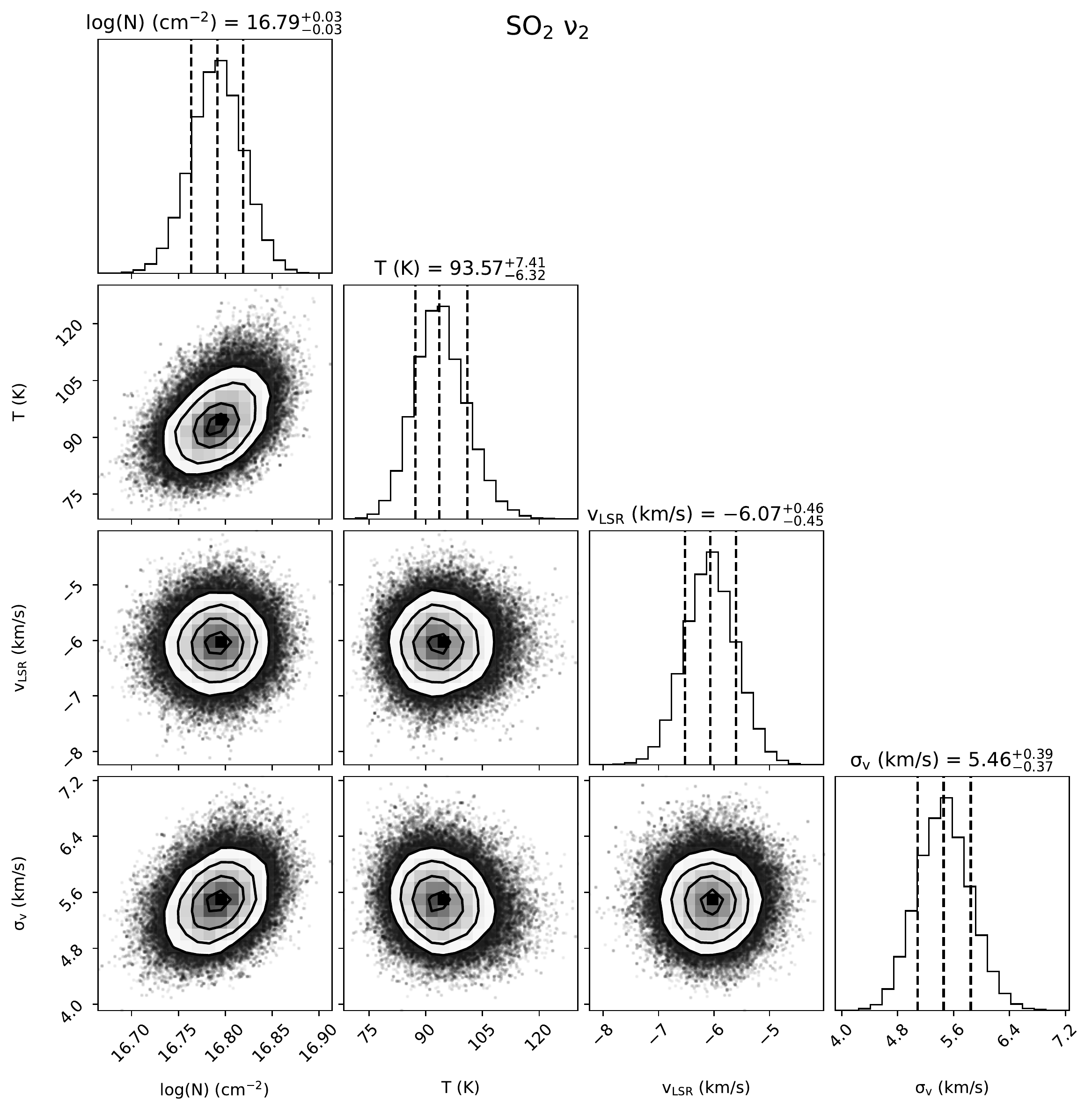}{0.8\textwidth}{}
\caption{The posterior distribution of the log of the total column density ($N$), temperature ($T$), central LSR velocity ($v_{\mathrm{LSR}}$), and line width ($\sigma_{\nu}$) for the wavelengths dominated by \sotwo\ $\nu_2$ resulting from an MCMC sampler. Left, central, and right dotted lines over the histograms correspond to the 16\textsuperscript{th}, 50\textsuperscript{th}, and 84\textsuperscript{th} percentiles respectively. \label{fig:so2cor2}}
\end{figure*}

\begin{figure*}
\centering
\fig{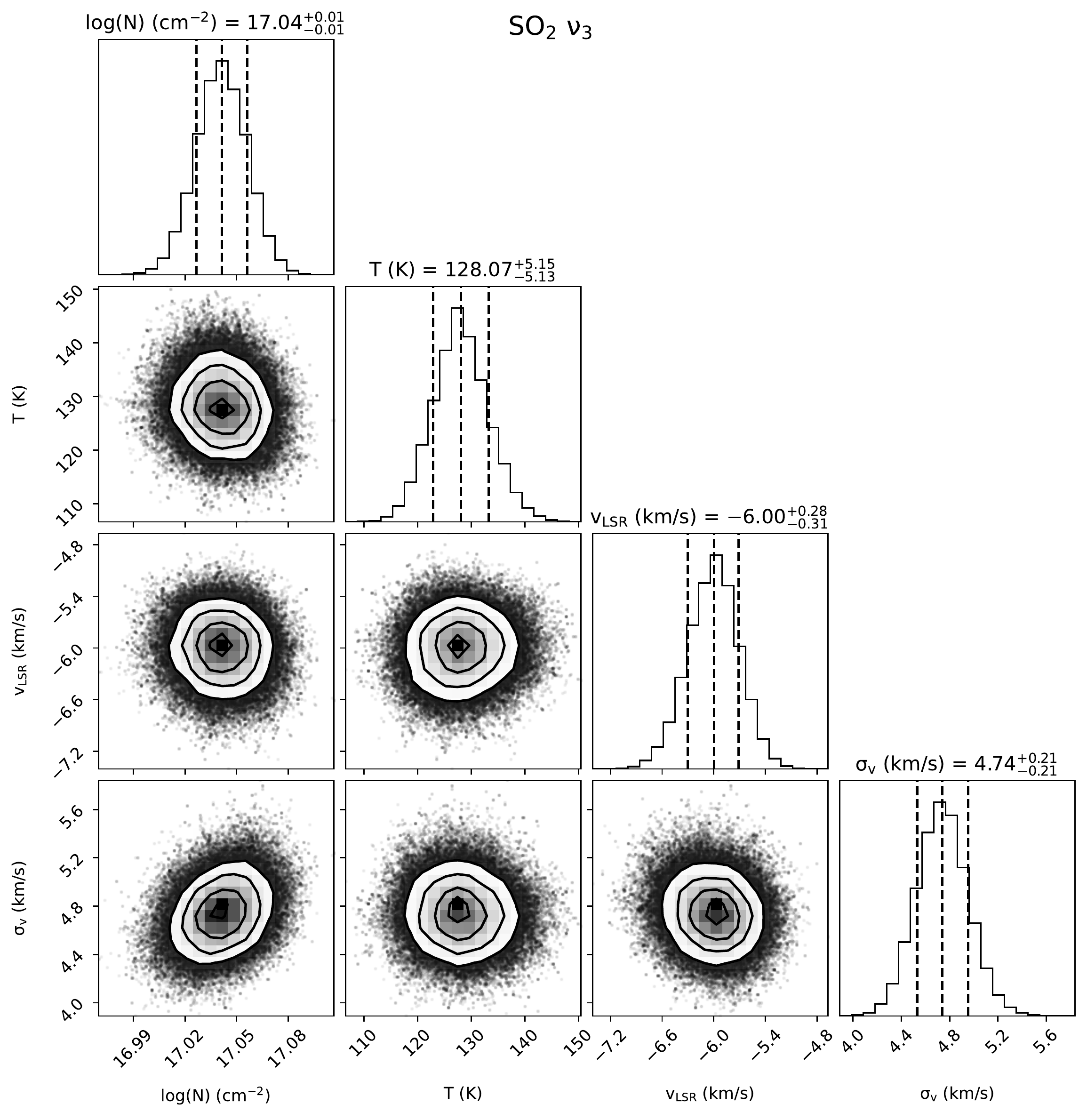}{0.8\textwidth}{}
\caption{Similar to Figure \ref{fig:so2cor2}, but for the wavelengths dominated by \sotwo\ $\nu_3$. \label{fig:so2cor3}}
\end{figure*}

Figures \ref{fig:so2cor2} and \ref{fig:so2cor3} give the resulting posterior distribution of log($N$), $T$, $v_{\mathrm{LSR}}$, and $\sigma_{\nu}$ from the MCMC sampler for the $\nu_2$ and $\nu_3$ bands respectively. For each variable, we take the 16\textsuperscript{th} and 84\textsuperscript{th} percentiles of the posterior distribution as the uncertainties around the median, as reported in Table \ref{tab:rot}. The bottom two panels of Figure \ref{fig:flux3} in Appendix \ref{ap:galflux} shows the simulated spectra for each band. We tested both bands to see if a double Gaussian fit the data better, but it did not. Each band has a single absorption component that fits with the blue clump seen in other absorption lines.

\subsection{Atomic Forbidden Transitions}
\label{sec:ation}

The survey includes four features from atomic forbidden transitions: [NeII] (12.81 \micron), [SIII] (18.71 \micron), [SI] (25.25 \micron), and [FeII] (25.99 \micron). These lines are all in emission, and, except for [FeII], are in emission at both the off-source position and the on-source position towards IRc2 (see \S \ref{sec:obs} for descriptions of the on- and off-source positions). [FeII] is not found in emission or absorption towards IRc2, but is in emission at the off-source position leading to apparent absorption at the on-source position. These lines are shown in Figure \ref{fig:flux4} in Appendix \ref{ap:galflux}.

The [NeII] and [SIII] features are very strong even compared to the background emission, and the lines are fit with Gaussians using the separate on- and off-source  data before subtraction. The [SIII] profile towards the off-source position is not well fit by a single Gaussian and has an extended blue wing.

The [FeII] profile shows emission in the off-source data, but because the emission is so weak compared to the background emission, the best fit is obtained by using the difference spectrum and inverting the apparent absorption spectrum to the actual emission spectrum. 

The [SI] line is also weak compared to the background, and the instrumental bandpass is impossible to accurately remove in the individual on- and off-source spectra. The on-off difference spectrum removes the instrumental bandpass quite well and displays an ``apparent'' reverse P-Cygni profile, with the absorption and emission peaks well separated. Parameters for the [SI] Gaussian fits are obtained by individually fitting the ``absorption'' and emission features in the on-off difference spectrum, converting the absorption values to the equivalent emission values, and confirming that these fits are consistent with the more erratic profiles extracted from the individual on- and off-source spectra.

Table \ref{tab:ationlines} in Appendix \ref{ap:line} gives the observed transitions and inferred parameters for the atomic and ion transition lines.

\section{Discussion} \label{sec:dicu}

\subsection{Kinematic Components}\label{sec:kin}

\begin{figure}
\centering
          \includegraphics[height=0.29\textheight]{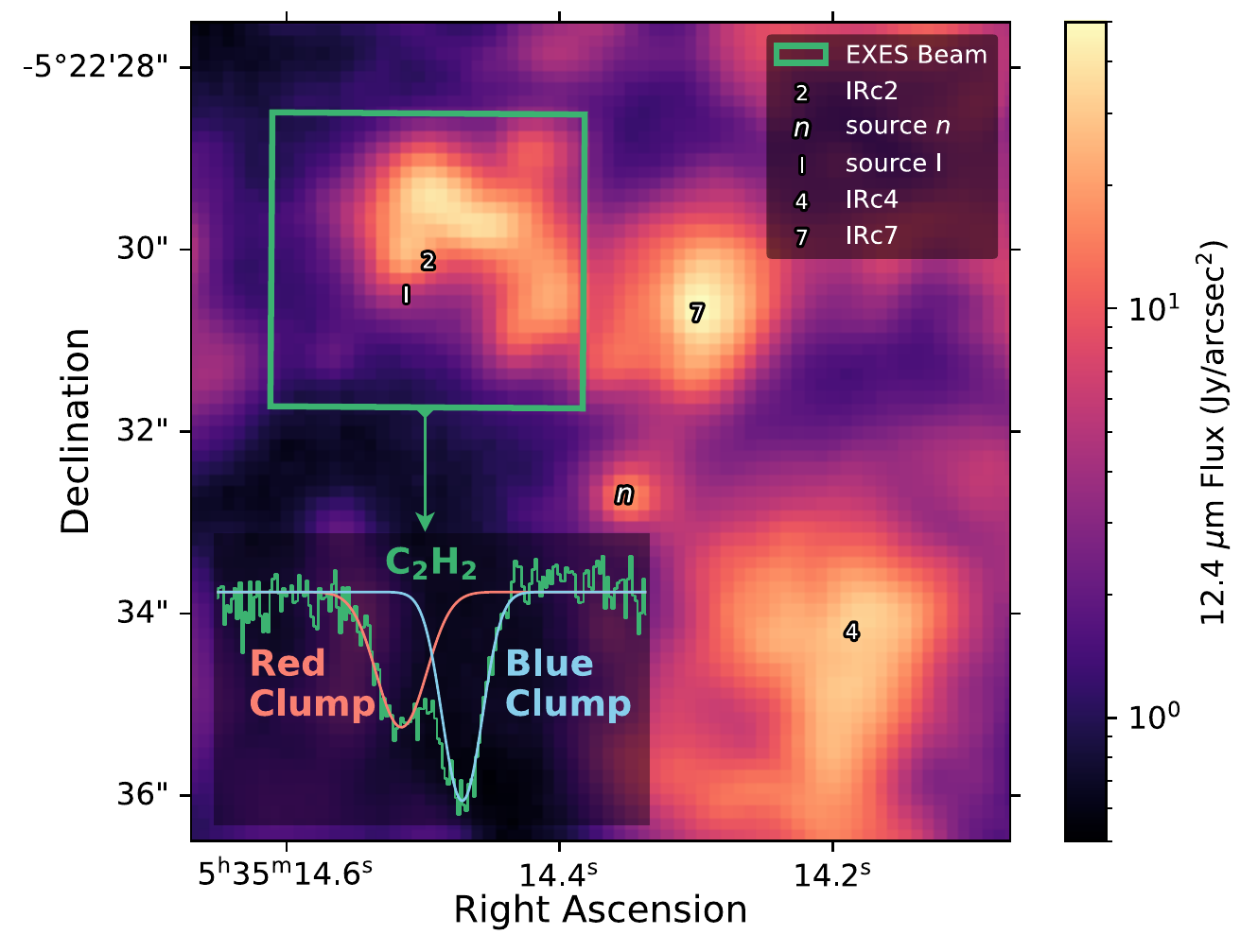}
\caption{Closeup of the IRc2 region from Figure \ref{fig:bigmap}. The \acet\ transition ($\nu_4+\nu_5$ band, P8) clearly has two components, the red clump and the blue clump, as also seen in other species. Colour map is the 12.4 \micron\ flux \citep[SUBARU/COMICS,][]{Okumura2011}. The green box is the EXES beam for the 7.6a \micron\ setting. Symbols refer to regional features discussed in this work placed as described in Figure \ref{fig:bigmap}. \label{fig:mapzoom}}
\end{figure}

Our work covers two new and distinct velocity components in the region, observed by absorption lines in the MIR: the blue clump with \vlsr $\;=-7.1\pm0.7$ \kms\ and the red clump with $1.4\pm0.5$ \kms\ (Figure \ref{fig:mapzoom}).  The top half of Table \ref{tab:comp} shows for these two components the average \vlsr, \vfwhm, and temperatures as well as the species found in this work. While the blue and red clumps have a similar temperature ($\approx140 $ K) and \vfwhm\ ($\approx 8$ \kms), the \vlsr\ of each is distinct. The other striking difference between them is that species column densities in the blue clump are 1.37 to 4.19 times higher than in the red clump (Table \ref{tab:clumpr}). The column density ratio variance shows that the molecular ratios themselves vary in each clump, suggesting different locations and chemical evolution for the blue and red clumps.  We discuss the \htwo\ emission line and its link to the blue and red clumps in \S \ref{sec:h2}. The bottom half of Table \ref{tab:comp} summarizes the properties of the classic components in the region as observed from emission lines in longer wavelength surveys: the hot core, the extended ridge, the compact ridge, and the plateau, as discussed in the introduction.

Because this survey is only pointed towards IRc2, it does not give any information on the spatial extent of the red and blue clumps. However, spectroscopic studies towards other targets offer clues.

Figure 7 in \citet{Lacy2002} illustrates the  HCN R(11) and \acet\ R(7) absorption lines towards IRc2, IRc4, and IRc7 without analysis. The absorption lines towards IRc2 are clearly a double Gaussian with the blue and red clumps. The lines for IRc4 and IRc7 appear asymmetric, but because they are noisier it is unclear if they show the red clump component. What is clear, however, is that the central velocities of the lines towards IRc4 and IRc7 closely match that of those of the blue clump towards IRc2. It suggests the blue clump is at least extended enough to cover IRc4 and IRc7. Figure \ref{fig:mapzoom} shows a map of the morphology of IRc2, IRc4, IRc7, and source \textit{n} as a segment of Figure \ref{fig:bigmap}.

There is evidence that the blue clump extends towards source \textit{n} \citep{Beuther2010}. High resolution spectroscopic observations towards source \textit{n} with VLT/CRIRES from 4.59 to 4.72 \micron\ revealed $^{13}$CO absorption lines with \vlsr$=\;-7$ \kms\ and by rotation diagram analysis  a temperature of $163\pm20$ K and column density $4.3\times10^{18}$\csi. The \vlsr\ matches the average for the blue clump almost exactly ($-7.1\pm0.7$ \kms), while the temperature  falls within the range of temperatures we measured with other species in the blue clump and the column density of $^{13}$CO is higher compared to our species, which is expected. The $^{12}$CO lines were saturated towards source \textit{n} and was not analyzed in \citet{Beuther2010}. We should also note that from the $^{13}$CO measurements they estimated the total \htwo\ towards source \textit{n} to be $1.4\times10^{23}$ \csi, which falls within the range estimated by \citet{Evans1991} towards IRc2 ($N_{\mathrm{H}_2}=(1.9\pm1.1)\times10^{23}$ \csi), showing little variation in the \htwo\ column density towards these two objects. The same work covers IRc3 and there is no evidence for absorption lines matching the blue and red clumps towards this object.

We compared our data to previous observations towards BN, and found that the blue clump does not extend towards BN. From $^{12}$CO and $^{13}$CO absorption lines covering 2 to 5 \micron, \citet{Scoville1983} and \citet{Beuther2010} estimate the \vlsr\ of three components to be $-18$/$-15$, $-3$, and 8/9 \kms. Further observations of \water\ with EXES at 6 \micron\ estimate the \vlsr\ of three similar absorption components to be  $-17$, 0.5, and 8 \kms\ \citep{Indriolo2018}. Though the 0.5 \kms\ \water\ component is quantitatively close to the \vlsr\ of the red clump, it is outside the error bars and it is therefore unlikely that the red clump extends towards BN.

\begin{deluxetable*}{lrrrl}
\tablecaption{Overview of Kinematic Components in Orion BN/KL\label{tab:comp}}
\tablehead{\colhead{Component} & \colhead{\vlsr} & \colhead{\vfwhm} & \colhead{T}&\colhead{Species Detected in This Work}\\
\colhead{} & \colhead{(\kms)} & \colhead{(\kms)} & \colhead{(K)} & \colhead{}}
\startdata
\multicolumn{5}{c}{MIR Components (This Work)}\\
Blue Clump&$-7.1\pm0.7$&$8.9\pm1.8$&$135\pm47$&\acet, \acetiso, \meth, CS, HCN, \hcniso, HNC, \htwo$^*$, \water, \amm, OH$^?$, \sotwo\\
Red Clump&$1.4\pm0.5$&$7.7\pm0.5$&$146\pm52$&\acet, \acetiso, \meth, \htwo$^*$, HCN\\
\hline
\multicolumn{5}{c}{\tablenotemark{a}Classic Components (Sub-mm to Radio Surveys)}\\
Hot Core&\tablenotemark{c}2.5--7.5&5--15&\tablenotemark{b}150--400&---\\
Extended Ridge&\tablenotemark{c}7--11&3--5&55--70&---\\
Compact Ridge&\tablenotemark{c}7--9&3--5&80--150&---\\
Plateau&6--9&$>$20&95--150&---\\
\enddata
\tablecomments{Columns are from left to right: central local standard of rest velocity, line full-width half-maximum, temperature, and species detected in this work only. Numbers are averages for this present work, and a typical range from other works. \tablenotemark{a}{Ranges are compiled from combining \citet{Blake1987,Genzel1989,Tercero2010,Tercero2011,Esplugues2013}} with supplementary data from: \tablenotemark{b}{\citet{Wilson2000}} and \tablenotemark{c}{\citet{Wright1996}}. \htwo, \water, OH, and $2\nu_2$ HCN are not counted towards the average \vlsr\ and \vfwhm\ in this work due to only two or one lines analyzed per species. $^*$ denotes emission lines. $^?$ denotes the tentative detection of OH.}
\end{deluxetable*}

\begin{deluxetable}{llr}
\tablecaption{Column Density Ratio for Blue and Red Clumps \label{tab:clumpr}}
\tablehead{\colhead{Species} & \colhead{Band} & \colhead{$N_{\mathrm{blue\,clump}}/N_{\mathrm{red\,clump}}$}}
\startdata
ortho-\acet&$\nu_5$ &4.19$\,\pm\,$0.93\\
para-\acet&$\nu_5$ &3.98$\,\pm\,$0.88\\
\acetiso&$\nu_5$ &3.80$\,\pm\,$0.45\\
ortho-\acet&$\nu_4+\nu_5$ &1.77$\,\pm\,$0.50\\
para-\acet&$\nu_4+\nu_5$ &1.37$\,\pm\,$0.31\\
\meth &$\nu_4$ &2.26$\,\pm\,$0.56\\
HCN&$\nu_2$ &2.91$\,\pm\,$0.65\\
\enddata
\end{deluxetable}

\subsubsection{The Blue Clump}
\label{sec:blueclump}
The blue clump has an average \vlsr\ of $-7.1\pm0.7$ \kms, \vfwhm\ of $8.9\pm1.8$ \kms, and temperature of $135\pm47$ K. Every species and band observed in absorption in this survey is present in the blue clump: \acet, \acetiso, \meth, CS, HCN, \hcniso, HNC, \water, \amm, \sotwo, and tentative OH. It has higher abundances of each species it shares in common with the red clump (\acet, \acetiso, \meth, and HCN), ranging from $4.19\pm0.93$ more ortho-\acet\ in the $\nu_5$ band, to $1.37\pm0.31$ more para-\acet\ in the $\nu_4+\nu_5$ band (Table \ref{tab:clumpr}). We have previously published EXES blue clump observations of \acet, \acetiso\ \citep{Rangwala2018}, HCN, \hcniso, and HNC \citep{Nickerson2021}. 

Previous studies have also measured these species in absorption towards IRc2 \citep{Evans1991,Wright2000,Boonman2003}. With \textit{ISO}, \citet{Wright2000} measured \water\ absorption in the MIR towards IRc2 at \vlsr$\;=-8$ \kms\ from 24 to 45 \micron. These lines, except for one, are obscured in this survey by the Earth's atmosphere where coverage overlaps. The \vfwhm\ ($\approx 30$ \kms) is much wider than this work's ($8.9\pm1.8$ \kms), possibly because \textit{ISO}'s large beam included more MIR sources than just those towards IRc2 (Figure \ref{fig:bigmap}). In a high resolution study towards IRc2 with IRTF/IRSHELL, \citet{Evans1991} detected the same bands of \acet\ ($\nu_5$ and $\nu_4+\nu_5$), HCN ($\nu_2$), and \amm\ ($\nu_2$) as in this work, as well as  OCS ($\nu_1$) and CO. The \vlsr, however, varied widely across the species ranging from $-10$ \kms\ for CO to 7.1 \kms\ for \acet, and was also inconsistent across multiple transitions for each individual species. This was possibly due to lower resolution and imprecise wavelength calibration. From that work it was unclear if these species belonged to the same component. \citet{Boonman2003} analyzed the strongest $\nu_5$ band of \acet\ and $\nu_2$ band of HCN absorption features towards IRc2 with \textit{ISO}, and estimated abundances comparable to hot core models and other massive protostars.

Because of EXES's higher resolution and smaller beam size, we have for the first time accurately measured the \vlsr\ and the \vfwhm\ of the blue clump consistently across several molecular species. The evidence presented here suggests that the blue clump is a distinct component that has no relationship with the classic components in the region. As laid out in Table \ref{tab:comp}, the blue clump's central velocity, line width, and temperature are distinct from these classic components as defined by previous molecular emission line surveys carried out in the sub-mm to radio wavelengths. 

However, if the blue clump is related to previously studied components in the region, there are two possibilities. The first possibility is that it is part of the low velocity flow, an outflow belonging to the plateau whose \vlsr\ ranges from about 8 to 20 \kms\ with respect to the ambient cloud velocity of 9 \kms\ \citep{Wright1996}. However, this gas is characterized by shocks and we do not find evidence that the blue clump \sotwo\ must be formed by shocks (\S \ref{sec:sulphur}). The other possibility is that the blue clump is associated with the hot core in some capacity. The blue clump abundances are similar to those estimated by longer wavelength line surveys (Table \ref{tab:abunhc} in Appendix \ref{ap:abcompare}), but the velocity of the blue clump is about 9 \kms\ blue-ward of the hot core. A possible scenario for the blue clump is that it may be gas that once belonged to the hot core (\citealt{Boonman2003}), separated by either outflows from Source I or the explosive event that tore the region apart 500 years ago \citep{Bally2017}, that has cooled from former higher hot core temperatures. Regardless of the blue clump's membership with the hot core, its high abundances means that it mimics a hot core-like chemistry with lower temperatures.

\break 
\subsubsection{The Red Clump}
\label{sec:redclump}


The red clump is unique to this EXES survey, having been undetected in previously published MIR studies with other instruments. The red clump has an average \vlsr\ of $1.4\pm0.5$ \kms, \vfwhm\ of $7.7\pm0.5$ \kms, and temperature of $146\pm52$ K, within error of the blue clump. It has lower relative column densities of each species shared with the blue clump (Table \ref{tab:clumpr}). The red clump HCN was first reported in in \citet{Nickerson2021} and in this survey, we find more molecules in the red clump:  \acet\ (both $\nu_5$ and $\nu_4+\nu_5$), \acetiso, and \meth. HCN and \acet\ on average have higher temperatures observed towards the red clump compared to the blue clump, while \acetiso\ and \meth\ have lower temperatures. \meth\ in particular, however, is difficult to measure due to atmospheric interference and its temperatures have large errors. 

The five absorption species that we detect in the blue clump, but not in the red clump are: CS, \hcniso, HNC, \amm, and \sotwo\ (our one \water\ line and tentative OH detection are not enough to comment on with regard to the red clump). Of those, HNC and both \sotwo\ bands have linewidths ($\approx$ 10 \kms) much wider compared to the blue clump species that have separate and resolved red clump components. \citet{Rangwala2018} observed a subset of the HCN and \acet\ transitions with EXES, but they only resolved the blue clump component with linewidths $> 9 $ \kms, due to a lower integration time. In this work with the same transitions, we are able to resolve the two components and the linewidths for each species are smaller. This indicates that HNC and \sotwo\ may have an unresolved red clump component. 

CS lines have small enough widths that there could be no measurable red clump component in this survey, while the \hcniso\ lines are faint enough that its red clump component is not abundant enough for us to measure. This \amm\ data was taken with a different instrument, TEXES, than the other molecules. Unpublished TEXES archival data includes a small subset of the \acet\ and HCN transitions present in this survey. We found that their line shapes are identical to the two-component double Gaussians published in this survey, meaning that TEXES is capable of resolving the red clump (also see Figure 7 in \citealt{Lacy2002}). On the other hand, the TEXES \amm\ observations had different beam centres compared to the EXES observations, which can affect the molecular linewidths in the region. It is inconclusive whether \amm\ is measurable in the red clump or not.

The red clump may, as we suggest for the blue clump, be independent of the region's classic components and is entirely new to this work. The results of analysis show that the central velocity, linewidth, and temperature do not neatly match any of these classic components (Table \ref{tab:comp}). The red clump is especially distinct from the extended ridge, compact ridge, and plateau.

The classic component that the red clump is possibly be related to is the hot core. It could be the hot core's outer edge as probed by IRc2 (Figure \ref{fig:bigmap}), exhibiting lower column densities and temperatures. In VLA radio maps, the column density of \amm\ at the line of sight towards IRc2 is $30\pm10$\% the column density towards the \amm\ peak of the hot core \citep{Genzel1982,Wynn-Williams1984}. The red clump's \vfwhm\ is within the hot core's range as measured by emission lines at longer wavelengths (Table \ref{tab:comp}), while it's temperature falls on the cooler end of the range. However, the red clump's \vlsr\ is about 1 \kms\ blueward of the lower limit of the hot core's.

\subsection{Individual Species}

\subsubsection{C\texorpdfstring{\textsubscript{2}}{2}H\texorpdfstring{\textsubscript{2}}{2} and \texorpdfstring{\textsuperscript{13}}{13}CCH\texorpdfstring{\textsubscript{2}}{2}}

For a detailed introduction on \acet, refer to \citet{Rangwala2018}. The blue clump \acet\ has been observed towards IRc2 with the ground-based NASA IRTF \citep{Lacy1989,Evans1991,Carr1995} and space-based \textit{ISO} \citep{VanDishoeck1998,Boonman2003} observatories. In higher resolution, \citet{Lacy2002} presented the  \acet\ R7 $\nu_5$ line with TEXES, and \citet{Rangwala2018} \acet\ R8 to R17  $\nu_5$ and \acetiso\ R9, R10, and R12 with EXES. Note that the \citet{Rangwala2018} SOFIA/EXES observations were centred over the 13 \micron\ peak which could lead to slight differences in line profiles between the observations in \citet{Rangwala2018} and the present work. Furthermore, their integration time was lower than this survey's, likely explaining why they did not detect the red clump component.

This present work expands on \citet{Rangwala2018}. Here we resolve \acet\ into two components, the blue and red clumps. For $\nu_5$ band of \acet, we detect both ortho and para transitions in the \textit{R}, \textit{P}, and \textit{Q} branches, and we also detect $\nu_5$ \acetiso\ \textit{P} and \textit{R} branch transitions. For $\nu_4+\nu_5$ \acet\ we detect ortho and para \textit{P} as well as para \textit{R} branch transitions. We should also note that we observe the R17e and R19e transitions in the 7.3 \micron\ settings, but they are mixed with \sotwo\ lines and omitted from the analysis.

We expect that the ortho-to-para ratio (OPR) of \acet\ will follow similar trends as \htwo. $^{12}$C has a nuclear spin of zero, which means that the degeneracy of the nuclear spin wavefunction is 1. Thus for \acet, the carbon atoms do not contribute to the spin degeneracy. The ortho and para states arise solely from the \htwo\ part of this linear molecular and thus \acet\ is expected to exhibit the following trends with \htwo. The OPR of \htwo\ in LTE with temperatures $\gtrsim$ 200 K converges to 3. The OPR for \htwo\ upon formation on dust grains is also 3 \citep{Takahashi2001,Gavilan2012}, while in the cold gas of protostellar cores and star-forming regions \htwo\ is mostly in para form with OPRs $< 0.001$ \citep{Troscompt2009,Pagani2013,Lupi2021}. 

For both bands of \acet\ towards IRc2, the ortho- and para-\acet\ ladders are not in equilibrium, tracing separate temperatures and column densities (Figures \ref{fig:rot1} and \ref{fig:rot2}, Table \ref{tab:rot}), as was found in \citet{Rangwala2018}. Table \ref{tab:opr} gives the OPR for both bands and velocity components. For the $\nu_5$ band we find a similar ratio in the blue and red clumps, $1.22\pm0.19$ and $1.16\pm0.31$ respectively. This is similar to the result in \citet{Rangwala2018}, $1.7 \pm 0.1$ in the blue clump. The OPR in the $\nu_4+\nu_5$ band is closer to the equilibrium value of 3 in the blue and red clumps, $2.45\pm0.67$ and $1.89\pm0.45$ respectively. This indicates that the $\nu_4+\nu_5$ band is located in a physically different region that is approaching equilibrium. 

This survey's values for OPR in both components are much closer to the formation value of 3 than that found in cold, star-forming gas. This may reflect the recent liberation of molecules with a high OPR from dust grains into the gas phase by shocks \citep{Rangwala2018} or protostellar heating \citep{Dickens1999}. 

In the blue clump, the ortho ladder is systematically tracing hotter gas for both bands ($\nu_5$ has $175\pm12$ K for ortho and $145\pm9$ K for para; $\nu_4+\nu_5$ has $124\pm13$ for ortho and $73\pm14$ K  for para). In the red clump, the ortho ladder traces hotter gas in $\nu_5$ ($229\pm27$ K for ortho and $158\pm16$ K for para) and cooler gas in $\nu_4+\nu_5$ ($111\pm14$ K for ortho and $140\pm18$ K for para).

Towards the hot cores AFGL 2591 and 2136, with EXES \citet{Barr2020} measured an \acet\ OPR of $\approx$2 and attribute this lower value to optical depth effects. We do not see evidence of optically thick lines in our own data and have found other explanations, though we cannot rule  out this possibility.

\begin{deluxetable}{lrr}
\tablecaption{\acet\ Ortho-to-Para Column Density Ratios Towards IRc2 \label{tab:opr}}
\tablehead{\colhead{Band} & \colhead{Blue Clump} & \colhead{Red Clump}}
\startdata
$\nu_5$&$1.22\pm0.19$&$1.16\pm0.31$\\
$\nu_4+\nu_5$&$2.45\pm0.67$&$1.89\pm0.45$\\
\enddata
\end{deluxetable}
We discuss the isotope ratio  $^{12}$C/$^{13}$C in \S \ref{sec:iso} and the band ratio $N_{\nu_4+\nu_5}/N_{\nu_5}$ in \S \ref{sec:band}. 

\subsubsection{CH\texorpdfstring{\textsubscript{4}}{4}}
\meth\ is amongst the simplest organic molecules and, along with \acet, gas-phase \meth\ is an important carbon reservoir \citep{Markwick2000,Aikawa2008}. Its detection remains scarce in astrochemical surveys because its lack of a permanent dipole moment dictates that its transitions are not accessible at longer wavelengths. Due to its presence in Earth's atmosphere, it is only accessible from space or the ground in objects with a high Doppler shift. Experimental \citep{Qasim2020}, observational \citep{Lacy1991,Boogert1998,Boogert2004,Oberg2008}, and modelling \citep{Hasegawa1992} studies suggest that \meth\ forms from the hydrogenation of carbon atoms on dust grains in the polar (i.e. \water-rich) ices of dark molecular clouds. Evolving protostars then heat their natal cloud and sublimate \meth\ from the solid- into the gas-phase \citep{Boogert2004} at 90 K \citep{Tielens1997}.

\meth\ has been observed in both the solid- and gas-phase towards the hot cores W3, W33A, NGC 7538 IRS 1 and 9, and AFGL 2591 \citep{Lacy1991,Boogert1996, Boogert1998,Boogert2004, Carr1995, Knez2009, Barentine2012}. Ground-based spectroscopy with IRSHELL first tentatively detected the \textit{R} branch \meth\ $\nu_4$ towards Orion IRc2 \citep{Lacy1991}. Later, \textit{ISO} observed  gas phase \textit{Q} branch \meth\ $\nu_2$/$\nu_4$ at 7.66 \micron\ in emission towards IRc2, but did not resolve individual lines \citep{VanDishoeck1998}.

We measure one \textit{Q} branch and five \textit{R} branch lines of the $\nu_4$ band of \meth\ in both the blue and red clumps. The 7.68 \micron\ solid  \meth\ transition feature found towards other hot cores \citep{Boogert1997} is not present in this survey due to strong atmospheric interference. We also do not observe the 7.66 \micron\ gas phase emission feature \citep{VanDishoeck1998}, likely because \textit{ISO}'s beam size is much larger than EXES and the feature they found may originate from elsewhere in the region. 

The temperatures of the gas phase \meth\ in both the blue and red clumps ($193\pm42$ and $141\pm33$ K respectively) are higher compared to the hot cores NGC 7538 IRS 9 (55--70K) and W33A (110 K) \citep{Boogert1998, Boogert2004}. This suggests that the blue and red clumps had a stronger heating source, liberating more \meth\ into the gas phase. This is despite the fact that W33A and NGC 7538 IRS 9 are conventional hot cores with embedded massive protostars \citep{Capps1987,Campbell1988,Mitchell1990}, while the Orion hot core region and IRc2 itself are externally heated. \meth\ is hotter towards NGC 7538 IRS 1 \citep[$\approx$ 670 K,][]{Knez2009} compared to this survey, however, which is suitable given that NGC 7538 IRS 1 has an evolved embedded protostar \citep{Campbell1988,Knez2009}.

\subsubsection{CS and SO\texorpdfstring{\textsubscript{2}}{2}}
\label{sec:sulphur}

H$_2$S is a likely candidate for the dominant sulphur reservoir in the icy mantels of dark clouds \citep{Charnley1997,Hatchell1998,Wakelam2004,Woods2015,Vidal2018}, while alternative suggestions include OCS \citep{Charnley1997,VanDerTak2003}, atomic sulphur, and HS \citep{Vidal2018}. As protostars heat up the ice grains in hot core regions, these reservoir species evaporate and subsequent reactions produce sulphur molecules, including CS and \sotwo. 

Shocks enhance \sotwo\ abundances making it a typical tracer of shocked gas \citep{Hartquist1980,PineauDesForets1993,Bachiller1997,Burkhardt2019}. In a radio survey of the Orion BN/KL region, \citet{Esplugues2013} identified \sotwo\ emission as an excellent tracer of shocked gas in the plateau and warm, dense gas in the hot core. \citet{Blake1987} also associated the plateau with high-velocity shocked gas, while \citet{Charnley1997} found no evidence of shock chemistry in the hot core. 

We have observed the $\nu_2$ and $\nu_3$ bands of \sotwo\ towards IRc2 in the blue clump only. The velocities of both bands ($-6.1\pm0.5$ \kms\ and $-6.0\pm0.3$ \kms\ respectively) are somewhat red-ward of the other blue clump species. The  \sotwo\ linewidths are much narrower than those of the plateau shocked gas (Table \ref{tab:comp}). The line widths of both bands ($\approx$12 \kms) are slightly larger than the blue clump's average ($8.9\pm1.8$ \kms) but are similar to the line widths of HNC, and as discussed in \S \ref{sec:redclump}, may be wide due to an unresolved red clump component. Therefore, we attribute this survey's \sotwo\ to the blue clump and not shocked plateau gas. The temperatures we measure, $\approx 100$ K, are high enough to form \sotwo\ from hot core-like chemistry, typically 50 to 300 K in models \citep{Charnley1997,Hatchell1998,Wakelam2004}. The \sotwo\ therefore likely arose by liberation from icy dust grains as the blue clump heated up, along with the other molecules we observe, and not shock chemistry. 

At least six other hot cores do not have shock signatures associated with \sotwo\ \citep{Keane2004,Dungee2018}.  Towards Orion IRc2 itself, \textit{ISO} observed $\nu_3$ \sotwo\ in emission \citep{VanDishoeck1998}, while we observe the same band in absorption. This difference is likely because \textit{ISO}'s larger beam measured the \sotwo\ emission from a region other than IRc2. 

Sub-mm to radio observations of sulphur-bearing molecules in hot cores and molecular clouds have found a sulphur deficit several orders of magnitude lower than solar sulphur abundance \citep{Charnley1997,Tieftrunk1994,Herpin2009,Woods2015}. One possible solution is that these wavelengths do not probe the hottest, inner regions of hot cores where more sulphur has evaporated into the gaseous phase \citep{Tieftrunk1994,Herpin2009}. MIR observations can, however, probe these regions. The $\nu=1\mbox{--}0$ band of CS has been detected in the MIR towards the hot cores NGC 7538 IRS 1 \citep{Knez2009}, AFGL 2136, and AFGL 2591 \citep{Barr2018,Barr2020}, with abundances 2 to 3 orders of magnitude higher than sub-mm to radio observations of the same hot cores.

Towards Orion IRc2, we observe the $\nu=1\mbox{--}0$ band's \textit{R} branch of CS in the blue clump with a temperature $175\pm34$ K. The CS abundance, $(3.67\pm2.15)\times10^{-8}$, falls within the range measured from FIR--radio emission lines of the Orion
hot core,  $2.90\times10^{-9}$ to $1.4\times10^{-7}$ \citep[][see Table \ref{tab:abunhc} in Appendix \ref{ap:abcompare}]{Sutton1995,Persson2007,Tercero2010,Crockett2014}. Similarly, the \sotwo\ abundances in each band ($(3.25\pm1.89)\times10^{-7}$ for $\nu_2$ and $(5.79\pm3.36)\times10^{-7}$ for $\nu_3$) are close to the upper limit of longer wavelength observations of the hot core, $4.7\times10^{-8}$ to $6.2\times10^{-7}$ \citep[][see Table \ref{tab:abunhc}]{Sutton1995,Crockett2014,Gong2015,Feng2015,Luo2019}. It appears, at least where \sotwo\ and CS are concerned, that the MIR measurements towards IRc2 do not uncover a repository of missing sulphur. The missing sulphur may still be hiding at infrared wavelengths, but in yet-to-be detected species \citep{Bilalbegovic2015}.

For a discussion on the ratio of the \sotwo\ $\nu_2$ and $\nu_3$ bands, refer to \S \ref{sec:band}.

\subsubsection{H\texorpdfstring{\textsubscript{2}}{2}}
\label{sec:h2}

\htwo, being the the most abundant molecule in the Universe, is the main species from which molecular gas is composed. It is an important coolant of the interstellar medium \citep{Gnedin2011}, and its presence is strongly correlated with star formation \citep[e.g.][]{Leroy2013}. However, it is difficult to detect directly. Its lack of a dipole moment limits its available transitions to FIR wavelengths and shorter \citep{Bolatto2013}, with strong transitions starting in the MIR. Cooler \htwo\ may be observed in absorption against a background ultra violet source, while the infrared is critical to detecting \htwo\ in high temperature environments \citep{Wakelam2017}. 

In the Orion region, \htwo\ emission has been directly measured in both the near- and mid-IR \citep{Beckwith1978,Beck1984,Brand1989,Parmar1994,Burton1997,Rosenthal2000,Allers2005}. Towards IRc2 in particular, \textit{ISO} detected the \htwo\ pure rotational transitions from S(0) to S(17) \citep{VanDishoeck1998}. However, given \textit{ISO}'s large beam size, further localization of the \htwo\ gas was not possible.

In this survey, we detected the S(1) pure rotational \htwo\ transition at 17.03 \micron\ in emission. Similar to the survey's absorption lines, our \htwo\ emission line fits to a double Gaussian. The \vlsr\ of each \htwo\ component, $-10.7\pm 2.6$ and $0.5\pm0.5$ \kms, is within the error close to the average \vlsr\ of the blue and red clumps, respectively (Table \ref{tab:comp}). Thus, our \htwo\ emission is likely associated with the same gas as the blue and red clumps.  We note that the \vfwhm\ of each \htwo\ component is wider than that of the absorption lines, especially in the blue clump. 

Our observation of the \htwo\ S(1) transition in both the red and blue clumps supports our conclusion that these could be two unique components that are independent of the classic components in region previously measured by molecular emission line surveys at longer wavelengths.

Because we cannot measure \nhtwo\ from a single line, we choose a range \nhtwo$\;=(1.9\pm1.1)\times10^{23}$ \csi, as estimated by \citet{Evans1991} in the line of sight towards Orion IRc2 to calculate the abundances (Table \ref{tab:abundance}, discussed in \S \ref{sec:singlelines}). As discussed in \S \ref{sec:kin}, the blue and red clumps may or may not be associated with the hot core itself. Given that we cannot conclusively match the blue and red clumps to other components in the region found at longer wavelengths, the \citet{Evans1991} value is the best option. We do note the large uncertainty for finding \nhtwo\ in this region. Estimations for the hot core alone range between $10^{23}$ to $5 \times 10^{24}$ \csi \citep{Blake1987,Schilke1992,Sutton1995,Persson2007,Tercero2010,Favre2011,Plume2012,Crockett2014,Hirota2015,Feng2015,Peng2019}. 

We do not observe the \htwo\ S(0) transition at 28.22 \micron\ even though this region is covered by this survey's 28.1 \micron\ setting. Assuming the same \vlsr\ as S(1), S(0) falls into an atmosphere-free region of the spectrum. S(0) is intrinsically weaker by an order of magnitude compared to S(1), and IRc2 is much brighter at the longer wavelengths requiring a higher continuum signal-to-noise than achieved, both of which can explain this line's absence. The other \htwo\ transitions fell outside of this survey's coverage. 

The Orion BN/KL region hosts several \htwo\ ``bullets’’ as detected by the $\nu=1$--0 S(1) transition at 2.121 \micron, caused by the explosive event that occurred about 500 years ago \citep{Nissen2007,Nissen2012}. However, most of IRc2, especially our beam centre, does not probe these \htwo\ structures. It is therefore unlikely that the \htwo\ in this work, as well as the blue and red clumps, are associated with these bullets.


\subsubsection{H\texorpdfstring{\textsubscript{2}}{2}O}
\label{sec:h2o}
We observe one well-defined \water\ absorption line in the wings of an atmosphere line (Figure \ref{fig:flux1}, upper left panel): a pure rotational transition, $5_{41}$--$4_{14}$, at 25.9 \micron. This line was previously also observed by \citet{Wright2000} towards IRc2 with \textit{ISO}, one of 19 pure rotational lines. That work observed two other lines covered by this survey, but we do not observe them as they are obscured by the atmosphere. We also observe a bump corresponding to the $7_{44}$--$6_{15}$ transition at 26.0 \micron\ that was not separated enough from the atmosphere to analyze. This transition was not seen in \citet{Wright2000}.

After dividing out the atmosphere, we fit $5_{41}$--$4_{14}$ to a Gaussian to obtain a rough estimate of the \vlsr$\;=-8.0\pm0.4$ \kms, which fits with the blue clump component. The \vfwhm$\;=17.0\pm1.0$ \kms\ is much wider than all other absorption lines in this survey. It may be an artefact of being blended with the atmosphere. If it is in fact real, there are two possible explanations. Firstly, that \water\ is much more abundant compared to the other molecular species and the line is broadened by saturation. Secondly, that the larger solid angle of the brighter continuum at 26 \micron\ causes us to observe molecules at more positions in the beam with slightly different central velocities. We did find that extracting only the interior 2.5\arcsec\ of the slit results in a narrower line.

Our \water\ observations were possible for the 25.3 through 28.1 \micron\ settings due to the Doppler shift in March, while the observation of any water absorption was impossible for the 16.3 to 24.7 \micron\ settings taken in October.

We also observe numerous $\nu_2$ band of \water\ emission lines in the 7.3 to and 7.9 \micron\ settings, all in the \textit{P} branch. Their average  \vlsr$\;=8.8\pm0.1$ \kms\ places them in a region that is similar to the SiO emission lines and separate from the absorption lines in the blue and red clumps. Monzon et al., in preparation, will provide a detailed analysis and treatment of the \water\ emission.

\subsubsection{HCN, H\texorpdfstring{\textsuperscript{13}}{13}CN, and HNC}

\citet{Nickerson2021} covered the species HCN, \hcniso, and HNC in detail but here we summarize the results. They measured the \textit{P}, \textit{Q}, and \textit{R} branch transitions of  HCN and HNC, as well as the \textit{R} branch transitions of \hcniso, all in the $\nu_2$ band. The blue clump hosts all three species, while only HCN has a clear red clump component. In the blue clump, they derived HCN/HNC=72$\pm$7, in line with sub-mm to radio measurements of the region. With chemical network modelling \citep{Acharyya2018} they found that the gas reaches this ratio after $10^6$ years. This suggests that the blue clump may predate the explosive event in the region that occurred 500 years ago \citep{Bally2011}. In \S \ref{sec:iso} we discuss the isotope ratio $^{12}$C/$^{13}$C.

Note that the temperatures and column densities resulting from rotation diagram analysis in \citet{Nickerson2021} differed from those reported here because in the previous work individual branches were fit, while here we fit all branches in a single species together (Table \ref{tab:rot}). The difference is only noteworthy for red clump HCN, where fitting all branches together results in a lower temperature. This did not affect the HCN/HNC column density ratio. Also, the nomenclature is different in \citet{Nickerson2021} in which they referred to the blue clump as the primary velocity component and the red clump as the secondary velocity component. They chose a different value for \nhtwo\ in \citet{Nickerson2021} than in this work, an ALMA measurement towards the hot core \citep{Peng2019}, while here in Table \ref{tab:abundance} we use an estimation towards Orion IRc2 itself \citep{Evans1991}.

New to this work, we report the detection of five $2\nu_2$ HCN lines in the 7.3 \micron\ setting,  P10e to P14e. Due to the forest of \sotwo\ lines in this setting, it is difficult to measure other species. We divide the setting by the simulated \sotwo\ spectra (\S \ref{sec:crowd}) and satisfactorily fit P10e and P12e to double Gaussians with central velocities close to the blue and red clumps. The other three lines were mixed with atmospheric and deep \sotwo\ lines. The spectral region after division was not robust enough to fit. With only two lines fit, we do not have enough data points to draw a reliable rotation diagram and estimate the temperature and the column density of $2\nu_2$ HCN with confidence.

\subsubsection{NH\texorpdfstring{\textsubscript{3}}{3}}

\amm\ is a commonly-used tracer of gas temperature in star-forming regions \citep{Ho1983}. Previous measurements of \amm\ in the region have been conducted mostly in radio, and limited to emission lines. Radio mapping of \amm\ in the Orion BN/KL region showed that \amm\ emission lines peak towards the hot core \citep{Ho1979,Murata1990,Wilson2000,Goddi2011,Friedel2017} and pinpoint source I as the main heating source of the hot core \citep{Wilson2000,Goddi2011}. Rotation diagrams of the hot core \amm\ reveal two temperature components in radio, 130 K for transitions $<1000$ K, and 400 K for higher energy transitions \citep{Hermsen1988b,Wilson1993,Wilson2000}. 

In the TEXES spectra, we observe five \textit{P} branch \amm\ transitions in the $\nu_2$ band, two para and three ortho, in the blue clump They were not numerous enough to fit to separate ortho and para rotation diagrams and we fit all five lines as a single species. This is reasonable, however, because at temperatures over 30 K, \amm\ that has formed via gas-phase reactions is expected to have an OPR of 1. Radio \amm\ mapping of the Orion hot core shows an OPR of 0.9--1.6 \citep{Goddi2011}. Elsewhere, TEXES observations of the hot cores AFGL 2591 and 2136, \citep{Barr2020} fit ortho- and para-\amm\ to separate rotation diagrams, but find that they both agree on temperature and column density.

Towards Orion IRc2 we obtain a temperature of $230\pm86$ K, which falls into the spread of hot core temperatures measured in FIR to radio from 118 K to 400 K \citep{Hermsen1988b,Wilson1993,Crockett2014,Gong2015,Friedel2017,Wilson2000}. The abundance ($(8.32\pm6.29)\times10^{-8}$, Table \ref{tab:abundance}) is close to the lower limit of of hot core  measurements at longer wavelengths, $(4.0\pm2.1)\times10^{-7}$--$(6.0\pm3.5)\times 10^{-6}$ \citep[][see Table \ref{tab:abunhc} in Appendix \ref{ap:abcompare}]{Persson2007,Gong2015}. This fits with the possibility discussed in \S \ref{sec:blueclump} that the blue clump may exhibit hot core-like chemistry.

\subsubsection{OH}\label{sec:oh}
OH is important to the formation of water \citep[e.g.][]{Keto2014}. We observe a tentative OH line from the two pure rotational R(7/2) transitions at 24.6417 and 24.6419 \micron. The two transitions are close enough to be blended into a doublet, and we estimate \vlsr $\approx -8$ 
\kms, situating it in the blue clump. There are other strong transitions covered by the 24.7 to 28.1 \micron\ settings, but at the typical temperatures of the other blue clump molecules ($\approx$100--200 K), these transitions are weaker than the observed doublet.  Another strong transition pair falls between orders. 

Previously in the FIR, \citet{Lerate2006} detected OH with \textit{ISO} towards IRc2 largely attributed to the plateau. They found absorption and P-Cygni lines at shorter wavelengths and emission lines at longer wavelengths.

\subsubsection{SiO}\label{sec:sio}
We detect several $\nu=1\mbox{--}0$ band of SiO emission lines with an average \vlsr $\;=9.8\pm0.1$ \kms\ in the \textit{R} branch. This is a similar velocity to the \water\ emission and is different than the blue and red clumps in which we observe the absorption lines and \htwo\ emission. Monzon et al., in preparation will provide a detailed analysis for the SiO lines.

\subsubsection{[FeII], [NeII], [SI], and [SIII]}
Forbidden line emission from [NeII] (12.81 \micron), [SIII] (18.71 \micron), [SI] (25.25 microns), and [FeII] (25.99 \micron) are measured at the off-source position and, except for [FeII], at the on-source position towards IRc2. 

The line brightnesses are consistent with \textit{ISO} \citep{Rosenthal2000} and \textit{Spitzer} measurements (position I4 of \citealt{Rubin2011b}, which is at the most similar distance from from the ionizing star theta 1C as are our two positions), though both studies averaged over much larger areas. The agreement with larger area measurements and the detection of emission at both the on- and off-source positions indicates that the EXES spectra are sampling extended emission. 

Ne$^+$ and S$^{++}$ require ~22 eV photons and the [NeII] and [SIII] lines have slightly blue-shifted LSR velocities, so this emission probably originates from the extended Huygen's region of the Orion Nebula \citep{ODell2020}, where hydrogen is fully ionized. The [SI] and [FeII] lines have similar, redshifted LSR velocities at the off-source position where both are detected. Based on the velocities and the extended emission, these lines likely originate from the photodissoication region lying behind the Huygen's region, but between it and OMC-1 (e.g. \citealt{ODell2017}).


\subsection{\texorpdfstring{\textsuperscript{12}}{12}C/\texorpdfstring{\textsuperscript{13}}{13}C}\label{sec:iso}
For \acet, we find $^{12}$C/$^{13}$C$\;=21.3\pm2.2$ and $19.8\pm3.4$ in the blue and red clumps, and for  HCN, $12.5\pm2.1$ in the blue clump (Table \ref{tab:iso}). This is similar to the original \acet\ blue clump measurement in \citet{Rangwala2018}, $14\pm1$. 

\citet{Nickerson2021} discussed this ratio in detail in the context of HCN, providing a comparison between the ratio we find and those at longer wavelengths. By Galactocentric distance, this ratio is expected to be 50--90 \citep{Favre2014,Milam2005}. Sub-mm to radio studies towards IRc2 have measured $^{12}$C/$^{13}$C$\;=20$ to 79.6 \citep{Schilke1997,Tercero2010,Favre2014,Feng2015} and we match the lower end of those measurements. EXES's small beam size in the MIR means that we are able to probe different gas compared to these previous measurements. Another possible explanation for this low ratio is that the HCN and \acet\ lines are optically thick. However, the lines do not display flat bottoms, a typical sign of optical thickness.

Other star-forming regions have reported a lower $^{12}$C/$^{13}$C compared to expectations based on Galactocentric distance \citep{Daniel2013,Jorgensen2018,Magalhaes2018} as well as planetary nebulae \citep{Ziurys2020}. This may hint at new chemical processes \citep{Colzi2020}. Ultimately, we require more measurements of this isotope towards hot cores in the MIR in order to determine if these results are an anomaly or part of a wider trend.

\begin{deluxetable}{lrr}
\tablecaption{$^{12}$C/$^{13}$C Column Density Ratios \label{tab:iso}}
\tablehead{\colhead{Species} & \colhead{Blue Clump} & \colhead{Red Clump}}
\startdata
\acet&$21.3\pm2.2$&$19.8\pm3.4$\\
HCN&$12.5\pm2.1$&---\\
\enddata
\end{deluxetable}

\subsection{Band Ratios}
\label{sec:band}

\begin{deluxetable}{lllrr}
\setlength{\tabcolsep}{1.5pt}
\tablecaption{Band Column Density Ratios\label{tab:bandratio}}
\tablehead{\colhead{Species} &\colhead{Band Ratio}& \colhead{Ladder} & \colhead{Blue Clump} & \colhead{Red Clump}}
\startdata
\acet&$\nu_4+\nu_5/\nu_5$ &Ortho&$5.59\pm1.11$&$13.2\pm 3.95$\\
&&Para&$2.78\pm0.68$&$8.09\pm1.63$\\
\sotwo&$\nu_3/\nu_2$&---&$1.78\pm0.5$&---\\
\enddata
\tablecomments{The central wavelengths of species' bands: \acet\ $\nu_4+\nu_5$ at 7.6 \micron, \acet\ $\nu_5$ at 13.5 \micron, \sotwo\ $\nu_3$ at 7.2 \micron, and \sotwo\ $\nu_2$ at 19 \micron.}
\end{deluxetable}

We have column density measurements of two different bands for two species: \acet\ and \sotwo. Each absorption band for these two species probe the ground state of the gas.  We would expect, therefore, the same temperature and column density if they were probing the same material. The fact that the bands have different column densities and temperatures (Table \ref{tab:rot}), means that the different bands are probing different material. Firstly, because the brightest point in the continuum of IRc2 at different wavelengths changes position, the bands probe slightly different positions in right ascension and declination (as discussed in \S\ \ref{sec:obs}). Secondly, and perhaps more importantly, the different bands probe different depths along the line of sight. As explained below, shorter wavelengths tend to probe deeper material. This effect is also seen in the conventional hot cores AFGL 2591 and 2136 \citep{Barr2020}.

%

For the blue clump \acet, we estimate $N_{\nu_4+\nu_5}/N_{\nu_5}=5.59\pm1.11$ and $2.78\pm0.68$ for the ortho and para respectively. We measure higher ratios in the red clump, $13.2\pm3.95$ and $8.09\pm1.63$ for ortho and para respectively. Another way to highlight the difference is that individual quantum states of the different bands of \acet\ also have different column densities (e.g. P3e, see Table \ref{tab:abslines}). We only measure \sotwo\ in the blue clump and find $N_{\nu_3}/N_{\nu_2}=1.78\pm0.5$ (Table \ref{tab:bandratio}; these ratios are given with the shorter wavelength bands in the numerator). 

Towards IRc2 \citet{Evans1991} (with corrections in \citealt{Carr1995}) previously measured this ratio for \acet\ in the blue clump to be $\approx\,$5--6, which is similar to this survey's numbers. This present work's more extensive measurements support the original explanation \citet{Evans1991} and \citet{Carr1995} put forth for the band ratio: emitting dust is mixed with these molecular species observed in absorption. The optical depth of dust is greater at 13.5 \micron\ ($\nu_5$) than at 7.6 \micron\ ($\nu_4+\nu_5$) \citep{Draine1989} (though the extent of their difference depends on the model, see \citealt{Xue2016}). Therefore the $\nu_4+\nu_5$ band samples gas deeper into each clump, which results in higher column densities. This effect is apparent in both the blue and red clumps, with the greater ratio in the red clump. Similarly for \sotwo\ we also find that the shorter wavelength band, $\nu_3$ at 7.2 \micron, has a larger column density compared to the longer wavelength band $\nu_2$ at 19 \micron, though the ratio is much lower compared to \acet.

Interestingly, in both clumps the gas probed by the $\nu_5$ band of \acet\ is hotter (ortho blue clump $175\pm12$ K; para blue clump $145\pm9$ K; ortho red clump $229\pm27$; para red clump $158\pm16$) than that by the $\nu_4+\nu_5$ band (ortho blue clump $124\pm13$ K; para blue clump $73\pm14$ K; ortho red clump $111\pm14$; para red clump $140\pm18$), while the $\nu_3$ band of \sotwo\ probes hotter gas ($128\pm5$ K) than the $\nu_2$ band ($94^{+7}_{-6}$ K).  With the evidence here, at least, it is clear that along the line of sight we are probing different physical structure at different bands. However, we also cannot rule out that some of the differences are caused by the changing position of the IRc2 continuum at different wavelengths.

Though we have many lines of $\nu_2$ HCN at 13.5 \micron, we only obtain two clean lines of $2\nu_2$ HCN at 7.0 \micron, which is not enough for a linear fit. If we draw a line between the two points given, we obtain column densities of roughly $3\times10^{17}$ and $5\times10^{16}$ \csi, giving $N_{2\nu_2}/N_{\nu_2}=6$ and 2.5 in the blue and red clumps respectively. This agrees with the \acet\ and \sotwo\ measurements where the shorter wavelength band is more abundant. We must stress, however, the highly uncertain nature of this calculation with HCN.

\section{Conclusions} \label{sec:con}

With SOFIA/EXES we present the first MIR survey in the Orion BN/KL region with high enough spectral resolution to resolve individual molecular transitions of multiple species from 7.2 to 28.3 \micron. We target IRc2, the second brightest MIR source in the region, which coincides with the outer edge of the Orion hot core (Figure \ref{fig:bigmap}).  This work builds on publication of previous SOFIA/EXES observations of \acet, \acetiso, HCN, \hcniso, and HNC towards IRc2 \citep{Rangwala2018,Nickerson2021}. We supplement this survey with a small amount of IRTF/TEXES data around 11.8 \micron, where ground-based observations are comparable to SOFIA.

Our main results follow:
\begin{enumerate}
    \setlength\itemsep{-5pt} 
    \item For the first time two new kinematic components in the region are unambiguously identified in the MIR with multiple species, which we refer to as the blue clump and the red clump. These two components are characterized by molecular absorption lines. Their temperatures ($\approx140 $ K) and \vfwhm\ ($\approx 8$ \kms) are similar. Their central velocities differ with $-7.1\pm0.7$ \kms\ for the blue clump and $1.4\pm0.5$ \kms\ for the red clump (Table \ref{tab:comp}). The blue clump has higher column densities of each species it shares in common with the red clump, ranging from about 1.4 to 4.2 (Table \ref{tab:clumpr}). In the blue clump we detect every molecular species in absorption in this work: \acet, \acetiso, \meth, CS, \water, HCN, \hcniso, HNC, \amm, \sotwo, and tentatively OH. The red clump contains a subset of these species: \acet, \acetiso, \meth, and HCN. 
    \item The blue and red clumps could be their own distinct components with no relationship with the classic components previously identified in the region by emission line surveys at longer wavelengths. IRTF/TEXES \citep{Lacy2002} and VLT/CRIRES \citep{Beuther2010} spectra show that the blue clump is extended to cover IRc7, IRc4, and source \textit{n}, while the extent of the red clump is unclear. Future MIR mapping and spectroscopy in the Orion BN/KL region could clarify the nature of the blue and red clumps.
    \item We observe one pure rotational transition of \htwo, S(1), in emission with two kinematic components with \vlsr$\;=-10.7\pm2.6$ \kms\ and $0.5\pm0.5$ \kms. These components fall within error of the central velocities for the blue and red clumps respectively. It is likely that the S(1) \htwo\ emission traces the blue and red clumps. This supports our supposition that the blue and red clumps are independent of other regional components and future observations of \htwo\ in the region may help us further understand the blue and red clumps.
    \item We also observe numerous \water\ and SiO lines in emission with a \vlsr\ of around 9 \kms. They belong to neither blue nor red clumps, and detailed analysis of these lines will appear in a future publication. We provided a limited analysis of the atomic forbidden transitions [FeII], [SI], [SIII], and [NeII] observed both on-source towards IRc2, and towards the off-source position.
    \item Ortho- and para-\acet\ are not in equilibrium, with separate temperatures and column densities (Table \ref{tab:opr}). This suggests that \acet\ has been recently liberated from dust grains and has little in common with the cold gas in star-forming regions. In the blue and red clumps, the $\nu_5$ band (13.5 \micron) has an OPR of $1.22\pm0.19$ and $1.16\pm0.31$ respectively, while in the $\nu_4+\nu_5$ band (7.6 \micron) has an OPR of $2.45\pm0.67$ and $1.89\pm0.45$ respectively. This difference in OPR indicates that the gas traced by each band is located in physically different regions within the clumps, and the $\nu_4+\nu_5$ band is closer to equilibrium in both clumps.
    \item With HCN and \hcniso\ in the blue clump and \acet\ and \acetiso\ in both clumps, we find $^{12}$C/$^{13}$C$\approx$10--20 in both clumps. This is much lower than what is expected given the Galactocentric distance of Orion BN/KL.
    \item Numerous lines are observed in two different bands at two different wavelengths for \acet\ (blue and red clumps) and \sotwo\ (blue clump only). The ratio of column densities of these bands reveals that in both clumps, the shorter wavelength band has the higher column density. This suggests that shorter wavelengths probe material deeper into the clumps along the line of sight.
\end{enumerate}

Our survey highlights the importance of the MIR. While the Orion BN/KL region is the closest and best-studied massive star-forming region, a paucity of MIR observations has left much unobserved until this work. The MIR provides an observational window to rovibrational transitions and many molecules, including \acet, \meth, and \htwo, that lack a permanent dipole moment. In this survey the MIR probes kinematic components not visible to longer wavelengths. The fact that the blue and red clumps have different abundances despite having similar temperatures and line widths suggests that the they have followed separate paths of chemical evolution. Further MIR observations in the region, along with testing chemical networks against this survey's results,  will broaden our understanding of the unique molecular inventory presented in this work.

\acknowledgments
This survey is the result of amazing support from the entire SOFIA team including people who supported multiple flights to complete this survey: the PI team led by Matt Richter, mission directors, telescope operators, and pilots. The authors thank John Lacy for his wonderful assistance with the analysis of IRTF/TEXES data, Shin Okumura for the MIR map of the region,  Ciriaco Goddi for \amm\ maps of the region, and the anonymous reviewer who provided valuable insights. 

This work made use of the following Python packages: \texttt{Astropy} \citep{Price-Whelan2018,Robitaille2013}, \texttt{corner} \citep{Foreman-Mackey2013}, \texttt{emcee} \citep{Foreman-Mackey2013}, \texttt{Matplotlib} \citep{Hunter2007}, \texttt{Numpy} \citep{VanDerWalt2011}, \texttt{Scipy} \citep{Virtanen2020}, and \texttt{HAPI} \citep{Kochanov2016}. 

This work was based on observations made with the NASA/DLR Stratospheric Observatory for Infrared Astronomy (SOFIA). SOFIA is jointly operated by the Universities Space Research Association, Inc. (USRA), under NASA contract NNA17BF53C, and the Deutsches SOFIA Institut (DSI) under DLR contract 50 OK 2002 to the University of Stuttgart. S.N. gratefully acknowledges NASA GO funding from the BAERI Cooperative Agreement \#80NSSC20M0198 for Cycle 5 Program 0043 and Cycle 6 Program 0061 observations. M.N.D. is supported by the Swiss National Science Foundation (SNSF) Ambizione grant 180079, the Center for Space and Habitability (CSH) Fellowship, and the IAU Gruber Foundation Fellowship. TJL and XH gratefully acknowledge support from the NASA Grants 10-APRA10-0096, 17-APRA17-0051, and 18-APRA18-0013. XH acknowledges the support by NASA/SETI Institute Co-operative Agreement NNX17AL03G. 

\bibliography{mendeley,slnrefs,confproc}{}
\bibliographystyle{aasjournal}

\appendix

\section{Gallery of Observed Spectral Lines}\label{ap:galflux}

\begin{figure}[hb]
\centering
\begin{tabular}{cc}
\includegraphics[scale=0.39]{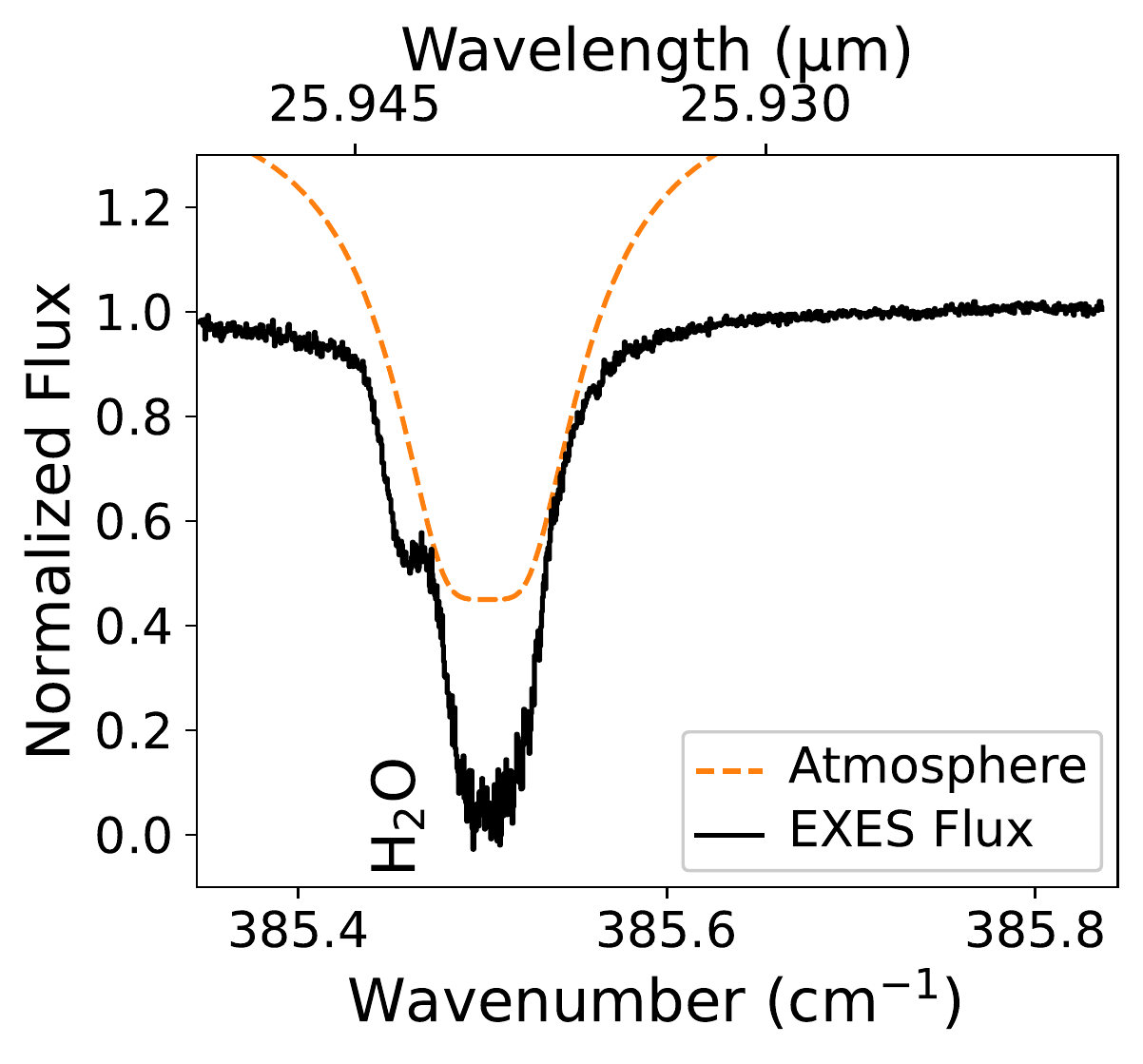}&
\includegraphics[scale=0.39]{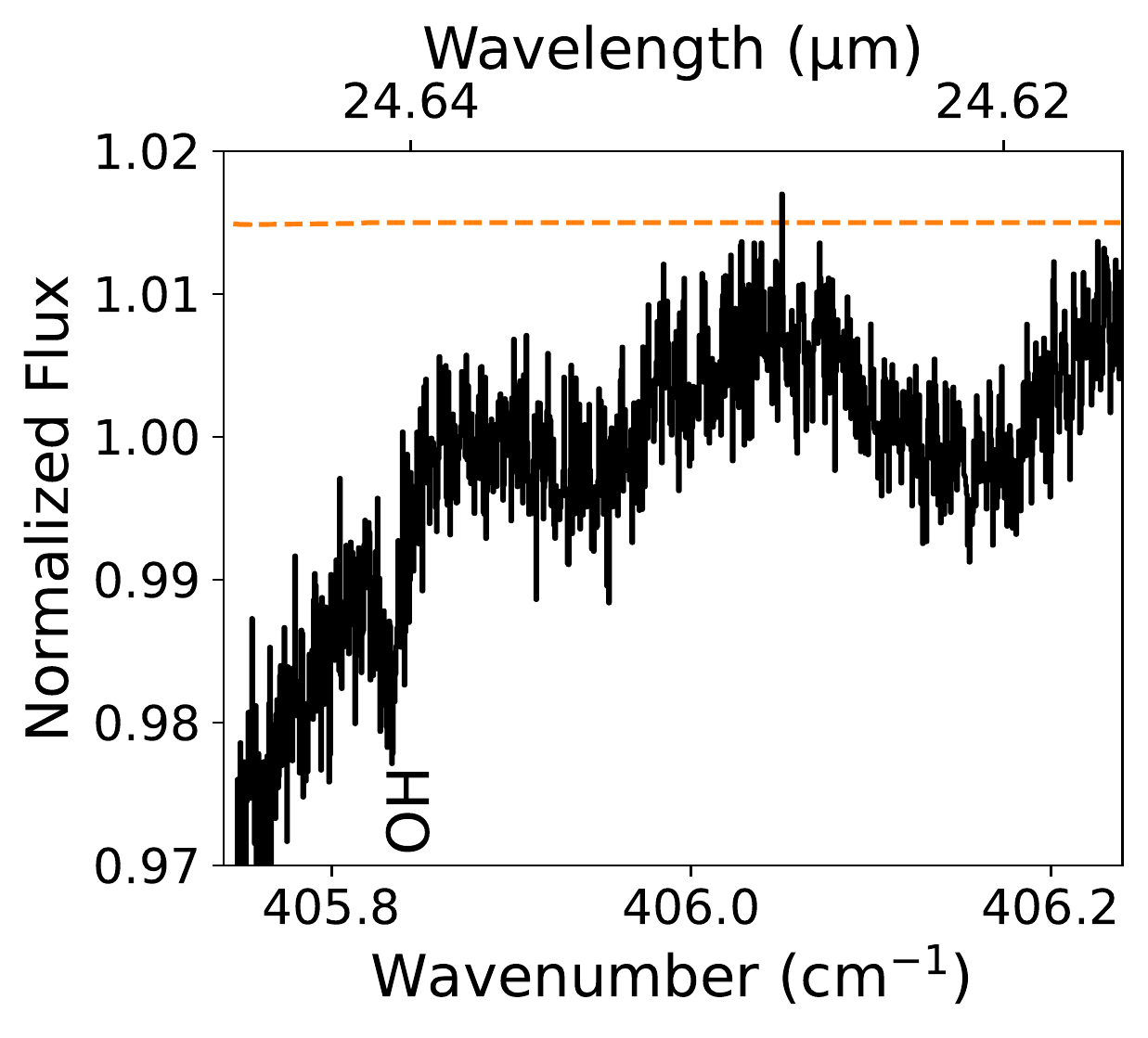}\\
\includegraphics[scale=0.39]{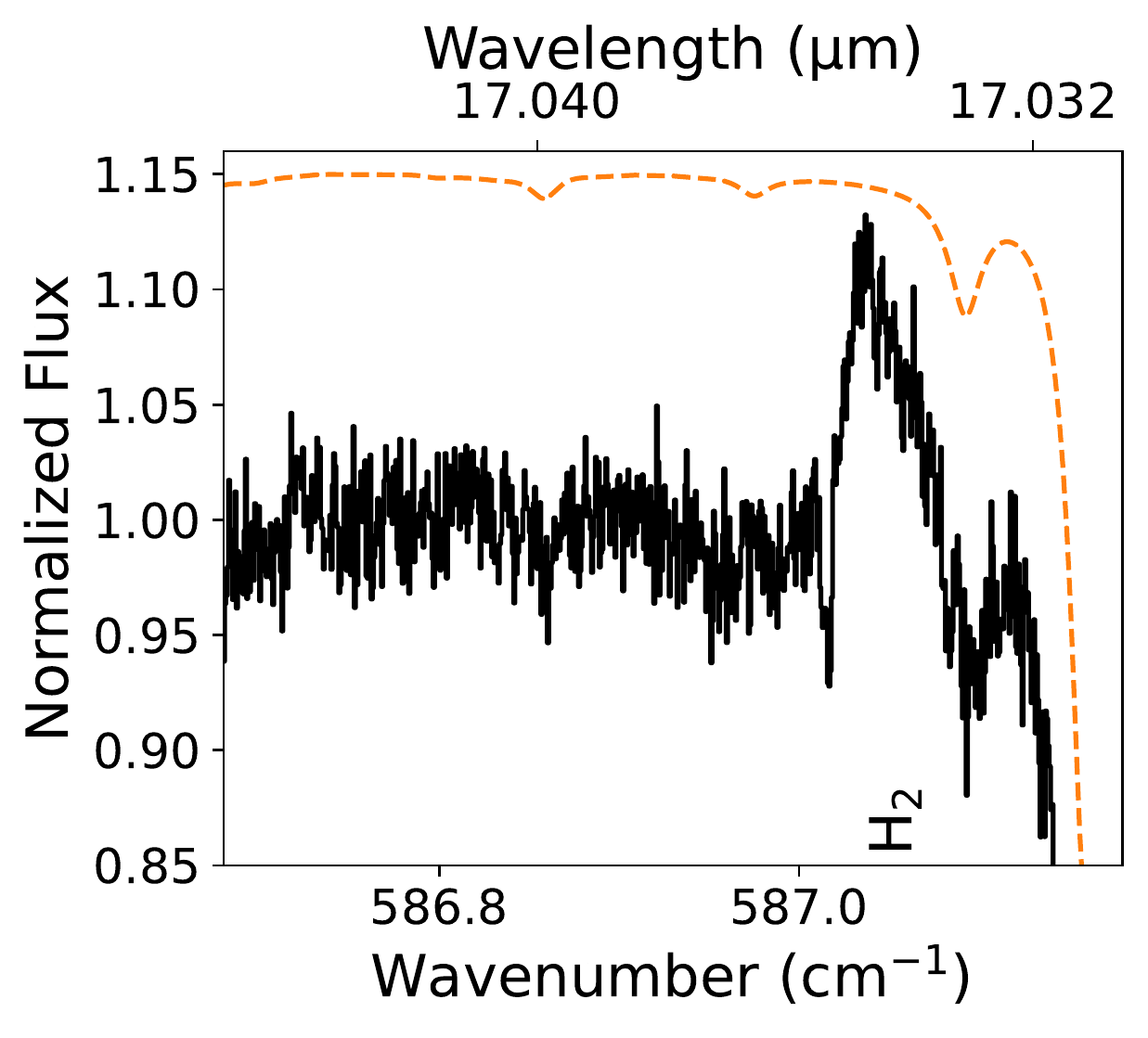}&
\includegraphics[scale=0.39]{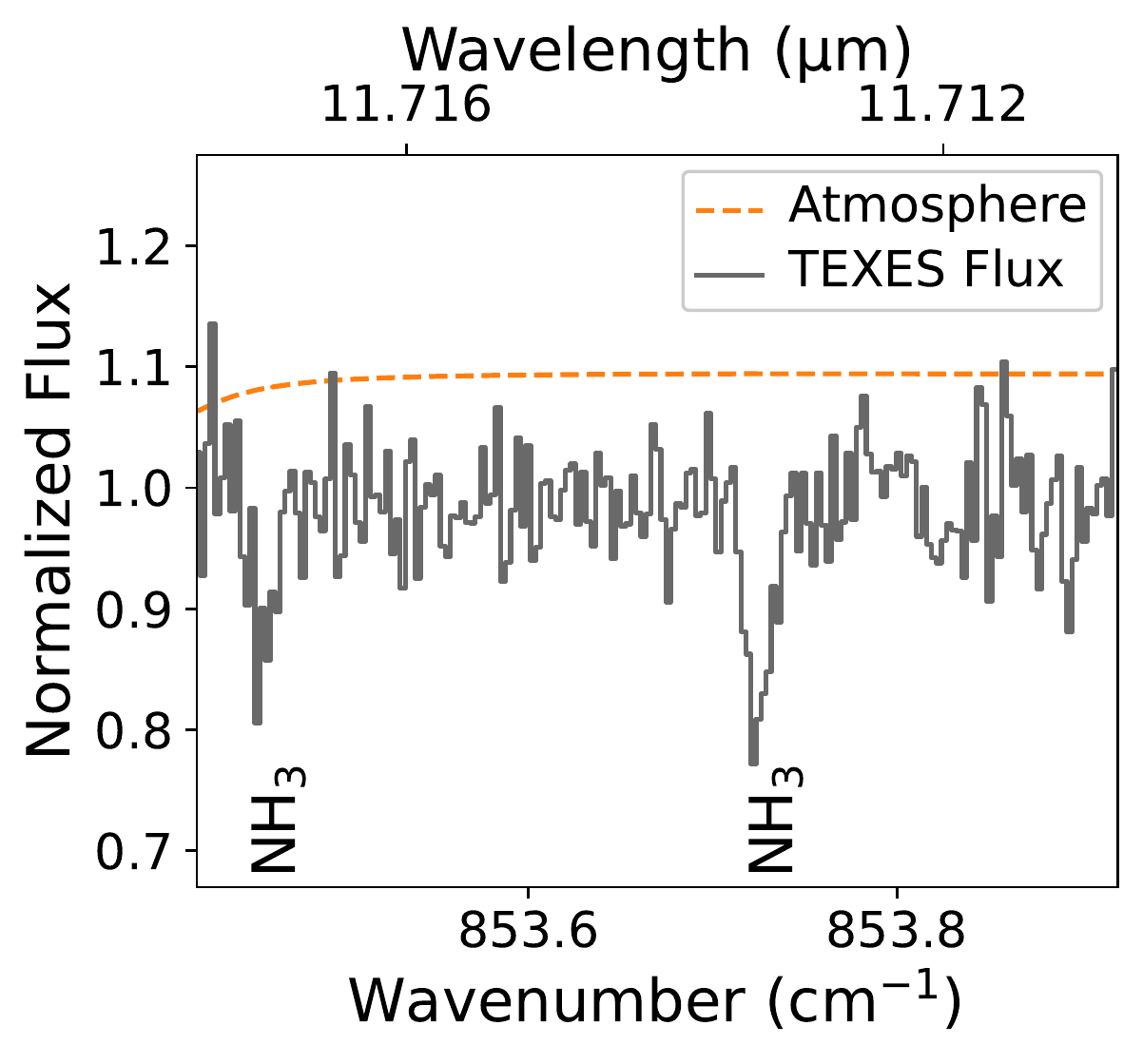}\\
\includegraphics[scale=0.39]{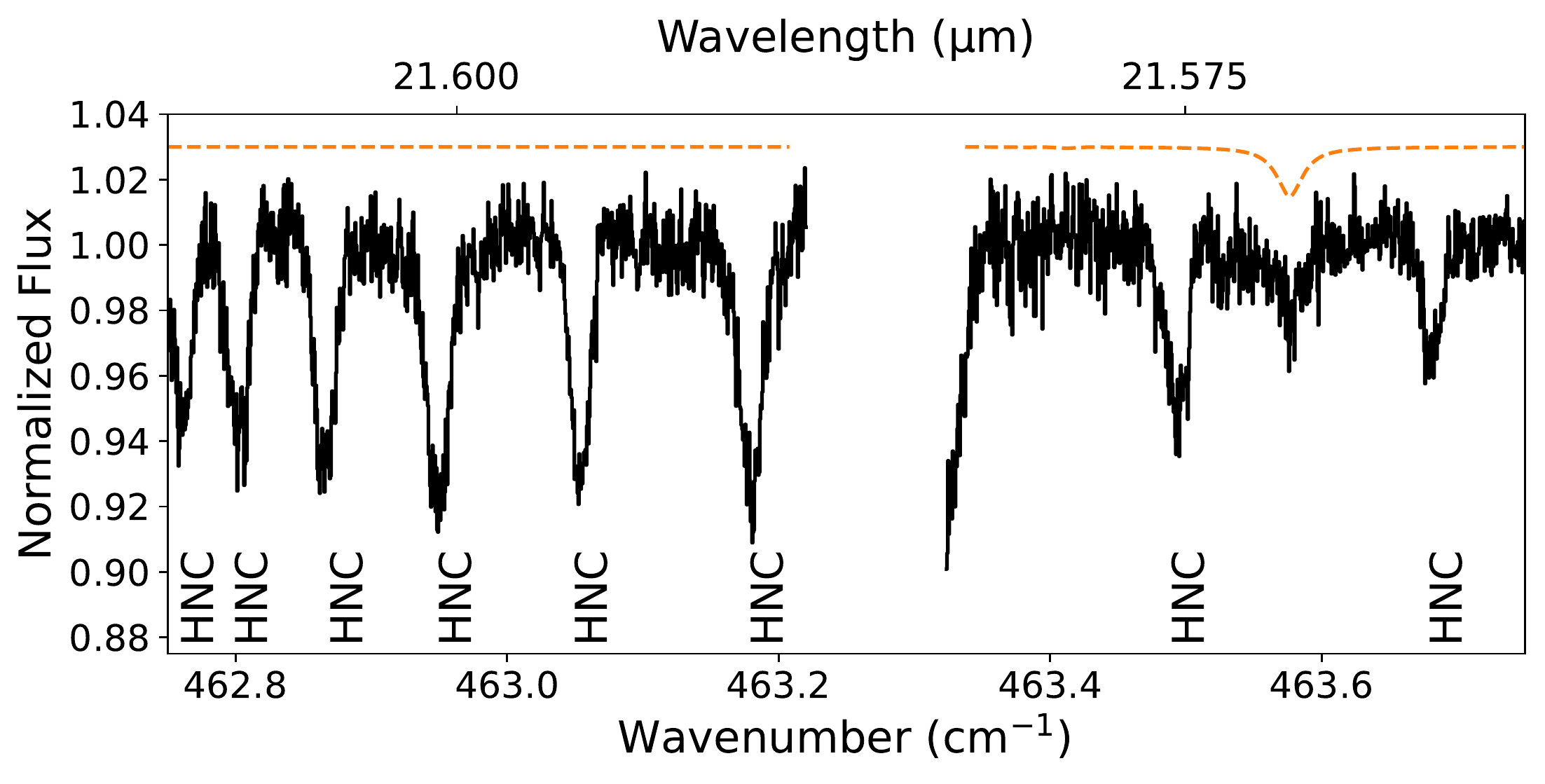}&
\includegraphics[scale=0.39]{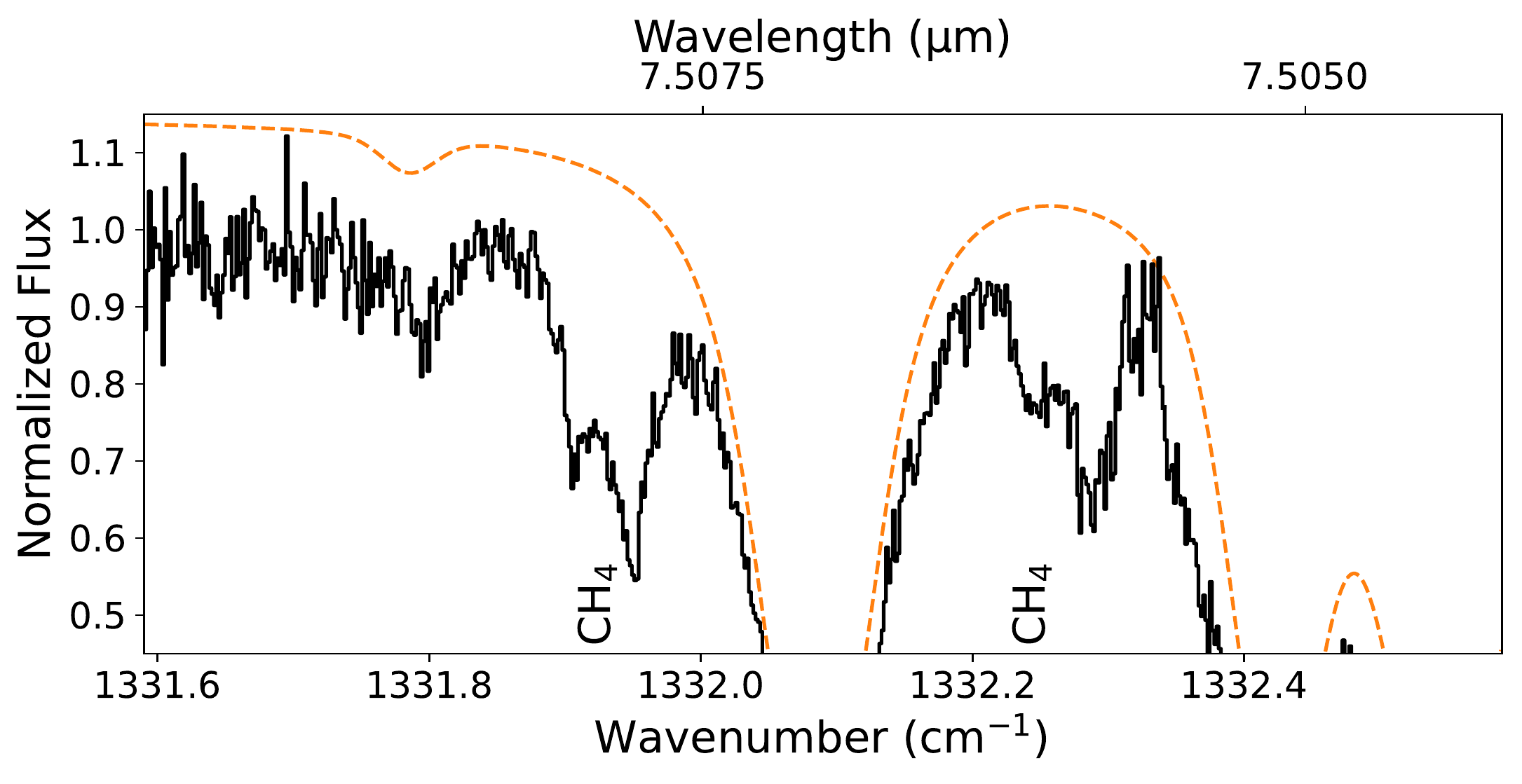}\\ 
\end{tabular}
\caption{Examples of molecular transition lines observed with SOFIA/EXES (solid black) and TEXES (solid grey) with offset atmospheric transmission (dotted orange). Species left to right, top to bottom: \water; OH (grey); \htwo; \amm; HNC; and \meth}
\label{fig:flux1}
\end{figure}


\begin{figure*}
\centering
\includegraphics[scale=0.39]{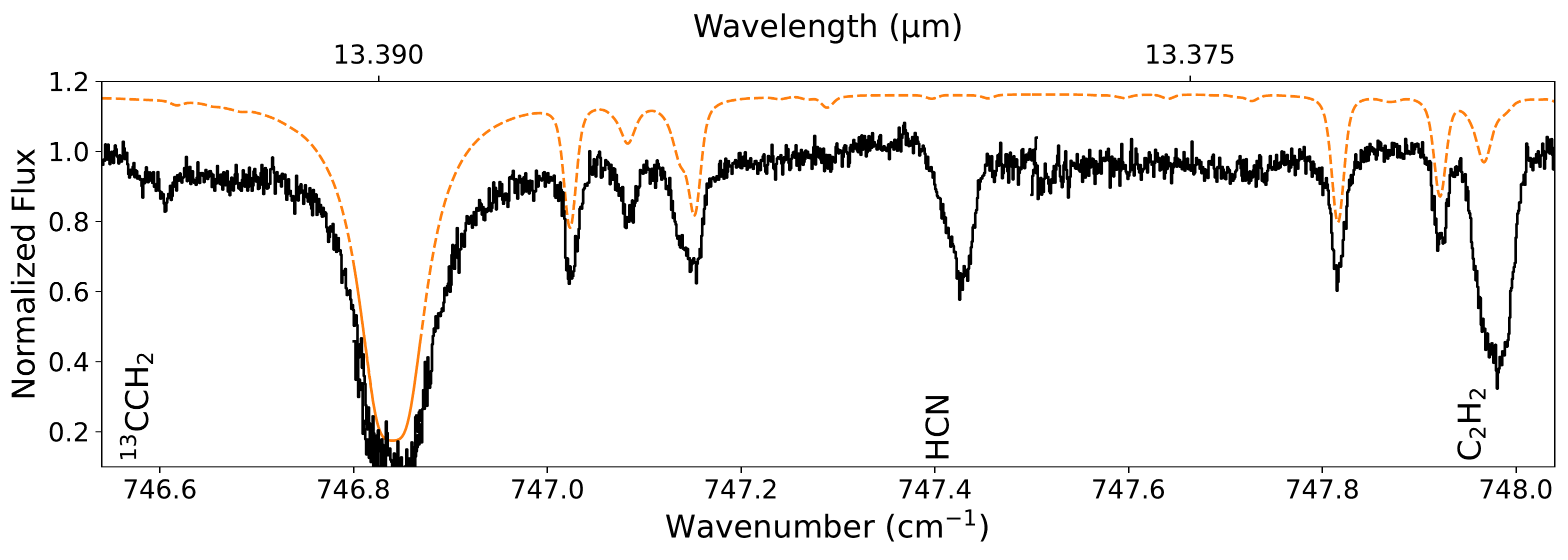}\\
\includegraphics[scale=0.39]{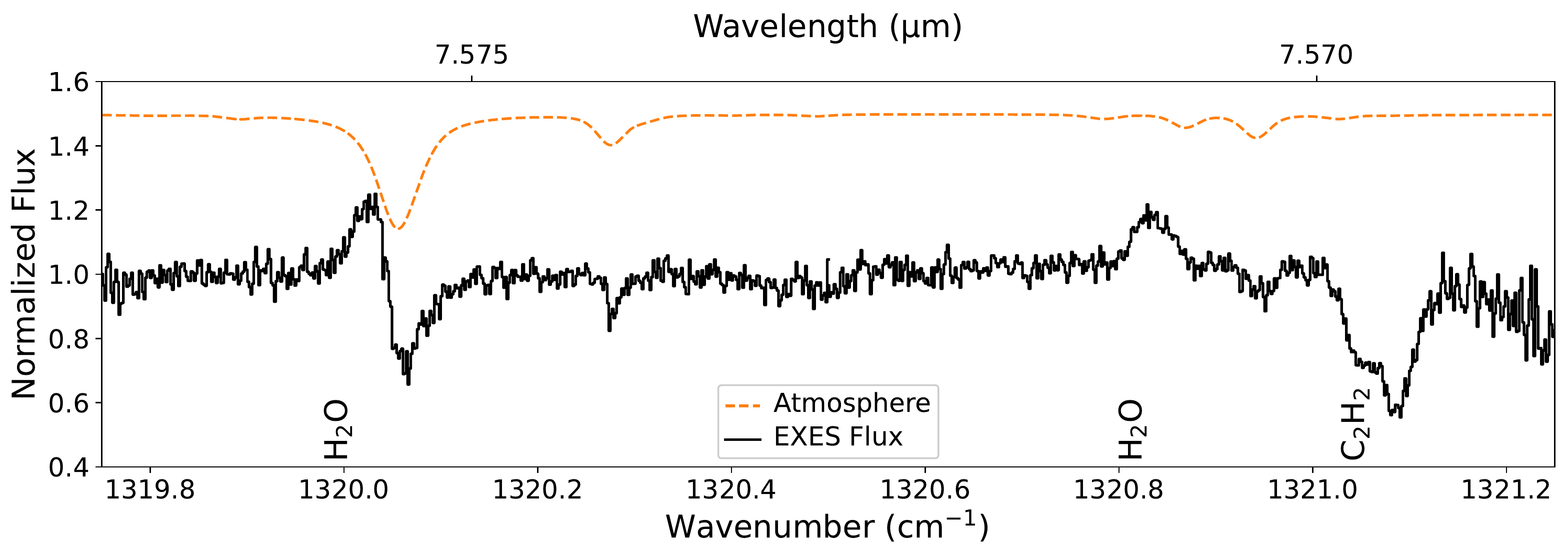}\\
\includegraphics[scale=0.39]{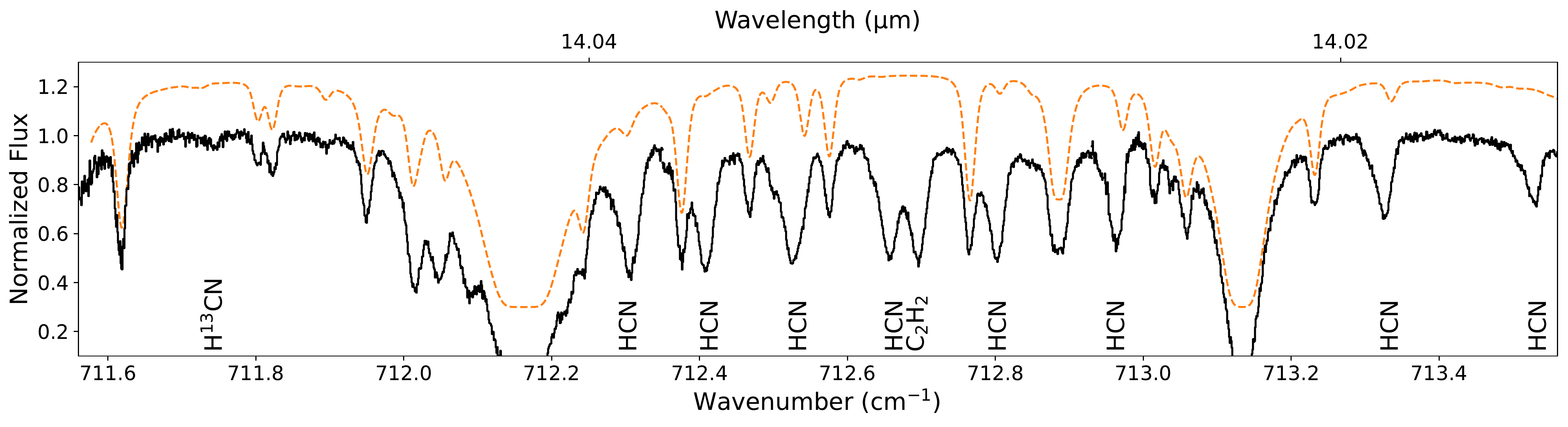}\\
\caption{Examples of molecular transition lines observed with SOFIA/EXES (solid black) with offset atmospheric transmission (dotted orange). Species top to bottom: \acetiso, HCN, and the $\nu_5$ band of \acet; \water\ and the $\nu_4 +\nu_5$ band of \acet; and \hcniso, HCN, and the $\nu_5$ band of \acet.}
\label{fig:flux2}
\end{figure*}

\begin{figure*}
\centering

\includegraphics[scale=0.39]{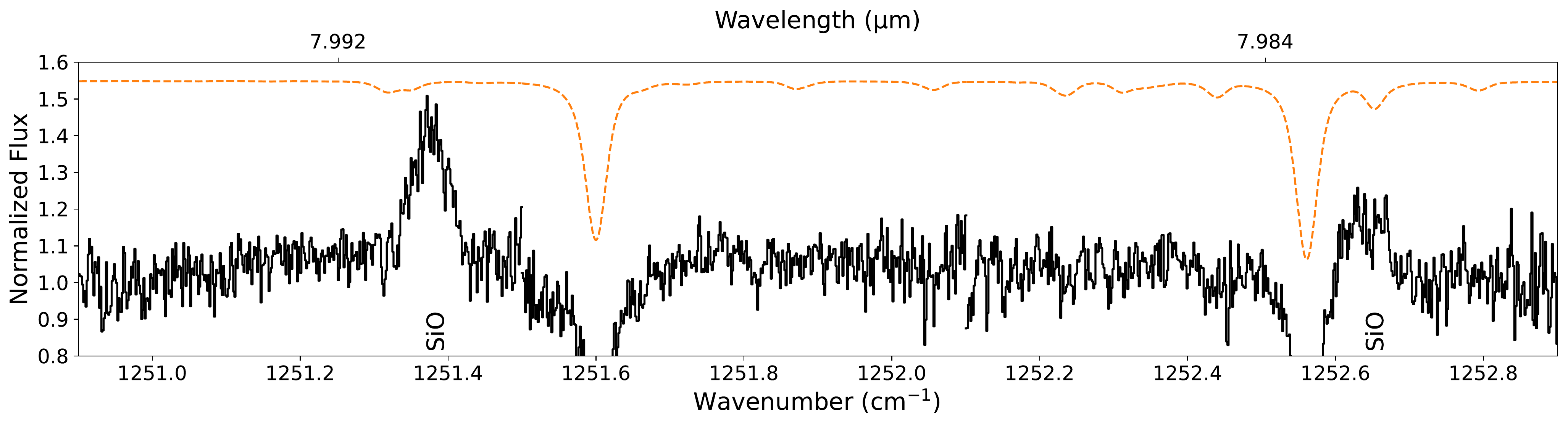}\\
\includegraphics[scale=0.39]{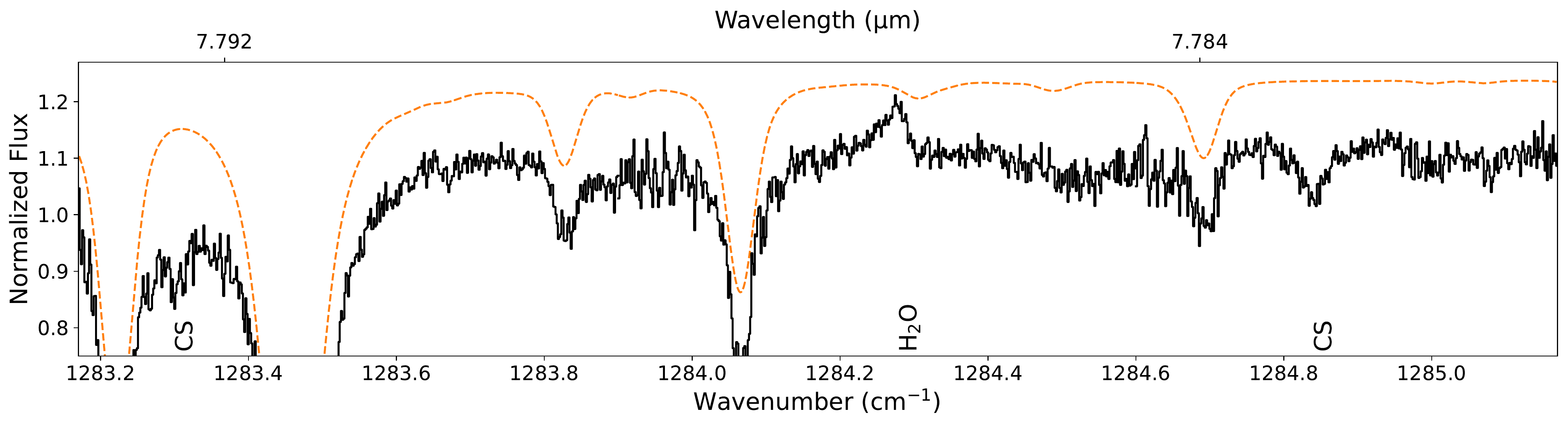}\\
\includegraphics[scale=0.39]{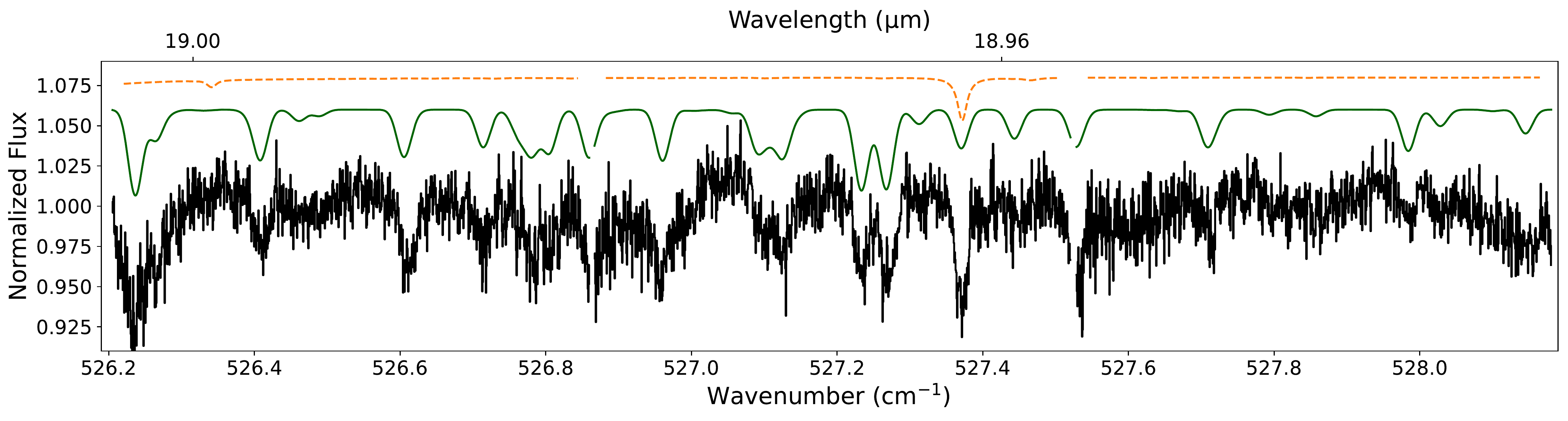}\\
\includegraphics[scale=0.39]{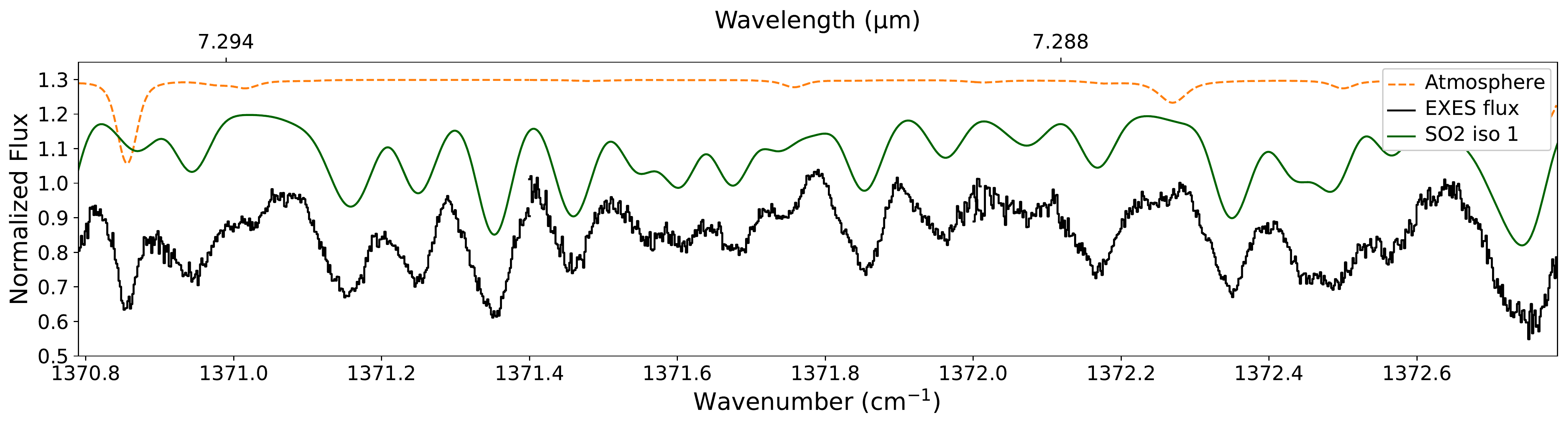}\\
\caption{Examples of molecular transition lines observed with SOFIA/EXES (solid black) with offset atmospheric transmission (dotted orange). Species top to bottom:  SiO; CS and \water; the $\nu_2$ band of \sotwo; and the $\nu_3$ band of \sotwo. The bottom two plots also include simulated spectra resulting from a fit to the data (solid green, \S \ref{sec:crowd}).}
\label{fig:flux3}
\end{figure*}

\begin{figure}[hb]
\centering
\begin{tabular}{cc}
\includegraphics[scale=0.39]{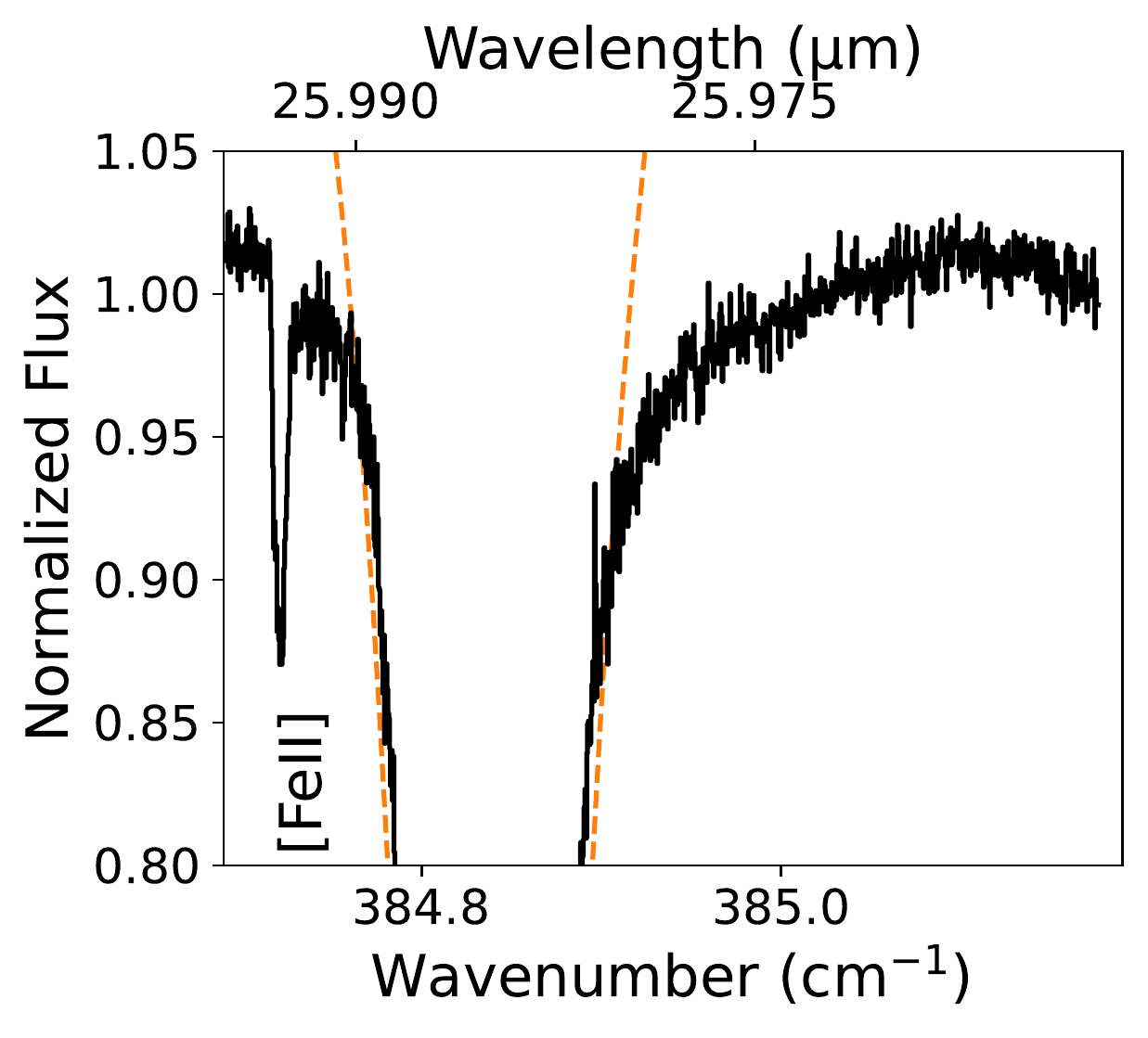}&
\includegraphics[scale=0.39]{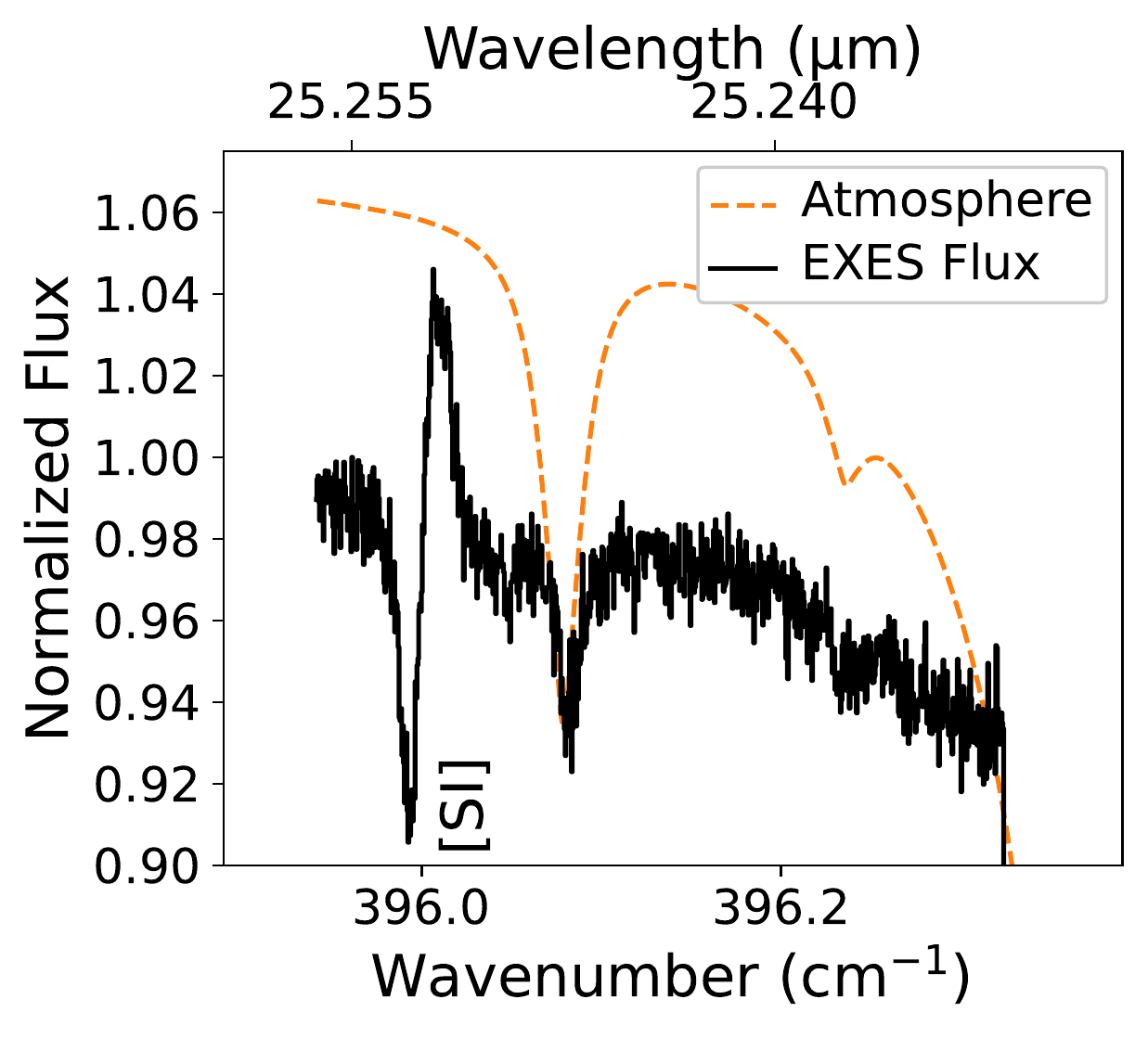}\\
\includegraphics[scale=0.39]{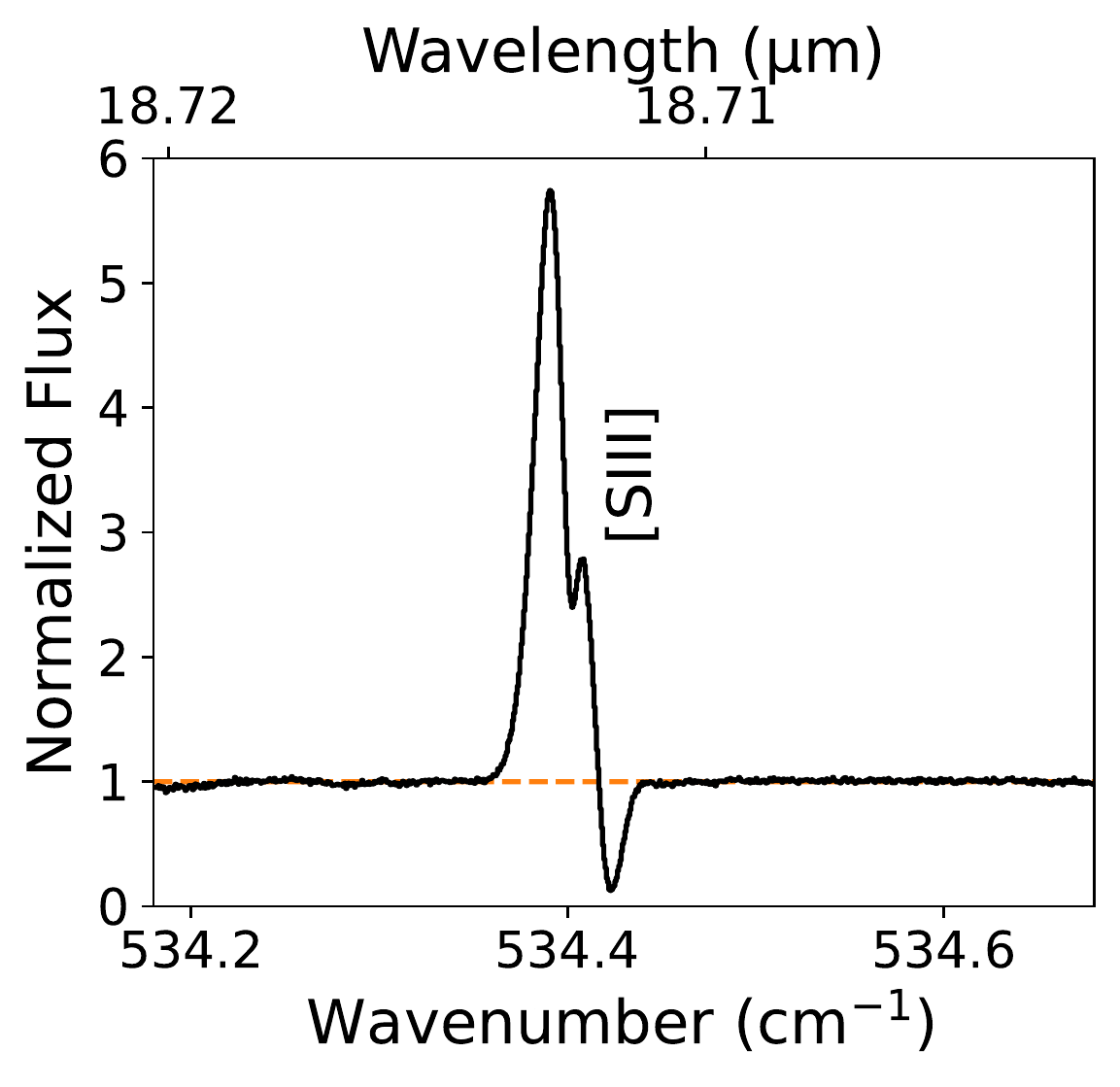}&
\includegraphics[scale=0.39]{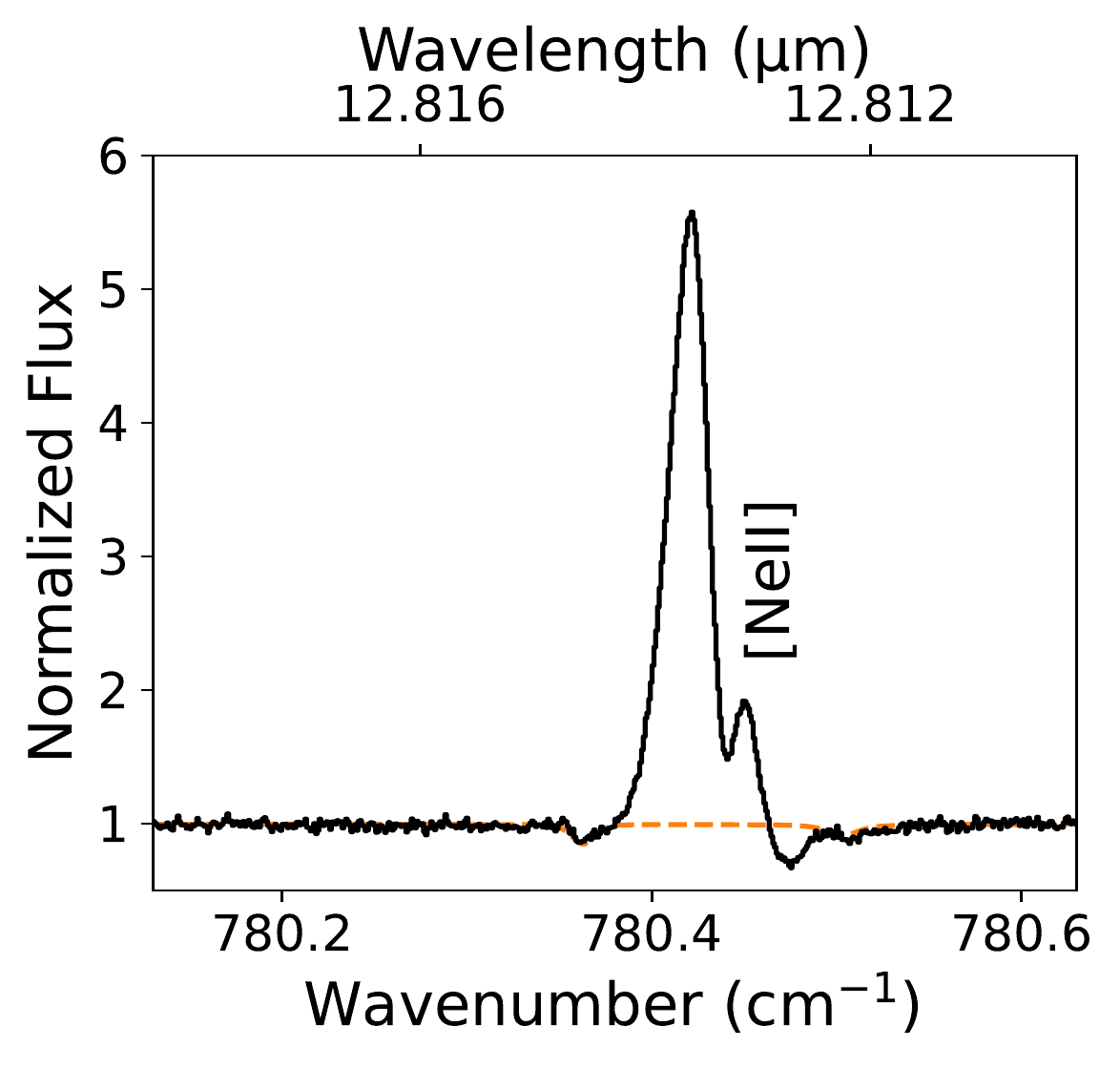}\\
\end{tabular}
\caption{Examples of atomic forbidden transition lines observed with SOFIA/EXES (solid black) with offset atmospheric transmission (dotted orange). Transitions left to right, top to bottom: [FeII], [SI], [SIII], and [NeII]. The emission lines are on-source towards IRc2, while the apparent absorption features are off-source emission lines that resulted from the nod subtraction. This creates a spurious P-cygni feature. In \S \ref{sec:ation} we analyze the flux from the on- and off-source separately.}
\label{fig:flux4}
\end{figure}

\clearpage

\section{Gallery of EXES Beams}\label{ap:beams}

\begin{figure*}[b]
\centering
\begin{tabular}{lr}
\includegraphics[scale=0.64]{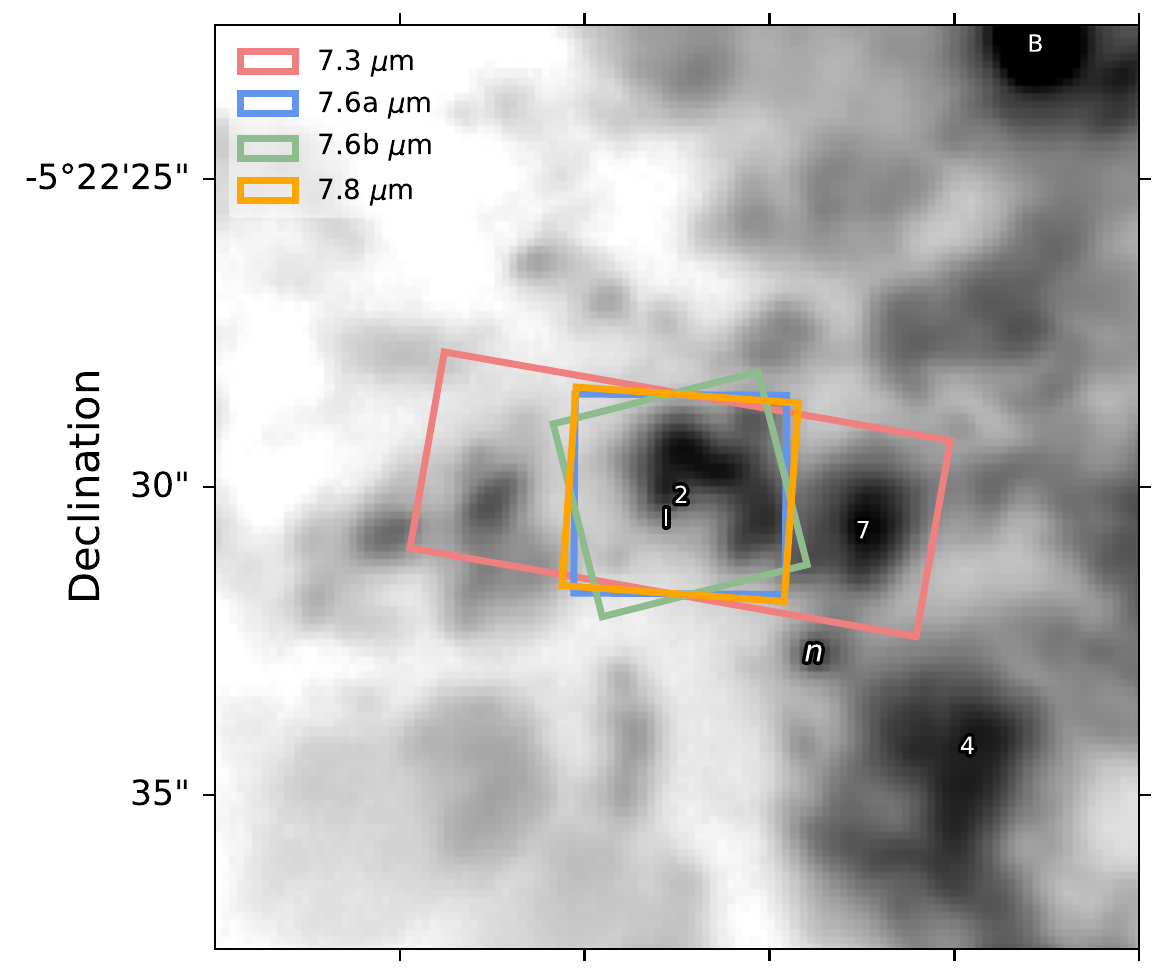}&
\includegraphics[scale=0.64]{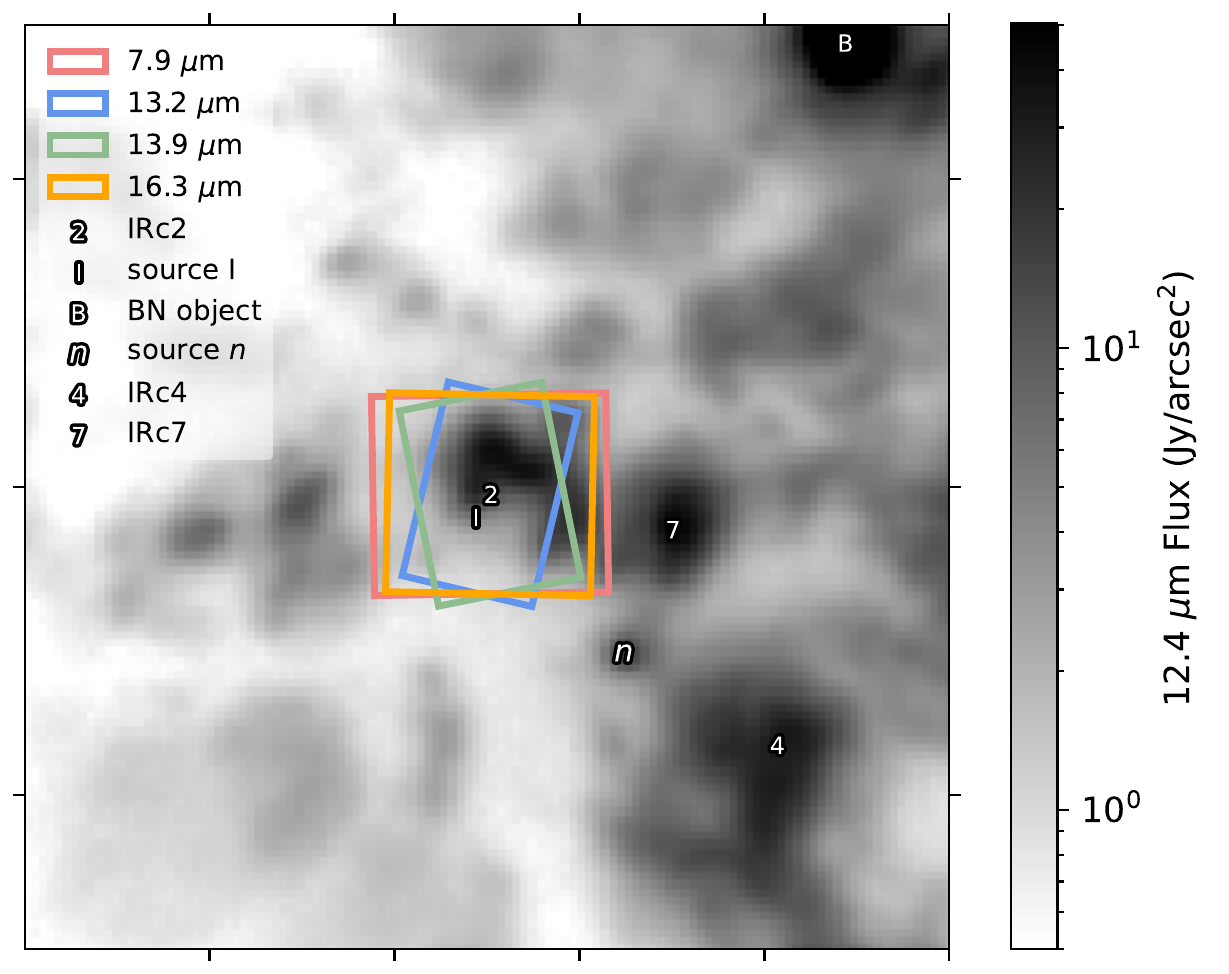}\\
\includegraphics[scale=0.64]{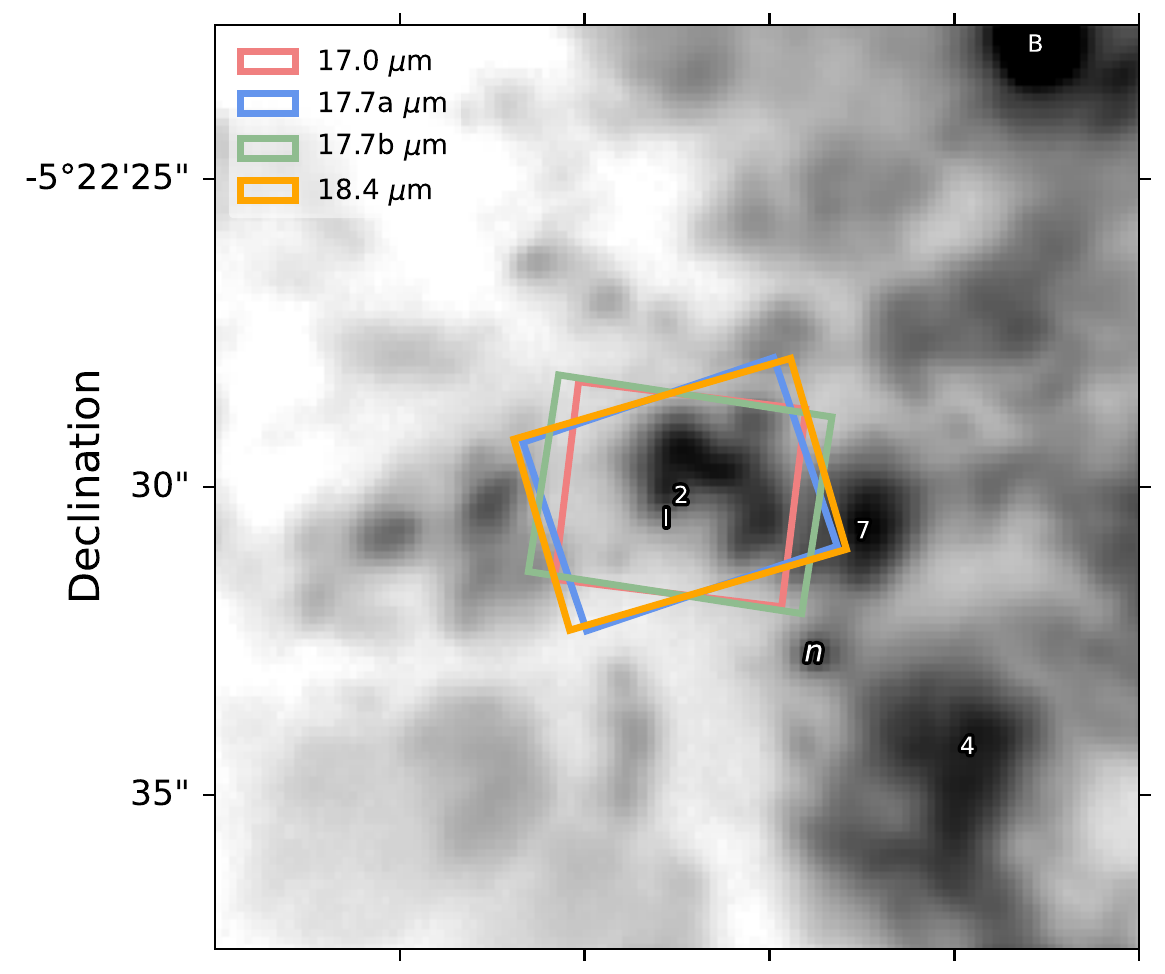}&
\includegraphics[scale=0.64]{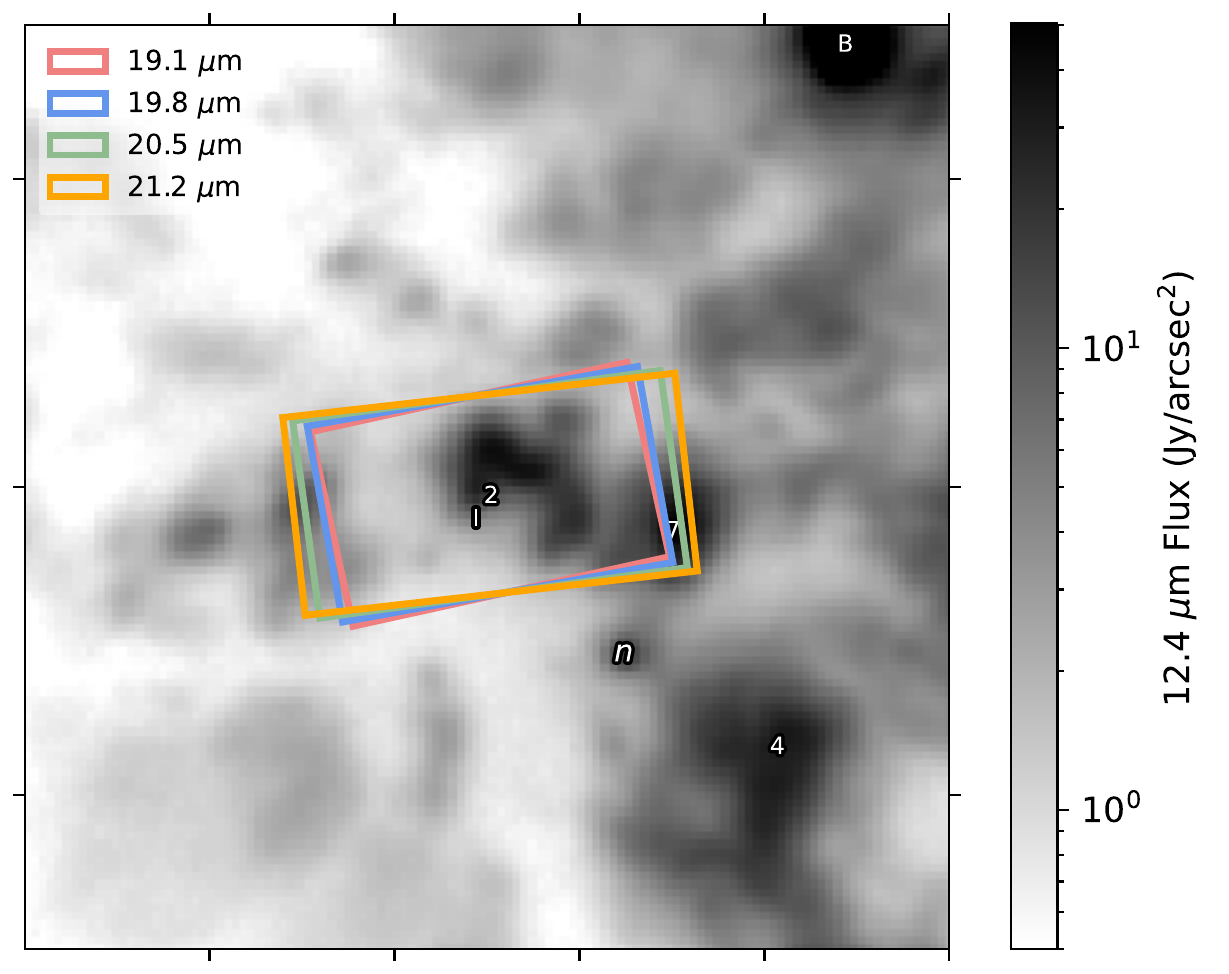}\\
\includegraphics[scale=0.64]{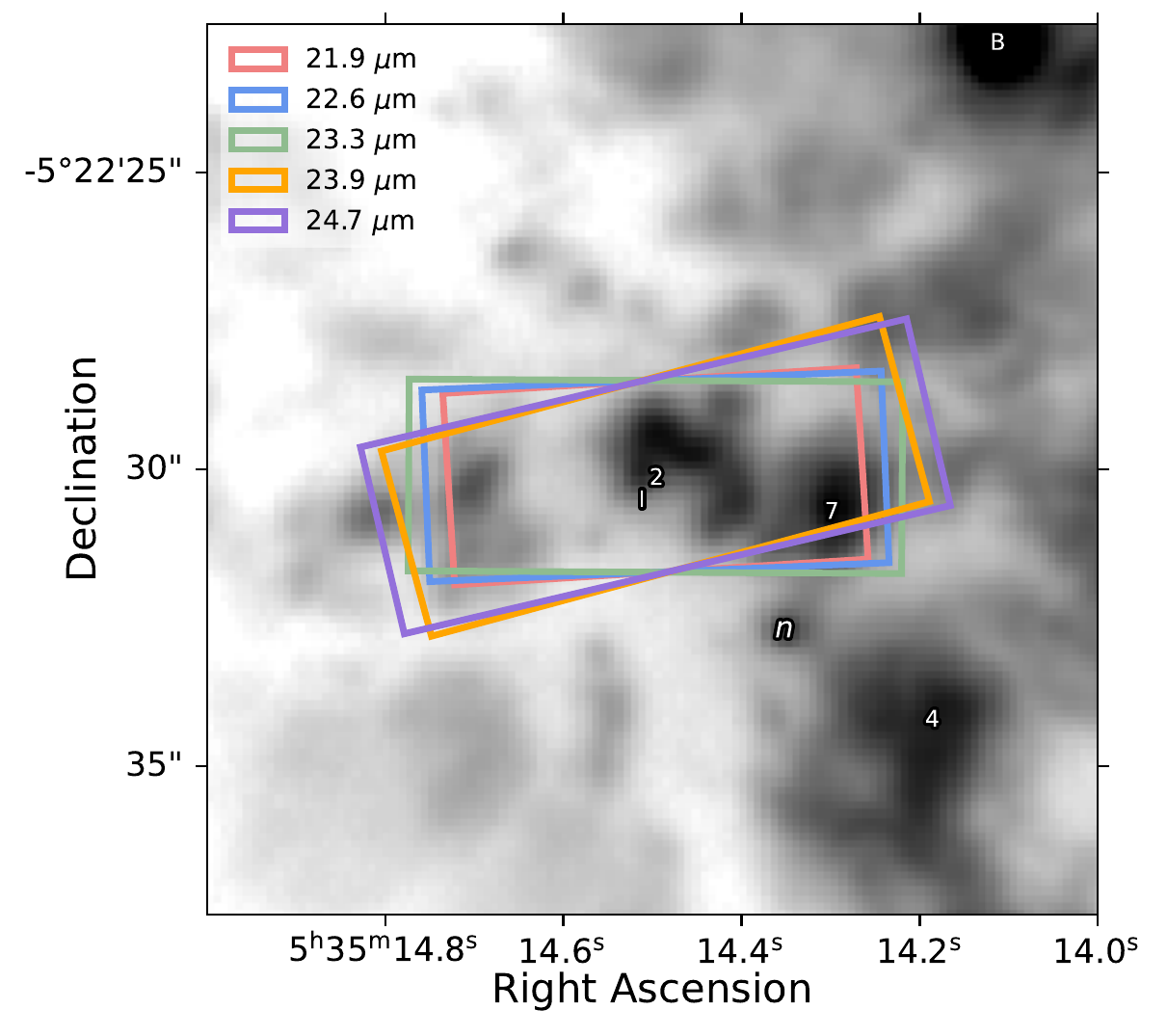}&
\includegraphics[scale=0.64]{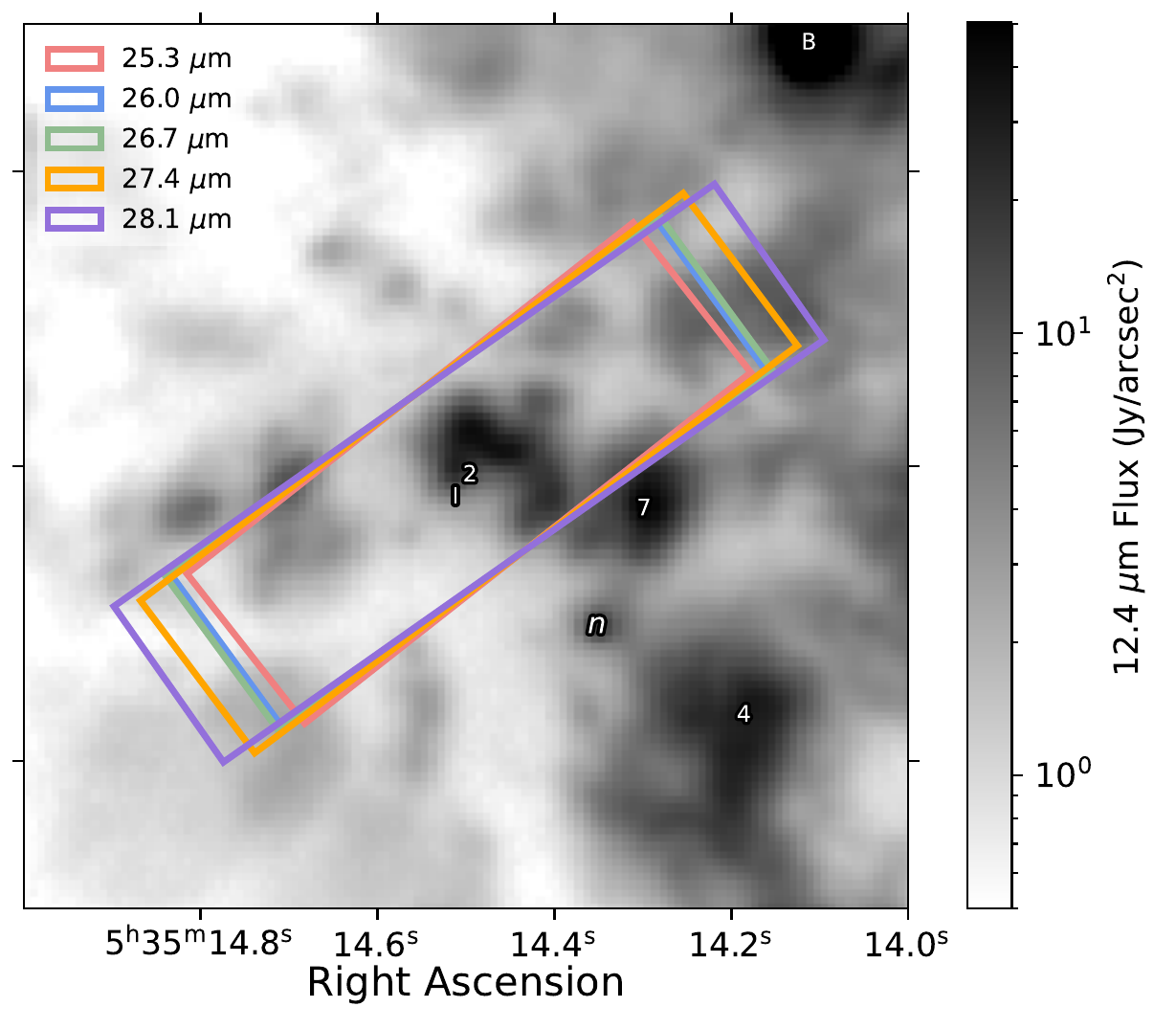}\\
\end{tabular}
\caption{Coloured boxes give the location, size, and orientation of the EXES beam for all settings in this work. The ``2'' for IRc2 is given at the beam's centre for this survey. The grey scale background is the 12.4 \micron\ map of the region from \citet{Okumura2011} taken with SUBARU/COMICS. Locations of other regional features are marked: source I, BN object, source \textit{n}, IRc4, and IRc7. BN is located according to \citet{Gomez2005} while all other objects are located according to \citep{Okumura2011}.  \label{fig:beams}}
\end{figure*}

\clearpage

\section{Comparison of Abundances}\label{ap:abcompare}

\placetable{tab:abunhc}
\begin{deluxetable}{llllll}[h]
\setlength{\tabcolsep}{1.25pt}
\tablecaption{Abundances with respect to molecular hydrogen estimated in this work compared to the Orion the hot core by select line surveys at longer wavelengths.\label{tab:abunhc}}
\tablehead{\colhead{CS} & \colhead{HCN} & \colhead{HNC} & \colhead{\amm} & \colhead{\sotwo} & Source}
\startdata
$(3.67\,\pm\,2.15)\times 10^{-8}$ &$(3.85\,\pm\,2.25)\times 10^{-7}$&$(3.90\,\pm\,2.28)\times 10^{-9}$&$(8.32\,\pm\,6.29)\times 10^{-8}$&$(3.25\,\pm\,1.89)\times 10^{-7}$ ($\nu_2$)&This Work\\
&&&&$(5.79\,\pm\,3.36)\times 10^{-7}$ ($\nu_3$)&\\
\hline
$1.40\times10^{-7}$&$6.40\times10^{-7}$&$8.70\times10^{-10}$&---&$6.20\times10^{-7}$& \citet{Crockett2014}\\
---&---& ---& ---&($3.50\pm0.69) \times10^{-7}$& \citet{Feng2015}\\
---& ---& ---&$(4.0\pm2.1)\times10^{-7}$ to &$(4.70\pm 0.80)\times10^{-8}$& \citet{Gong2015}\\
&& &$(6.0\pm3.5)\times10^{-6}$ &&\\
---& ---& ---& ---&$(2.90\pm1.50)\times10^{-7}$& \citet{Luo2019}\\
$2.90\times10^{-9}$& ---&$4.40\times10^{-11}$&$1.60\times10^{-6}$&---& \citet{Persson2007}\\
$6.00\times10^{-9}$&---&---&---&$1.20\times10^{-7}$& \citet{Sutton1995}\\
$5.00\pm1.19\times10^{-8}$&---&---&---&---& \citet{Tercero2010}\\
\enddata
\tablecomments{Our abundances are totalled along the line of sight, i.e. the sum of the blue and red clumps for HCN (Table \ref{tab:abundance}), while other species are blue clump only. For $N_{\mathrm{H}_2}$, we use the column density of \htwo\ along the line of sight towards IRc2, $1.9\pm1.1\times10^{23}$ \csi\ \citep{Evans1991}. Abundances from \citet{Crockett2014} are those estimated by MADEX modelling except for CS, which is estimated by XCLASS modelling. \amm\ abundances as estimated by \citet{Gong2015} are given as a range over all transitions.} 
\end{deluxetable}
\clearpage

\section{Observed Transitions}\label{ap:line}
\startlongtable
\begin{deluxetable*}{rrrrrrrrrl}
\tablecaption{Observed transitions and inferred parameters for molecular absorption lines. Wavelength and wavenumber are the rest value for each transition, $E_l$ is the energy level of the lower state, $k_B$ is the Boltzmann constant, $g_l$ is the lower statistical weight, $A$ is the Einstein coefficient, \vlsr\ is the observed local standard of rest velocity at the transition's centre, \vfwhm\ is the observed full-width half-maximum, $\tau_0$ is the observed optical depth, and $N_l$ is the observed column density of the transition as calculated in Equation \ref{eqn:nl}. Data in the first six columns are from the HITRAN database \citep{Gordon2017} for most species, and the GEISA database for HNC \citep{Jacquinet-Husson2016}. Superscript letters in the final column refer to which velocity component $N_l$ will be totalled towards for the rotation diagrams: $^\mathrm{b}$blue clump, $^\mathrm{r}$red clump. Velocity for lines with more than one Gaussian is taken to be the that of the line with the largest $\tau_0$ and the other Gaussian is recorded without velocity and in italics. Lines marked with $^*$ are blended with either the atmosphere (the majority of cases) or other transitions towards IRc2, and may be compromised to some degree. \label{tab:abslines}}
\tablehead{
\colhead{Transition} & \colhead{Wavelength} & \colhead{Wavenumber} & \colhead{$E_l/k_B$} & \colhead{$g_l$} & \colhead{$A$} & \colhead{\vlsr} & \colhead{\vfwhm} & \colhead{$\tau_0$} & \colhead{$N_l$} \\
\colhead{} & \colhead{(\micron)} & \colhead{(\ci)} & \colhead{(K)} & \colhead{} & \colhead{(\si)} & \colhead{(\kms)} & \colhead{(\kms)} & \colhead{} & \colhead{($\times10^{14}$\csi)}}
\startdata
\multicolumn{10}{c}{$\nu_5$ ortho-\acet}\\
R21e&12.80816&780.75233&781.6&129&3.838&--8.9$\,\pm\,$0.2&4.3$\,\pm\,$0.6&0.111$\,\pm\,$0.013&1.52$\,\pm\,$0.22$^\mathrm{b}$\\
R19e&12.88525&776.08101&643.0&117&3.761&--7.6$\,\pm\,$0.4&8.2$\,\pm\,$0.8&0.151$\,\pm\,$0.009&3.94$\,\pm\,$0.39$^\mathrm{b}$\\
&&&&&&1.9$\,\pm\,$0.5&6.9$\,\pm\,$1.1&0.094$\,\pm\,$0.009&2.04$\,\pm\,$0.36$^\mathrm{r}$\\
R17e&12.96336&771.40482&517.8&105&3.689&--7.5$\,\pm\,$0.2&7.8$\,\pm\,$0.4&0.283$\,\pm\,$0.010&6.94$\,\pm\,$0.34$^\mathrm{b}$\\
&&&&&&2.1$\,\pm\,$0.3&6.1$\,\pm\,$0.6&0.136$\,\pm\,$0.010&2.61$\,\pm\,$0.28$^\mathrm{r}$\\
R15e&13.04250&766.72409&406.2&93&3.622&--7.4$\,\pm\,$0.2&8.9$\,\pm\,$0.6&0.396$\,\pm\,$0.015&11.01$\,\pm\,$0.75$^\mathrm{b}$\\
&&&&&&2.0$\,\pm\,$0.5&6.8$\,\pm\,$1.1&0.145$\,\pm\,$0.015&3.06$\,\pm\,$0.60$^\mathrm{r}$\\
R13e&13.12269&762.03916&308.0&81&3.562&--7.6$\,\pm\,$0.2&7.3$\,\pm\,$0.4&0.541$\,\pm\,$0.021&12.20$\,\pm\,$0.64$^\mathrm{b}$\\
&&&&&&0.7$\,\pm\,$0.4&5.8$\,\pm\,$0.9&0.182$\,\pm\,$0.018&3.27$\,\pm\,$0.56$^\mathrm{r}$\\
R11e&13.20393&757.35035&223.4&69&3.509&--8.1$\,\pm\,$0.7&8.8$\,\pm\,$0.9&0.562$\,\pm\,$0.123&15.16$\,\pm\,$4.61$^\mathrm{b}$\\
&&&&&&---&\textit{11.3$\,\pm\,$2.8}&\textit{0.278$\,\pm\,$0.067}&\textit{9.54$\,\pm\,$4.46$^\mathrm{b}$}\\
R9e&13.28625&752.65799&152.3&57&3.467&--7.7$\,\pm\,$0.4&9.3$\,\pm\,$0.5&0.711$\,\pm\,$0.049&19.59$\,\pm\,$2.35$^\mathrm{b}$\\
&&&&&&0.7$\,\pm\,$1.0&9.7$\,\pm\,$1.2&0.317$\,\pm\,$0.041&9.12$\,\pm\,$2.18$^\mathrm{r}$\\
R7e&13.36966&747.96239&94.8&45&3.438&--7.0$\,\pm\,$0.3&10.1$\,\pm\,$0.4&0.887$\,\pm\,$0.022&25.69$\,\pm\,$1.49$^\mathrm{b}$\\
&&&&&&1.4$\,\pm\,$0.6&7.5$\,\pm\,$0.7&0.290$\,\pm\,$0.039&6.22$\,\pm\,$1.30$^\mathrm{r}$\\
R5e&13.45417&743.26389&50.8&33&3.433&--7.5$\,\pm\,$0.4&8.9$\,\pm\,$0.5&0.800$\,\pm\,$0.039&19.36$\,\pm\,$1.80$^\mathrm{b}$\\
&&&&&&1.3$\,\pm\,$1.0&9.0$\,\pm\,$1.3&0.304$\,\pm\,$0.036&7.41$\,\pm\,$1.78$^\mathrm{r}$\\
R3e&13.53981&738.56281&20.3&21&3.48&--7.4$\,\pm\,$0.3&7.0$\,\pm\,$1.1&0.807$\,\pm\,$0.100&13.67$\,\pm\,$3.69$^\mathrm{b*}$\\
&&&&&&--0.1$\,\pm\,$1.4&10.3$\,\pm\,$1.8&0.363$\,\pm\,$0.056&9.00$\,\pm\,$2.81$^\mathrm{r*}$\\
&&&&&&---&\textit{3.5$\,\pm\,$1.0}&\textit{0.224$\,\pm\,$0.107}&\textit{1.91$\,\pm\,$1.38$^\mathrm{b*}$}\\
R1e&13.62659&733.85945&3.4&9&3.693&--7.4$\,\pm\,$0.1&8.6$\,\pm\,$0.2&0.792$\,\pm\,$0.013&11.72$\,\pm\,$0.46$^\mathrm{b}$\\
&&&&&&0.6$\,\pm\,$0.4&7.4$\,\pm\,$0.7&0.197$\,\pm\,$0.013&2.48$\,\pm\,$0.37$^\mathrm{r}$\\
Q21e&13.67576&731.22084&781.6&129&6.111&--7.2$\,\pm\,$0.2&7.7$\,\pm\,$0.4&0.143$\,\pm\,$0.005&1.88$\,\pm\,$0.11$^\mathrm{b}$\\
&&&&&&3.2$\,\pm\,$0.5&8.6$\,\pm\,$1.2&0.055$\,\pm\,$0.004&0.80$\,\pm\,$0.11$^\mathrm{r}$\\
Q19e&13.68260&730.85539&643.0&117&6.102&--7.2$\,\pm\,$0.2&7.5$\,\pm\,$0.4&0.258$\,\pm\,$0.008&3.32$\,\pm\,$0.21$^\mathrm{b}$\\
&&&&&&1.3$\,\pm\,$0.5&6.0$\,\pm\,$1.2&0.068$\,\pm\,$0.008&0.70$\,\pm\,$0.15$^\mathrm{r}$\\
Q17e&13.68878&730.52509&517.8&105&6.094&--7.9$\,\pm\,$0.1&7.4$\,\pm\,$0.2&0.420$\,\pm\,$0.007&5.34$\,\pm\,$0.15$^\mathrm{b}$\\
&&&&&&0.5$\,\pm\,$0.2&6.1$\,\pm\,$0.5&0.131$\,\pm\,$0.007&1.36$\,\pm\,$0.14$^\mathrm{r}$\\
Q15e&13.69431&730.23009&406.2&93&6.086&--6.9$\,\pm\,$0.1&7.5$\,\pm\,$0.3&0.564$\,\pm\,$0.012&7.26$\,\pm\,$0.28$^\mathrm{b*}$\\
&&&&&&---&\textit{2.7$\,\pm\,$0.5}&\textit{0.149$\,\pm\,$0.017}&\textit{0.70$\,\pm\,$0.14$^\mathrm{b*}$}\\
Q13e&13.69918&729.97053&308.0&81&6.08&--7.0$\,\pm\,$0.1&9.9$\,\pm\,$0.5&0.740$\,\pm\,$0.033&12.51$\,\pm\,$1.16$^\mathrm{b*}$\\
&&&&&&---&\textit{2.6$\,\pm\,$0.6}&\textit{0.117$\,\pm\,$0.015}&\textit{0.52$\,\pm\,$0.13$^\mathrm{b*}$}\\
Q11e&13.70339&729.74654&223.4&69&6.072&--6.6$\,\pm\,$0.1&10.4$\,\pm\,$0.2&0.890$\,\pm\,$0.012&15.81$\,\pm\,$0.38$^\mathrm{b*}$\\
Q9e&13.70693&729.55822&152.3&57&6.07&--7.0$\,\pm\,$0.1&9.4$\,\pm\,$0.1&0.973$\,\pm\,$0.009&15.67$\,\pm\,$0.24$^\mathrm{b}$\\
&&&&&&2.4$\,\pm\,$0.3&7.3$\,\pm\,$0.4&0.180$\,\pm\,$0.008&2.26$\,\pm\,$0.20$^\mathrm{r}$\\
Q7e&13.70979&729.40565&94.8&45&6.067&--7.1$\,\pm\,$0.1&9.0$\,\pm\,$0.2&0.991$\,\pm\,$0.015&15.31$\,\pm\,$0.49$^\mathrm{b}$\\
&&&&&&1.6$\,\pm\,$0.5&8.8$\,\pm\,$0.7&0.205$\,\pm\,$0.014&3.09$\,\pm\,$0.42$^\mathrm{r}$\\
Q5e&13.71199&729.28892&50.8&33&6.063&--6.4$\,\pm\,$0.1&8.0$\,\pm\,$0.4&0.688$\,\pm\,$0.030&9.42$\,\pm\,$0.71$^\mathrm{b*}$\\
P3e&13.84863&722.09321&20.3&21&2.364&--7.3$\,\pm\,$0.1&8.7$\,\pm\,$0.2&0.633$\,\pm\,$0.009&32.87$\,\pm\,$1.10$^\mathrm{b}$\\
&&&&&&0.4$\,\pm\,$0.3&6.4$\,\pm\,$0.9&0.151$\,\pm\,$0.012&5.74$\,\pm\,$1.15$^\mathrm{r}$\\
P5e&13.93953&717.3844&50.8&33&2.588&--7.1$\,\pm\,$0.1&8.5$\,\pm\,$0.2&0.719$\,\pm\,$0.011&28.53$\,\pm\,$0.86$^\mathrm{b}$\\
&&&&&&1.5$\,\pm\,$0.4&7.7$\,\pm\,$0.6&0.201$\,\pm\,$0.011&7.26$\,\pm\,$0.82$^\mathrm{r}$\\
P7e&14.03165&712.67472&94.8&45&2.649&--7.0$\,\pm\,$0.1&9.7$\,\pm\,$0.2&0.757$\,\pm\,$0.011&31.02$\,\pm\,$0.49$^\mathrm{b}$\\
P9e&14.12500&707.96448&152.3&57&2.663&--7.2$\,\pm\,$0.1&9.4$\,\pm\,$0.2&0.689$\,\pm\,$0.011&25.77$\,\pm\,$0.66$^\mathrm{b}$\\
&&&&&&2.5$\,\pm\,$0.4&7.0$\,\pm\,$0.8&0.136$\,\pm\,$0.010&3.78$\,\pm\,$0.54$^\mathrm{r}$\\
\hline
\multicolumn{10}{c}{$\nu_5$ para-\acet}\\
R20e&12.84658&778.4173&710.6&41&3.799&--8.6$\,\pm\,$0.4&5.8$\,\pm\,$1.3&0.062$\,\pm\,$0.010&1.13$\,\pm\,$0.30$^\mathrm{b}$\\
R16e&13.00280&769.065&460.3&33&3.654&--6.7$\,\pm\,$0.3&9.8$\,\pm\,$0.7&0.170$\,\pm\,$0.008&5.26$\,\pm\,$0.41$^\mathrm{b}$\\
&&&&&&3.6$\,\pm\,$0.4&6.4$\,\pm\,$0.8&0.100$\,\pm\,$0.009&2.03$\,\pm\,$0.30$^\mathrm{r}$\\
R14e&13.08246&764.38213&355.4&29&3.591&--7.4$\,\pm\,$0.2&8.3$\,\pm\,$0.5&0.309$\,\pm\,$0.011&8.02$\,\pm\,$0.44$^\mathrm{b}$\\
&&&&&&2.2$\,\pm\,$0.4&6.1$\,\pm\,$0.8&0.121$\,\pm\,$0.011&2.31$\,\pm\,$0.34$^\mathrm{r}$\\
R12e&13.16317&759.69522&264.0&25&3.535&--7.0$\,\pm\,$0.2&8.4$\,\pm\,$0.4&0.375$\,\pm\,$0.012&9.65$\,\pm\,$0.40$^\mathrm{b}$\\
&&&&&&2.3$\,\pm\,$0.4&6.0$\,\pm\,$0.8&0.135$\,\pm\,$0.011&2.50$\,\pm\,$0.34$^\mathrm{r}$\\
R10e&13.24495&755.00459&186.2&21&3.486&--7.6$\,\pm\,$0.2&7.8$\,\pm\,$0.3&0.486$\,\pm\,$0.014&11.36$\,\pm\,$0.64$^\mathrm{b}$\\
&&&&&&1.3$\,\pm\,$0.5&8.9$\,\pm\,$0.9&0.191$\,\pm\,$0.010&5.13$\,\pm\,$0.66$^\mathrm{r}$\\
R8e&13.32781&750.31057&121.9&17&3.45&--7.6$\,\pm\,$0.2&13.2$\,\pm\,$0.6&0.556$\,\pm\,$0.014&21.56$\,\pm\,$0.79$^\mathrm{b*}$\\
&&&&&&3.7$\,\pm\,$0.4&6.0$\,\pm\,$1.0&0.128$\,\pm\,$0.018&2.26$\,\pm\,$0.55$^\mathrm{r*}$\\
R6e&13.41177&745.61349&71.1&13&3.432&--7.2$\,\pm\,$0.1&8.9$\,\pm\,$0.3&0.677$\,\pm\,$0.017&16.84$\,\pm\,$0.56$^\mathrm{b}$\\
&&&&&&3.0$\,\pm\,$0.4&7.3$\,\pm\,$0.7&0.218$\,\pm\,$0.014&4.43$\,\pm\,$0.49$^\mathrm{r}$\\
R4e&13.49685&740.91365&33.9&9&3.447&--6.8$\,\pm\,$0.2&10.6$\,\pm\,$0.6&0.719$\,\pm\,$0.028&19.77$\,\pm\,$1.12$^\mathrm{b*}$\\
&&&&&&---&\textit{2.2$\,\pm\,$0.4}&\textit{0.417$\,\pm\,$0.059}&\textit{2.39$\,\pm\,$0.46$^\mathrm{b*}$}\\
Q24e&13.66429&731.83453&1014.9&49&6.127&--8.1$\,\pm\,$0.2&5.2$\,\pm\,$0.7&0.046$\,\pm\,$0.005&0.41$\,\pm\,$0.07$^\mathrm{b}$\\
R0e&13.67041&731.50702&0.0&1&4.07&--7.2$\,\pm\,$0.2&8.2$\,\pm\,$0.5&0.488$\,\pm\,$0.015&3.42$\,\pm\,$0.23$^\mathrm{b}$\\
&&&&&&0.7$\,\pm\,$0.5&5.5$\,\pm\,$1.0&0.120$\,\pm\,$0.015&0.56$\,\pm\,$0.13$^\mathrm{r}$\\
Q20e&13.67926&731.03373&710.6&41&6.108&--8.0$\,\pm\,$0.2&5.3$\,\pm\,$0.5&0.094$\,\pm\,$0.007&0.85$\,\pm\,$0.10$^\mathrm{b}$\\
Q18e&13.68577&730.68584&578.7&37&6.099&--7.8$\,\pm\,$0.1&7.8$\,\pm\,$0.3&0.158$\,\pm\,$0.004&2.10$\,\pm\,$0.09$^\mathrm{b}$\\
&&&&&&0.9$\,\pm\,$0.4&6.1$\,\pm\,$0.8&0.049$\,\pm\,$0.004&0.51$\,\pm\,$0.08$^\mathrm{r}$\\
Q16e&13.69163&730.37317&460.3&33&6.091&--7.2$\,\pm\,$0.1&7.3$\,\pm\,$0.3&0.292$\,\pm\,$0.007&3.66$\,\pm\,$0.18$^\mathrm{b}$\\
&&&&&&1.2$\,\pm\,$0.4&6.6$\,\pm\,$0.9&0.081$\,\pm\,$0.006&0.92$\,\pm\,$0.14$^\mathrm{r}$\\
Q12e&13.70137&729.85409&264.0&25&6.077&--7.6$\,\pm\,$0.2&8.1$\,\pm\,$0.3&0.594$\,\pm\,$0.028&8.19$\,\pm\,$0.63$^\mathrm{b*}$\\
&&&&&&---&\textit{10.0$\,\pm\,$1.3}&\textit{0.221$\,\pm\,$0.013}&\textit{3.79$\,\pm\,$0.65$^\mathrm{b*}$}\\
Q8e&13.70844&729.47746&121.9&17&6.067&--8.4$\,\pm\,$0.2&6.5$\,\pm\,$1.0&0.466$\,\pm\,$0.176&5.22$\,\pm\,$2.74$^\mathrm{b*}$\\
&&&&&&---&\textit{12.1$\,\pm\,$3.7}&\textit{0.323$\,\pm\,$0.083}&\textit{6.70$\,\pm\,$3.75$^\mathrm{b*}$}\\
Q6e&13.71097&729.3428&71.1&13&6.065&--7.5$\,\pm\,$0.1&8.2$\,\pm\,$0.3&0.694$\,\pm\,$0.018&9.72$\,\pm\,$0.53$^\mathrm{b*}$\\
&&&&&&1.8$\,\pm\,$0.6&9.6$\,\pm\,$1.7&0.153$\,\pm\,$0.007&2.51$\,\pm\,$0.46$^\mathrm{r*}$\\
Q4e&13.71283&729.244&33.9&9&6.061&--6.7$\,\pm\,$0.1&7.7$\,\pm\,$0.5&0.500$\,\pm\,$0.030&6.59$\,\pm\,$0.82$^\mathrm{b*}$\\
P2e&13.80363&724.4472&10.2&5&1.986&--7.5$\,\pm\,$0.1&8.1$\,\pm\,$0.2&0.328$\,\pm\,$0.006&22.80$\,\pm\,$0.62$^\mathrm{b}$\\
&&&&&&1.5$\,\pm\,$0.3&6.2$\,\pm\,$0.6&0.091$\,\pm\,$0.005&4.85$\,\pm\,$0.51$^\mathrm{r}$\\
P8e&14.07817&710.31965&121.9&17&2.659&--6.8$\,\pm\,$0.1&9.3$\,\pm\,$0.3&0.472$\,\pm\,$0.011&17.84$\,\pm\,$0.63$^\mathrm{b}$\\
P10e&14.17215&705.60925&186.2&21&2.663&--8.0$\,\pm\,$0.4&8.0$\,\pm\,$0.5&0.342$\,\pm\,$0.025&10.74$\,\pm\,$1.37$^\mathrm{b}$\\
&&&&&&--0.1$\,\pm\,$1.4&8.9$\,\pm\,$2.0&0.108$\,\pm\,$0.019&3.75$\,\pm\,$1.39$^\mathrm{r}$\\
\hline
\multicolumn{10}{c}{$\nu_5$ \acetiso}\\
R10e&13.27221&753.45375&181.7&168&3.071&--8.2$\,\pm\,$0.4&7.0$\,\pm\,$1.0&0.072$\,\pm\,$0.008&1.73$\,\pm\,$0.27$^\mathrm{b}$\\
R9e&13.31267&751.16424&148.7&152&3.056&--6.6$\,\pm\,$0.3&5.6$\,\pm\,$1.0&0.071$\,\pm\,$0.009&1.34$\,\pm\,$0.30$^\mathrm{b}$\\
R7e&13.39436&746.58267&92.5&120&3.033&--7.7$\,\pm\,$0.3&7.5$\,\pm\,$0.8&0.099$\,\pm\,$0.009&2.40$\,\pm\,$0.25$^\mathrm{b}$\\
&&&&&&2.0$\,\pm\,$0.5&3.9$\,\pm\,$1.1&0.049$\,\pm\,$0.011&0.62$\,\pm\,$0.16$^\mathrm{r}$\\
R6e&13.43561&744.29068&69.4&104&3.026&--6.0$\,\pm\,$0.5&10.2$\,\pm\,$1.7&0.091$\,\pm\,$0.011&2.95$\,\pm\,$0.62$^\mathrm{b}$\\
R5e&13.47713&741.99794&49.6&88&3.027&--7.9$\,\pm\,$0.2&7.0$\,\pm\,$0.4&0.128$\,\pm\,$0.006&2.75$\,\pm\,$0.21$^\mathrm{b*}$\\
&&&&&&4.6$\,\pm\,$3.3&10.6$\,\pm\,$5.8&0.035$\,\pm\,$0.008&1.15$\,\pm\,$0.82$^\mathrm{r*}$\\
&&&&&&---&\textit{4.5$\,\pm\,$2.6}&\textit{0.027$\,\pm\,$0.029}&\textit{0.37$\,\pm\,$0.59$^\mathrm{b*}$}\\
R4e&13.51891&739.7045&33.0&72&3.038&--7.5$\,\pm\,$0.2&7.5$\,\pm\,$0.4&0.143$\,\pm\,$0.005&3.11$\,\pm\,$0.18$^\mathrm{b}$\\
&&&&&&2.1$\,\pm\,$0.7&8.1$\,\pm\,$1.6&0.038$\,\pm\,$0.004&0.91$\,\pm\,$0.22$^\mathrm{r}$\\
R3e&13.56097&737.41038&19.8&56&3.066&--7.7$\,\pm\,$0.3&6.8$\,\pm\,$0.5&0.119$\,\pm\,$0.014&2.22$\,\pm\,$0.40$^\mathrm{b}$\\
&&&&&&--0.2$\,\pm\,$2.0&10.0$\,\pm\,$3.3&0.035$\,\pm\,$0.006&0.95$\,\pm\,$0.44$^\mathrm{r}$\\
R2e&13.60330&735.11563&9.9&40&3.123&--7.4$\,\pm\,$0.1&6.5$\,\pm\,$0.4&0.122$\,\pm\,$0.006&1.93$\,\pm\,$0.17$^\mathrm{b}$\\
&&&&&&---&\textit{3.8$\,\pm\,$0.9}&\textit{0.033$\,\pm\,$0.006}&\textit{0.30$\,\pm\,$0.09$^\mathrm{b}$}\\
R1e&13.64591&732.82028&3.3&24&3.249&--7.4$\,\pm\,$0.2&4.5$\,\pm\,$0.8&0.072$\,\pm\,$0.014&0.63$\,\pm\,$0.21$^\mathrm{b*}$\\
&&&&&&---&\textit{12.9$\,\pm\,$3.5}&\textit{0.044$\,\pm\,$0.005}&\textit{1.11$\,\pm\,$0.40$^\mathrm{b*}$}\\
&&&&&&---&\textit{2.4$\,\pm\,$0.7}&\textit{0.031$\,\pm\,$0.007}&\textit{0.14$\,\pm\,$0.05$^\mathrm{b*}$}\\
P5e&13.95207&716.73929&49.6&88&2.253&--7.3$\,\pm\,$0.2&5.7$\,\pm\,$0.6&0.076$\,\pm\,$0.006&2.34$\,\pm\,$0.26$^\mathrm{b}$\\
\hline
\multicolumn{10}{c}{$\nu_4+\nu_5$ ortho-\acet}\\
P1e&7.54307&1325.720103&3.4&9&4.324&--8.7$\,\pm\,$0.4&7.7$\,\pm\,$0.6&0.265$\,\pm\,$0.017&88.86$\,\pm\,$10.26$^\mathrm{b}$\\
&&&&&&0.5$\,\pm\,$0.6&9.7$\,\pm\,$1.1&0.185$\,\pm\,$0.011&77.65$\,\pm\,$10.87$^\mathrm{r}$\\
P3e&7.56981&1321.036604&20.3&21&2.578&--7.7$\,\pm\,$0.4&9.4$\,\pm\,$0.7&0.532$\,\pm\,$0.021&168.32$\,\pm\,$14.55$^\mathrm{b}$\\
&&&&&&2.1$\,\pm\,$0.6&8.6$\,\pm\,$1.0&0.277$\,\pm\,$0.021&79.91$\,\pm\,$12.92$^\mathrm{r}$\\
P5e&7.59657&1316.383626&50.8&33&2.371&--7.2$\,\pm\,$0.4&12.0$\,\pm\,$0.7&0.568$\,\pm\,$0.018&214.35$\,\pm\,$13.50$^\mathrm{b}$\\
&&&&&&2.0$\,\pm\,$0.5&5.7$\,\pm\,$1.1&0.182$\,\pm\,$0.037&32.43$\,\pm\,$7.17$^\mathrm{r}$\\
P7e&7.62334&1311.760274&94.8&45&2.282&--7.9$\,\pm\,$0.2&6.6$\,\pm\,$0.3&0.551$\,\pm\,$0.018&111.34$\,\pm\,$6.65$^\mathrm{b}$\\
&&&&&&0.0$\,\pm\,$0.3&8.0$\,\pm\,$0.6&0.339$\,\pm\,$0.012&82.81$\,\pm\,$7.04$^\mathrm{r}$\\
P9e&7.65014&1307.16496&152.3&57&2.228&--7.6$\,\pm\,$0.3&7.7$\,\pm\,$0.5&0.454$\,\pm\,$0.031&105.10$\,\pm\,$12.31$^\mathrm{b}$\\
&&&&&&1.3$\,\pm\,$0.7&10.0$\,\pm\,$1.2&0.279$\,\pm\,$0.017&83.72$\,\pm\,$12.66$^\mathrm{r}$\\
P11e&7.67698&1302.595406&223.4&69&2.188&--6.8$\,\pm\,$0.2&6.1$\,\pm\,$0.4&0.356$\,\pm\,$0.017&64.63$\,\pm\,$5.38$^\mathrm{b}$\\
&&&&&&1.2$\,\pm\,$0.5&7.3$\,\pm\,$1.0&0.185$\,\pm\,$0.013&39.98$\,\pm\,$5.88$^\mathrm{r}$\\
P13e&7.70387&1298.048655&308.0&81&2.155&--7.5$\,\pm\,$0.2&7.7$\,\pm\,$0.6&0.217$\,\pm\,$0.012&49.19$\,\pm\,$3.56$^\mathrm{b*}$\\
&&&&&&0.9$\,\pm\,$0.3&4.8$\,\pm\,$0.6&0.152$\,\pm\,$0.014&21.25$\,\pm\,$2.69$^\mathrm{r*}$\\
P15e&7.73084&1293.521104&406.2&93&2.124&--6.6$\,\pm\,$0.2&7.9$\,\pm\,$0.6&0.160$\,\pm\,$0.008&36.58$\,\pm\,$2.97$^\mathrm{b}$\\
&&&&&&2.5$\,\pm\,$0.3&5.8$\,\pm\,$0.8&0.086$\,\pm\,$0.008&14.61$\,\pm\,$2.24$^\mathrm{r}$\\
\hline
\multicolumn{10}{c}{$\nu_4+\nu_5$ para-\acet}\\
R0e&7.51634&1330.434328&0.0&1&1.451&--8.5$\,\pm\,$0.4&12.4$\,\pm\,$1.4&0.146$\,\pm\,$0.011&26.27$\,\pm\,$3.64$^\mathrm{b}$\\
&&&&&&2.3$\,\pm\,$0.4&4.5$\,\pm\,$1.0&0.087$\,\pm\,$0.015&5.68$\,\pm\,$1.42$^\mathrm{r}$\\
P2e&7.55644&1323.374509&10.2&5&2.874&--7.6$\,\pm\,$0.2&8.8$\,\pm\,$0.6&0.238$\,\pm\,$0.012&74.79$\,\pm\,$4.75$^\mathrm{b}$\\
&&&&&&1.3$\,\pm\,$0.2&4.4$\,\pm\,$0.5&0.161$\,\pm\,$0.015&25.34$\,\pm\,$3.17$^\mathrm{r}$\\
P4e&7.58319&1318.706339&33.9&9&2.447&--7.6$\,\pm\,$0.2&6.3$\,\pm\,$0.4&0.280$\,\pm\,$0.014&56.91$\,\pm\,$4.88$^\mathrm{b}$\\
&&&&&&1.2$\,\pm\,$0.5&9.5$\,\pm\,$1.1&0.163$\,\pm\,$0.010&49.80$\,\pm\,$6.91$^\mathrm{r}$\\
P6e&7.60995&1314.06833&71.1&13&2.32&--8.3$\,\pm\,$0.6&7.1$\,\pm\,$1.4&0.253$\,\pm\,$0.033&55.84$\,\pm\,$15.92$^\mathrm{b}$\\
&&&&&&0.2$\,\pm\,$1.4&9.9$\,\pm\,$2.5&0.169$\,\pm\,$0.022&51.58$\,\pm\,$17.84$^\mathrm{r}$\\
P8e&7.63674&1309.459238&121.9&17&2.253&--7.3$\,\pm\,$0.2&6.4$\,\pm\,$0.3&0.274$\,\pm\,$0.009&53.05$\,\pm\,$2.92$^\mathrm{b}$\\
&&&&&&1.5$\,\pm\,$0.3&8.2$\,\pm\,$0.6&0.170$\,\pm\,$0.007&42.04$\,\pm\,$3.34$^\mathrm{r}$\\
\hline
\multicolumn{10}{c}{$\nu_4$ \meth}\\
R4(E)&7.50511&1332.424711&150.7&18&2.327&--7.9$\,\pm\,$0.3&7.8$\,\pm\,$0.6&0.381$\,\pm\,$0.020&65.79$\,\pm\,$5.60$^\mathrm{b}$\\
&&&&&&1.6$\,\pm\,$0.4&7.2$\,\pm\,$1.0&0.219$\,\pm\,$0.018&35.27$\,\pm\,$5.22$^\mathrm{r}$\\
R4(F2)&7.50703&1332.085188&150.8&27&2.326&--7.7$\,\pm\,$0.2&7.7$\,\pm\,$0.4&0.442$\,\pm\,$0.016&75.71$\,\pm\,$4.15$^\mathrm{b}$\\
&&&&&&1.0$\,\pm\,$0.3&6.1$\,\pm\,$0.5&0.282$\,\pm\,$0.016&38.25$\,\pm\,$3.59$^\mathrm{r}$\\
R3(A2)&7.53537&1327.07402&90.5&35&2.322&--7.1$\,\pm\,$0.3&8.9$\,\pm\,$1.3&0.574$\,\pm\,$0.021&107.08$\,\pm\,$14.29$^\mathrm{b}$\\
&&&&&&1.4$\,\pm\,$0.6&6.7$\,\pm\,$0.8&0.321$\,\pm\,$0.044&45.03$\,\pm\,$10.80$^\mathrm{r}$\\
R2(F2)&7.56381&1322.085048&45.2&15&2.317&--8.6$\,\pm\,$0.2&6.9$\,\pm\,$0.8&0.572$\,\pm\,$0.031&75.72$\,\pm\,$11.60$^\mathrm{b*}$\\
&&&&&&--0.6$\,\pm\,$0.7&9.9$\,\pm\,$1.0&0.298$\,\pm\,$0.019&56.12$\,\pm\,$8.91$^\mathrm{r*}$\\
R1(F1)&7.59401&1316.82689&15.1&9&2.309&--8.2$\,\pm\,$0.2&6.8$\,\pm\,$0.4&0.518$\,\pm\,$0.023&55.67$\,\pm\,$4.47$^\mathrm{b}$\\
&&&&&&0.0$\,\pm\,$0.5&8.1$\,\pm\,$0.9&0.289$\,\pm\,$0.016&37.00$\,\pm\,$4.67$^\mathrm{r}$\\
Q4(F1)&7.66735&1304.23207&150.7&27&2.259&--8.1$\,\pm\,$0.4&7.2$\,\pm\,$0.6&0.372$\,\pm\,$0.028&70.52$\,\pm\,$9.17$^\mathrm{b}$\\
&&&&&&0.3$\,\pm\,$0.8&9.5$\,\pm\,$1.5&0.202$\,\pm\,$0.015&50.64$\,\pm\,$9.86$^\mathrm{r}$\\
\hline
\multicolumn{10}{c}{$\nu$ CS}\\
R11&7.74690&1290.839259&155.1&23&8.11&--4.2$\,\pm\,$0.7&13.0$\,\pm\,$3.2&0.049$\,\pm\,$0.009&4.18$\,\pm\,$1.57$^\mathrm{b}$\\
R10&7.75585&1289.349312&129.3&21&8.04&--7.3$\,\pm\,$0.3&7.8$\,\pm\,$0.8&0.083$\,\pm\,$0.007&4.23$\,\pm\,$0.46$^\mathrm{b}$\\
R8&7.77404&1286.332863&84.6&17&7.88&--4.1$\,\pm\,$0.5&12.1$\,\pm\,$1.4&0.070$\,\pm\,$0.007&5.48$\,\pm\,$0.66$^\mathrm{b}$\\
R7&7.78327&1284.806427&65.8&15&7.79&--6.3$\,\pm\,$0.3&9.9$\,\pm\,$0.7&0.083$\,\pm\,$0.005&5.28$\,\pm\,$0.40$^\mathrm{b}$\\
R6&7.79261&1283.267891&49.4&13&7.68&--6.9$\,\pm\,$0.3&5.7$\,\pm\,$0.9&0.109$\,\pm\,$0.012&3.96$\,\pm\,$0.66$^\mathrm{b}$\\
R5&7.80203&1281.717289&35.3&11&7.56&--7.3$\,\pm\,$0.4&7.2$\,\pm\,$1.0&0.075$\,\pm\,$0.009&3.41$\,\pm\,$0.50$^\mathrm{b}$\\
R4&7.81156&1280.154654&23.5&9&7.4&--5.7$\,\pm\,$0.4&8.1$\,\pm\,$1.3&0.065$\,\pm\,$0.008&3.26$\,\pm\,$0.60$^\mathrm{b}$\\
R3&7.82118&1278.580017&14.1&7&7.2&--7.7$\,\pm\,$0.5&7.1$\,\pm\,$1.3&0.079$\,\pm\,$0.012&3.36$\,\pm\,$0.66$^\mathrm{b}$\\
R2&7.83089&1276.99341&7.1&5&6.91&--8.1$\,\pm\,$0.3&6.5$\,\pm\,$0.8&0.059$\,\pm\,$0.005&2.20$\,\pm\,$0.29$^\mathrm{b}$\\
\hline
\multicolumn{10}{c}{Pure Rotational \water}\\
$5_{4,1}$--$4_{1,4}$&25.94015&385.502803&323.5&27&0.0264&--8.0$\,\pm\,$0.4&17.0$\,\pm\,$1.0&0.423$\,\pm\,$0.013&341.22$\,\pm\,$30.58$^\mathrm{b*}$\\
\hline
\multicolumn{10}{c}{$\nu_2$ HCN}\\
R15e&13.17206&759.182446&510.2&186&1.278&--7.3$\,\pm\,$0.6&7.6$\,\pm\,$1.0&0.143$\,\pm\,$0.011&9.31$\,\pm\,$1.58$^\mathrm{b}$\\
&&&&&&1.3$\,\pm\,$0.8&8.4$\,\pm\,$1.4&0.108$\,\pm\,$0.009&7.76$\,\pm\,$1.62$^\mathrm{r}$\\
R13e&13.27497&753.297564&386.9&162&1.254&--7.9$\,\pm\,$0.3&6.0$\,\pm\,$0.7&0.247$\,\pm\,$0.046&12.54$\,\pm\,$3.54$^\mathrm{b}$\\
&&&&&&--0.5$\,\pm\,$1.6&12.0$\,\pm\,$3.0&0.139$\,\pm\,$0.015&14.14$\,\pm\,$4.73$^\mathrm{r}$\\
R12e&13.32706&750.352972&331.7&150&1.242&--7.5$\,\pm\,$0.1&7.9$\,\pm\,$0.3&0.420$\,\pm\,$0.015&28.00$\,\pm\,$1.26$^\mathrm{b*}$\\
R11e&13.37959&747.407049&280.7&138&1.231&--7.3$\,\pm\,$0.3&9.9$\,\pm\,$0.5&0.468$\,\pm\,$0.013&38.86$\,\pm\,$2.47$^\mathrm{b}$\\
&&&&&&1.4$\,\pm\,$0.7&7.9$\,\pm\,$1.1&0.138$\,\pm\,$0.021&9.12$\,\pm\,$2.33$^\mathrm{r}$\\
R10e&13.43256&744.459871&233.9&126&1.221&--7.9$\,\pm\,$0.2&8.1$\,\pm\,$0.4&0.574$\,\pm\,$0.017&38.59$\,\pm\,$2.37$^\mathrm{b}$\\
&&&&&&0.5$\,\pm\,$0.6&7.9$\,\pm\,$0.9&0.208$\,\pm\,$0.017&13.56$\,\pm\,$2.32$^\mathrm{r}$\\
R7e&13.59413&735.611573&119.1&90&1.194&--7.5$\,\pm\,$0.2&8.3$\,\pm\,$0.2&0.760$\,\pm\,$0.018&49.65$\,\pm\,$2.16$^\mathrm{b}$\\
&&&&&&0.9$\,\pm\,$0.7&8.9$\,\pm\,$0.9&0.183$\,\pm\,$0.014&12.83$\,\pm\,$2.13$^\mathrm{r}$\\
R6e&13.64889&732.660136&89.3&78&1.187&--7.2$\,\pm\,$0.2&6.7$\,\pm\,$0.8&0.698$\,\pm\,$0.091&35.76$\,\pm\,$8.44$^\mathrm{b*}$\\
&&&&&&--1.0$\,\pm\,$1.4&11.0$\,\pm\,$1.6&0.272$\,\pm\,$0.042&22.99$\,\pm\,$6.66$^\mathrm{r*}$\\
&&&&&&---&\textit{3.5$\,\pm\,$0.7}&\textit{0.183$\,\pm\,$0.073}&\textit{4.97$\,\pm\,$2.84$^\mathrm{b*}$}\\
R5e&13.70412&729.70782&63.8&66&1.183&--7.3$\,\pm\,$0.1&11.6$\,\pm\,$0.2&0.755$\,\pm\,$0.013&65.02$\,\pm\,$1.25$^\mathrm{b*}$\\
R4e&13.75980&726.7547&42.5&54&1.184&--7.3$\,\pm\,$0.1&8.7$\,\pm\,$0.2&0.780$\,\pm\,$0.013&47.97$\,\pm\,$1.81$^\mathrm{b}$\\
&&&&&&1.1$\,\pm\,$0.6&7.5$\,\pm\,$1.1&0.154$\,\pm\,$0.014&8.20$\,\pm\,$1.69$^\mathrm{r}$\\
R3e&13.81596&723.800848&25.5&42&1.19&--7.2$\,\pm\,$0.1&9.4$\,\pm\,$0.3&0.779$\,\pm\,$0.019&48.37$\,\pm\,$2.25$^\mathrm{b}$\\
&&&&&&2.1$\,\pm\,$0.3&7.6$\,\pm\,$0.7&0.212$\,\pm\,$0.018&10.63$\,\pm\,$1.85$^\mathrm{r}$\\
R1e&13.92969&717.89124&4.3&18&1.251&--7.2$\,\pm\,$0.1&8.5$\,\pm\,$0.1&0.700$\,\pm\,$0.007&28.23$\,\pm\,$0.54$^\mathrm{b}$\\
&&&&&&1.4$\,\pm\,$0.2&6.5$\,\pm\,$0.4&0.152$\,\pm\,$0.007&4.66$\,\pm\,$0.40$^\mathrm{r}$\\
R0e&13.98727&714.935627&0.0&6&1.371&--6.7$\,\pm\,$0.3&9.7$\,\pm\,$0.5&0.438$\,\pm\,$0.011&10.12$\,\pm\,$0.63$^\mathrm{b*}$\\
&&&&&&2.4$\,\pm\,$0.8&8.0$\,\pm\,$1.4&0.112$\,\pm\,$0.014&2.13$\,\pm\,$0.57$^\mathrm{r*}$\\
Q14e&14.01526&713.5078&446.5&174&2.026&--7.8$\,\pm\,$0.2&6.1$\,\pm\,$0.3&0.309$\,\pm\,$0.010&9.08$\,\pm\,$0.65$^\mathrm{b*}$\\
&&&&&&0.2$\,\pm\,$0.7&8.0$\,\pm\,$1.2&0.109$\,\pm\,$0.007&4.18$\,\pm\,$0.81$^\mathrm{r*}$\\
&&&&&&---&\textit{3.2$\,\pm\,$0.8}&\textit{0.086$\,\pm\,$0.032}&\textit{1.33$\,\pm\,$0.79$^\mathrm{r*}$}\\
Q13e&14.01926&713.304602&386.9&162&2.026&--7.3$\,\pm\,$0.2&7.2$\,\pm\,$0.3&0.419$\,\pm\,$0.015&14.52$\,\pm\,$0.99$^\mathrm{b}$\\
&&&&&&--0.1$\,\pm\,$0.7&7.6$\,\pm\,$1.0&0.134$\,\pm\,$0.013&4.87$\,\pm\,$1.00$^\mathrm{r}$\\
Q11e&14.02640&712.941315&280.7&138&2.027&--7.4$\,\pm\,$0.1&8.1$\,\pm\,$0.3&0.592$\,\pm\,$0.011&23.04$\,\pm\,$0.75$^\mathrm{b*}$\\
&&&&&&2.7$\,\pm\,$0.7&5.9$\,\pm\,$1.8&0.140$\,\pm\,$0.011&3.96$\,\pm\,$1.23$^\mathrm{r*}$\\
&&&&&&---&\textit{3.2$\,\pm\,$1.0}&\textit{0.117$\,\pm\,$0.049}&\textit{1.78$\,\pm\,$1.21$^\mathrm{r*}$}\\
Q10e&14.02955&712.781294&233.9&126&2.027&--7.0$\,\pm\,$0.1&8.5$\,\pm\,$0.2&0.673$\,\pm\,$0.007&27.23$\,\pm\,$0.53$^\mathrm{b}$\\
&&&&&&---&\textit{4.6$\,\pm\,$0.6}&\textit{0.117$\,\pm\,$0.009}&\textit{2.58$\,\pm\,$0.37$^\mathrm{b}$}\\
Q9e&14.03241&712.635726&191.4&114&2.028&--7.4$\,\pm\,$0.1&12.4$\,\pm\,$0.2&0.692$\,\pm\,$0.010&41.06$\,\pm\,$0.68$^\mathrm{b}$\\
Q8e&14.03500&712.504639&153.1&102&2.028&--7.0$\,\pm\,$0.3&8.3$\,\pm\,$0.3&0.772$\,\pm\,$0.018&30.55$\,\pm\,$1.60$^\mathrm{b*}$\\
&&&&&&1.1$\,\pm\,$0.5&7.7$\,\pm\,$0.8&0.189$\,\pm\,$0.017&6.97$\,\pm\,$1.22$^\mathrm{r*}$\\
&&&&&&---&\textit{3.8$\,\pm\,$1.1}&\textit{0.097$\,\pm\,$0.041}&\textit{1.76$\,\pm\,$1.18$^\mathrm{b*}$}\\
Q7e&14.03729&712.388056&119.1&90&2.028&--6.9$\,\pm\,$0.1&8.9$\,\pm\,$0.2&0.801$\,\pm\,$0.011&34.08$\,\pm\,$0.71$^\mathrm{b}$\\
&&&&&&---&\textit{3.4$\,\pm\,$0.8}&\textit{0.081$\,\pm\,$0.017}&\textit{1.29$\,\pm\,$0.43$^\mathrm{b}$}\\
Q6e&14.03930&712.286&89.3&78&2.028&--7.0$\,\pm\,$0.1&9.7$\,\pm\,$0.3&0.750$\,\pm\,$0.017&34.76$\,\pm\,$1.11$^\mathrm{b}$\\
&&&&&&---&\textit{4.4$\,\pm\,$0.6}&\textit{0.180$\,\pm\,$0.019}&\textit{3.75$\,\pm\,$0.63$^\mathrm{b}$}\\
P2e&14.16297&706.0664&12.8&30&0.6576&--7.5$\,\pm\,$0.2&7.8$\,\pm\,$0.3&0.352$\,\pm\,$0.008&65.46$\,\pm\,$3.32$^\mathrm{b}$\\
&&&&&&0.7$\,\pm\,$0.3&7.5$\,\pm\,$0.6&0.173$\,\pm\,$0.009&30.99$\,\pm\,$3.46$^\mathrm{r}$\\
P3e&14.22254&703.109429&25.5&42&0.778&--6.8$\,\pm\,$0.3&11.1$\,\pm\,$1.0&0.429$\,\pm\,$0.029&79.92$\,\pm\,$8.57$^\mathrm{b*}$\\
\hline
\multicolumn{10}{c}{$\nu_2$ \hcniso}\\
R8e&13.66420&731.839651&149.2&204&1.157&--5.7$\,\pm\,$0.3&7.0$\,\pm\,$0.8&0.054$\,\pm\,$0.005&3.05$\,\pm\,$0.40$^\mathrm{b}$\\
R6e&13.77224&726.098321&87.0&156&1.143&--6.5$\,\pm\,$0.2&8.2$\,\pm\,$0.6&0.083$\,\pm\,$0.005&5.30$\,\pm\,$0.44$^\mathrm{b}$\\
R1e&14.05041&711.72312&4.1&36&1.206&--7.5$\,\pm\,$0.4&8.7$\,\pm\,$0.9&0.059$\,\pm\,$0.005&2.46$\,\pm\,$0.27$^\mathrm{b}$\\
\hline
\multicolumn{10}{c}{$2\nu_2$ HCN}\\
P10e&7.23320&1382.513753&233.9&126&1.117&--6.3$\,\pm\,$0.3&9.6$\,\pm\,$0.6&0.192$\,\pm\,$0.006&129.23$\,\pm\,$9.42$^\mathrm{b}$\\
&&&&&&2.4$\,\pm\,$0.6&6.8$\,\pm\,$1.1&0.072$\,\pm\,$0.010&33.96$\,\pm\,$8.18$^\mathrm{r}$\\
P12e&7.26273&1376.892237&331.7&150&1.1&--5.3$\,\pm\,$1.1&7.8$\,\pm\,$1.9&0.108$\,\pm\,$0.015&58.49$\,\pm\,$19.94$^\mathrm{b}$\\
&&&&&&1.2$\,\pm\,$1.8&6.1$\,\pm\,$2.5&0.050$\,\pm\,$0.025&21.17$\,\pm\,$17.72$^\mathrm{r}$\\
\hline
\multicolumn{10}{c}{$\nu_2$ HNC}\\
R10e&20.16417&495.92916&239.3&126&2.197&--5.9$\,\pm\,$0.5&8.4$\,\pm\,$1.3&0.026$\,\pm\,$0.004&0.30$\,\pm\,$0.07$^\mathrm{b}$\\
R9e&20.28742&492.91629&195.8&114&2.163&--7.8$\,\pm\,$0.3&12.9$\,\pm\,$0.9&0.043$\,\pm\,$0.002&0.75$\,\pm\,$0.06$^\mathrm{b}$\\
R8e&20.41225&489.902&156.6&102&2.131&--6.2$\,\pm\,$0.3&9.2$\,\pm\,$1.0&0.042$\,\pm\,$0.003&0.51$\,\pm\,$0.08$^\mathrm{b}$\\
R7e&20.53867&486.88636&121.8&90&2.102&--7.7$\,\pm\,$0.3&12.4$\,\pm\,$1.0&0.048$\,\pm\,$0.003&0.78$\,\pm\,$0.07$^\mathrm{b}$\\
R5e&20.79645&480.85136&65.3&66&2.055&--8.2$\,\pm\,$0.2&9.3$\,\pm\,$0.5&0.071$\,\pm\,$0.003&0.81$\,\pm\,$0.04$^\mathrm{b}$\\
R3e&21.06097&474.8119&26.1&42&2.037&--8.0$\,\pm\,$0.2&11.3$\,\pm\,$0.6&0.081$\,\pm\,$0.003&1.01$\,\pm\,$0.08$^\mathrm{b}$\\
R2e&21.19584&471.7907&13.1&30&2.053&--8.1$\,\pm\,$0.2&12.3$\,\pm\,$0.8&0.076$\,\pm\,$0.003&0.92$\,\pm\,$0.08$^\mathrm{b}$\\
R1e&21.33249&468.76863&4.4&18&2.113&--8.0$\,\pm\,$0.2&14.6$\,\pm\,$0.7&0.069$\,\pm\,$0.003&0.79$\,\pm\,$0.05$^\mathrm{b}$\\
R0e&21.47094&465.74576&0.0&6&2.298&--8.3$\,\pm\,$0.3&11.9$\,\pm\,$1.3&0.037$\,\pm\,$0.003&0.17$\,\pm\,$0.03$^\mathrm{b}$\\
Q12e&21.53525&464.35494&339.3&150&3.404&--5.9$\,\pm\,$0.6&8.6$\,\pm\,$1.6&0.013$\,\pm\,$0.002&0.09$\,\pm\,$0.02$^\mathrm{b}$\\
Q11e&21.54687&464.10446&287.1&138&3.4&--8.1$\,\pm\,$0.5&7.7$\,\pm\,$2.5&0.022$\,\pm\,$0.005&0.13$\,\pm\,$0.07$^\mathrm{b}$\\
Q9e&21.56727&463.66551&195.8&114&3.393&--7.7$\,\pm\,$0.3&10.5$\,\pm\,$1.0&0.037$\,\pm\,$0.002&0.30$\,\pm\,$0.04$^\mathrm{b}$\\
Q8e&21.57604&463.47714&156.6&102&3.39&--7.7$\,\pm\,$0.2&12.6$\,\pm\,$1.0&0.055$\,\pm\,$0.003&0.54$\,\pm\,$0.06$^\mathrm{b}$\\
Q6e&21.59068&463.16287&91.4&78&3.385&--8.4$\,\pm\,$0.2&13.2$\,\pm\,$0.5&0.078$\,\pm\,$0.003&0.81$\,\pm\,$0.04$^\mathrm{b}$\\
Q5e&21.59654&463.03705&65.3&66&3.383&--8.3$\,\pm\,$0.1&10.7$\,\pm\,$0.4&0.082$\,\pm\,$0.002&0.69$\,\pm\,$0.04$^\mathrm{b}$\\
Q4e&21.60144&462.93214&43.5&54&3.382&--8.4$\,\pm\,$0.2&11.6$\,\pm\,$0.6&0.080$\,\pm\,$0.003&0.72$\,\pm\,$0.05$^\mathrm{b}$\\
Q3e&21.60536&462.84818&26.1&42&3.38&--8.9$\,\pm\,$0.1&11.7$\,\pm\,$0.6&0.080$\,\pm\,$0.003&0.73$\,\pm\,$0.06$^\mathrm{b}$\\
Q2e&21.60830&462.78519&13.1&30&3.379&--8.1$\,\pm\,$0.2&12.8$\,\pm\,$1.2&0.071$\,\pm\,$0.004&0.71$\,\pm\,$0.10$^\mathrm{b}$\\
Q1e&21.61026&462.74319&4.4&18&3.379&--8.9$\,\pm\,$0.2&10.3$\,\pm\,$0.7&0.057$\,\pm\,$0.003&0.46$\,\pm\,$0.04$^\mathrm{b}$\\
P3e&22.04352&453.64798&26.1&42&1.269&--6.8$\,\pm\,$0.3&12.6$\,\pm\,$1.5&0.031$\,\pm\,$0.003&1.09$\,\pm\,$0.20$^\mathrm{b}$\\
P5e&22.34156&447.5964&65.3&66&1.352&--7.0$\,\pm\,$0.2&10.6$\,\pm\,$0.5&0.041$\,\pm\,$0.001&0.94$\,\pm\,$0.05$^\mathrm{b}$\\
P6e&22.49363&444.57022&91.4&78&1.353&--5.3$\,\pm\,$0.4&14.4$\,\pm\,$2.0&0.029$\,\pm\,$0.003&0.86$\,\pm\,$0.19$^\mathrm{b}$\\
P7e&22.64781&441.54388&121.8&90&1.345&--9.1$\,\pm\,$0.3&10.5$\,\pm\,$1.3&0.031$\,\pm\,$0.003&0.65$\,\pm\,$0.11$^\mathrm{b}$\\
P8e&22.80411&438.51747&156.6&102&1.33&--7.9$\,\pm\,$0.3&11.8$\,\pm\,$1.1&0.028$\,\pm\,$0.002&0.64$\,\pm\,$0.08$^\mathrm{b}$\\
\hline
\multicolumn{10}{c}{$\nu_2$ \amm}\\
$^{\mathrm{Q}}$P(4,0)a&11.71210&853.817812&285.6&108&6.559&--6.4$\,\pm\,$0.3&5.2$\,\pm\,$0.7&0.222$\,\pm\,$0.026&3.79$\,\pm\,$0.44$^\mathrm{b}$\\
$^{\mathrm{Q}}$P(4,1)a&11.71580&853.548188&280.3&54&6.255&--6.7$\,\pm\,$0.5&4.7$\,\pm\,$1.1&0.145$\,\pm\,$0.031&2.32$\,\pm\,$0.50$^\mathrm{b*}$\\
$^{\mathrm{Q}}$P(4,2)a&11.72711&852.72474&264.4&54&5.009&--5.3$\,\pm\,$1.1&8.6$\,\pm\,$4.8&0.153$\,\pm\,$0.063&5.59$\,\pm\,$4.50$^\mathrm{b}$\\
$^{\mathrm{Q}}$P(4,3)a&11.74637&851.326935&237.8&108&2.928&--6.2$\,\pm\,$0.4&4.7$\,\pm\,$1.2&0.201$\,\pm\,$0.041&6.84$\,\pm\,$1.85$^\mathrm{b}$\\
$^{\mathrm{Q}}$P(6,3)s&11.79832&847.578094&550.4&156&4.36&--8.9$\,\pm\,$1.3&10.3$\,\pm\,$3.8&0.049$\,\pm\,$0.014&2.24$\,\pm\,$0.95$^\mathrm{b}$\\
\enddata
\end{deluxetable*}

\begin{deluxetable*}{rrrrrrrrrrl}
\tablecaption{Observed transitions and inferred parameters for molecular emission lines. Wavelength and wavenumber are the rest value for each transition, $E_u$ is the energy level of the lower state, $k_B$ is the Boltzmann constant, $g_u$ is the upper statistical weight, $A$ is the Einstein coefficient, \vlsr\ is the observed local standard of rest velocity, \vfwhm\ is the observed full-width half-maximum, $S_{Jy}\times S_{\nu0}$ is the amplitude in units of Jy arcsec$^{-2}$, and $N_u$ is the observed column density of the transition as calculated in Equation \ref{eqn:nu}. Superscripts in the final column for \htwo\ refer to which velocity component the transition is likely associated with: $^b$blue clump and $^r$red clump. Data in the first six columns are from the HITRAN database \citep{Gordon2017} for \htwo\ and \water, and from the ExoMol database for SiO \citep{Tennyson2012}. \label{tab:emilines}}
\tablehead{
\colhead{Transition} & \colhead{Wavelength} &\colhead{Wavenumber} & \colhead{$E_u/k_B$} & \colhead{$g_u$} & \colhead{$A$} & \colhead{\vlsr} & \colhead{\vfwhm} & \colhead{$S_{\mathrm{Ja}}\times S_{\nu0}$} & \colhead{$N_u$} \\
\colhead{} & \colhead{(\micron)} & \colhead{(\ci)} & \colhead{(K)} & \colhead{} & \colhead{(\si)} & \colhead{(\kms)} & \colhead{(\kms)} & \colhead{(Jy arcsec$^{-2}$)} & \colhead{}}
\startdata
\hline
\multicolumn{9}{c}{Pure Rotational \htwo}&($\times10^{19}$\csi)\\
S(1)&17.03485&587.032&1015.1&21&$4.758\times10^{-10}$&--10.7$\,\pm\,$2.6&19.2$\,\pm\,$5.1&0.740$\,\pm\,$0.089&8.55$\,\pm\,$2.49$^\mathrm{b}$\\
&&&&&&0.5$\,\pm\,$0.5&9.3$\,\pm\,$1.8&0.795$\,\pm\,$0.258&4.45$\,\pm\,$1.68$^\mathrm{r}$\\
\hline
\multicolumn{9}{c}{$\nu_2$ \water}&($\times10^{9}$\csi)\\
$8_{3,5}$--$9_{4,6}$&7.51934&1329.90473&3842.7&17&4.377&8.9$\,\pm\,$0.3&11.1$\,\pm\,$1.0&0.791$\,\pm\,$0.053&5.77$\,\pm\,$0.63\\
$9_{3,6}$--$10_{4,7}$&7.55667&1323.33442&4179.2&57&3.903&9.9$\,\pm\,$0.4&12.5$\,\pm\,$1.2&0.900$\,\pm\,$0.066&8.24$\,\pm\,$0.99\\
$5_{1,5}$--$6_{2,4}$&7.57544&1320.05555&2766.5&11&0.6821&8.1$\,\pm\,$0.1&9.3$\,\pm\,$0.3&3.381$\,\pm\,$0.096&131.77$\,\pm\,$5.98\\
$6_{2,5}$--$7_{3,4}$&7.58191&1318.92943&3109.6&39&2.174&7.2$\,\pm\,$0.1&10.7$\,\pm\,$0.2&10.005$\,\pm\,$0.171&141.65$\,\pm\,$3.99\\
$7_{4,3}$--$8_{5,4}$&7.59317&1316.9724&3700.7&45&7.502&8.1$\,\pm\,$0.1&9.4$\,\pm\,$0.3&3.315$\,\pm\,$0.081&11.88$\,\pm\,$0.48\\
$7_{4,4}$--$8_{5,3}$&7.61269&1313.59641&3696.9&15&7.423&10.0$\,\pm\,$0.2&9.2$\,\pm\,$0.6&1.264$\,\pm\,$0.067&4.50$\,\pm\,$0.39\\
$4_{0,4}$--$5_{3,3}$&7.61335&1313.48301&2614.9&9&0.2621&8.5$\,\pm\,$0.1&8.3$\,\pm\,$0.3&2.504$\,\pm\,$0.061&227.16$\,\pm\,$9.05\\
$7_{5,2}$--$8_{6,3}$&7.61872&1312.55566&3919.5&45&10.09&9.8$\,\pm\,$0.2&11.8$\,\pm\,$0.5&2.202$\,\pm\,$0.070&7.40$\,\pm\,$0.40\\
$7_{5,3}$--$8_{6,2}$&7.61963&1312.39954&3919.3&15&10.09&8.7$\,\pm\,$0.3&9.7$\,\pm\,$0.8&0.928$\,\pm\,$0.056&2.56$\,\pm\,$0.26\\
$7_{3,5}$--$8_{4,4}$&7.64421&1308.17886&3510.6&15&4.296&9.8$\,\pm\,$0.2&11.5$\,\pm\,$0.4&1.707$\,\pm\,$0.053&13.05$\,\pm\,$0.65\\
$3_{1,2}$--$4_{4,1}$&7.78631&1284.30575&2550.1&21&0.01255&8.3$\,\pm\,$0.3&13.7$\,\pm\,$0.9&0.651$\,\pm\,$0.031&2032.20$\,\pm\,$163.19\\
$5_{0,5}$--$6_{3,4}$&7.86295&1271.78782&2763.6&33&0.2945&8.7$\,\pm\,$0.1&10.2$\,\pm\,$0.4&2.936$\,\pm\,$0.080&291.17$\,\pm\,$13.26\\
$6_{1,6}$--$7_{2,5}$&7.93435&1260.34347&2939.1&39&0.4262&8.8$\,\pm\,$0.2&10.3$\,\pm\,$0.7&1.532$\,\pm\,$0.073&105.75$\,\pm\,$8.48\\
\hline
\multicolumn{9}{c}{$\nu$ SiO}&($\times10^{9}$\csi)\\
R27&7.89890&1265.998663&2608.4&57&3.2883&11.1$\,\pm\,$0.5&12.3$\,\pm\,$1.8&0.511$\,\pm\,$0.053&5.47$\,\pm\,$0.96\\
R25&7.91344&1263.672931&2494.8&53&3.288&10.1$\,\pm\,$0.6&13.8$\,\pm\,$2.1&0.532$\,\pm\,$0.059&6.41$\,\pm\,$1.22\\
R24&7.92083&1262.494014&2441.1&51&3.2874&11.9$\,\pm\,$0.4&14.8$\,\pm\,$1.4&0.585$\,\pm\,$0.041&7.57$\,\pm\,$0.90\\
R23&7.92830&1261.304429&2389.5&49&3.2865&10.4$\,\pm\,$0.3&13.6$\,\pm\,$1.1&0.712$\,\pm\,$0.042&8.42$\,\pm\,$0.82\\
R22&7.93585&1260.104199&2339.9&47&3.2853&12.1$\,\pm\,$0.4&10.6$\,\pm\,$1.6&0.831$\,\pm\,$0.088&7.71$\,\pm\,$1.44\\
R21&7.94348&1258.89335&2292.4&45&3.2837&9.3$\,\pm\,$0.5&12.8$\,\pm\,$1.5&0.733$\,\pm\,$0.064&8.19$\,\pm\,$1.21\\
R20&7.95120&1257.671903&2246.9&43&3.2817&10.2$\,\pm\,$0.4&14.9$\,\pm\,$1.4&0.771$\,\pm\,$0.050&10.00$\,\pm\,$1.12\\
R18&7.96687&1255.197323&2162.1&39&3.2762&9.2$\,\pm\,$0.2&12.8$\,\pm\,$0.8&0.958$\,\pm\,$0.042&10.71$\,\pm\,$0.80\\
R17&7.97484&1253.944235&2122.9&37&3.2727&7.9$\,\pm\,$0.4&13.4$\,\pm\,$1.2&0.875$\,\pm\,$0.048&10.27$\,\pm\,$1.10\\
R16&7.98288&1252.680647&2085.6&35&3.2685&10.1$\,\pm\,$0.4&17.1$\,\pm\,$2.3&0.942$\,\pm\,$0.100&14.12$\,\pm\,$2.42\\
R15&7.99101&1251.406585&2050.5&33&3.2636&9.0$\,\pm\,$0.2&13.8$\,\pm\,$0.6&1.175$\,\pm\,$0.042&14.27$\,\pm\,$0.80\\
R14&7.99922&1250.12207&2017.4&31&3.2578&6.7$\,\pm\,$0.4&13.3$\,\pm\,$1.8&1.105$\,\pm\,$0.105&12.94$\,\pm\,$2.15\\
\enddata
\end{deluxetable*}

\begin{deluxetable*}{rrrrrrrl}
\setlength{\tabcolsep}{4pt}
\tablecaption{Observed transitions and inferred parameters for atomic forbidden transitions. Position is the position of the line as explained in \S \ref{sec:ation} (on-source, off-source, and extended blue wing of off-source), wavelength and wavenumber are the rest value for each transition, $A$ is the Einstein coefficient, \vlsr\ is the observed local standard of rest velocity, \vfwhm\ is the observed full-width half-maximum, emission peak is the amplitude in units of Jy arcsec$^{-2}$, and $N_u$ is the observed column density of the transition. Wavelength, wavenumber, and $A$ are from the NIST ASD database \citep{Kramida2021}. The \vlsr\ errors are statistical only, and do not include the uncertainties in the rest wavelength.} \label{tab:ationlines}
\tablehead{
\colhead{Position}	&	\colhead{Wavelength}			&	\colhead{Wavenumber}			&	\colhead{$A$}	&	\colhead{\vlsr}			&	\colhead{\vfwhm}			&	\colhead{Emission Peak}			&	\colhead{$N_u$}			\\
&	\colhead{(\micron)}			&	\colhead{(\ci)}			&	\colhead{($\times10^{-3}$ \si)}	&	\colhead{(\kms)}			&	\colhead{(\kms)}			&	\colhead{(Jy arcsec$^{-2}$)}			&	\colhead{($\times10^{14}$\csi)}			\\}
\startdata
\hline
\multicolumn{8}{c}{[NeII]}\\
On	&	12.81355	$\,\pm\,$	0.00002	&	780.42397	$\,\pm\,$	0.00122	&	8.59	&	$-3.44	\,\pm\,$	0.06	&	11.81	$\,\pm\,$	0.16	&	60.247	$\,\pm\,$	0.653	&	2.37	$\,\pm\,$	0.02	\\
Off	&				&				&		&	$-5.06	\,\pm\,$	0.04	&	8.93	$\,\pm\,$	0.08	&	53.749	$\,\pm\,$	0.441	&	1.60	$\,\pm\,$	0.01	\\
\hline
\multicolumn{8}{c}{[SIII]}\\
On	&	18.713	$\,\pm\,$	0.001	&	534.388	$\,\pm\,$	0.029	&	2.06	&	$-6.48	\,\pm\,$	0.02	&	14.12	$\,\pm\,$	0.05	&	151.694	$\,\pm\,$	0.464	&	29.74	$\,\pm\,$	0.07	\\
Off	&				&				&		&	$-7.75	\,\pm\,$	0.04	&	10.89	$\,\pm\,$	0.10	&	120.605	$\,\pm\,$	0.855	&	18.24	$\,\pm\,$	0.10	\\
Off: Wing	&				&				&		&	$-19.65	\,\pm\,$	0.05	&	9.53	$\,\pm\,$	0.14	&	18.087	$\,\pm\,$	0.164	&	2.39	$\,\pm\,$	0.02	\\
\hline
\multicolumn{8}{c}{[SI]}\\
On	&	25.245	$\,\pm\,$	0.001	&	396.118	$\,\pm\,$	0.016	&	1.40	&	33.18	$\,\pm\,$	0.46	&	12.18	$\,\pm\,$	1.00	&	3.576	$\,\pm\,$	0.203	&	0.89	$\,\pm\,$	0.04	\\
Off	&				&				&		&	50.89	$\,\pm\,$	0.17	&	11.24	$\,\pm\,$	0.18	&	16.340	$\,\pm\,$	0.203	&	3.75	$\,\pm\,$	0.04	\\
\hline
\multicolumn{8}{c}{[FeII]}\\
Off	&	25.98839	$\,\pm\,$	0.00002	&	384.78721	$\,\pm\,$	0.00030	&	2.13	&	51.53	$\,\pm\,$	0.05	&	5.41	$\,\pm\,$	0.20	&	15.599	$\,\pm\,$	0.362	&	1.13	$\,\pm\,$	0.02	\\
\enddata
\end{deluxetable*}


\end{document}